\documentclass[preprint]{aastex62}

\usepackage{latexsym}
\usepackage{amssymb}
\usepackage{apjfonts}
\usepackage{epsfig}
\usepackage{float}
\usepackage{graphicx}
\usepackage{amsmath}
\usepackage{natbib}
\usepackage{multirow}
\usepackage{bm}

\begin{document}

\title{Characteristics of Long Gamma-Ray bursts in the Comoving Frame}


\correspondingauthor{Fu-Wen Zhang}
\email{fwzhang@pmo.ac.cn}

\author{Liang Xue}
\affil{College of Science, Guilin University of Technology, Guilin 541004, China}

\author{Fu-Wen Zhang}
\affiliation{College of Science, Guilin University of Technology, Guilin 541004, China}
\affiliation{Key Laboratory of Dark Matter and Space Astronomy, Chinese Academy of Sciences, Nanjing 210008, China}

\author{Si-Yuan Zhu}
\affiliation{College of Science, Guilin University of Technology, Guilin 541004, China}

\begin{abstract}

We compile a sample of 93 long gamma-ray bursts (GRBs) from Fermi satellite and 131 from Konus-Wind, which have measured redshifts and well determined spectra, and estimate their pseudo Lorentz factors (${\Gamma_0}$) using the tight ${L_{{\rm{iso}}}}$-${E_{\rm p}}$-${\Gamma_0}$ correlation. The statistical properties and pair correlations of temporal and spectral parameters are studied in the observer frame, rest frame and comoving frame, respectively. We find that the distributions of the duration, peak energy, isotropic energy and luminosity in the different frames are basically lognormal, and their distributions in the comoving frame are narrow, clustering around $T'_{\rm 90}\sim 4000$ s, $E'_{\rm p,c}\sim 0.7$ keV, $E'_{\rm iso,c} \sim 8\times10^{49}$ erg and $L'_{\rm iso,c}\sim 2.5\times10^{46}$ erg s$^{-1}$, where the redshift evolution effect has been taken into account. We also find that the values of ${\Gamma_0}$ are broadly distributed between few tens and several hundreds with median values $\sim 270$. We further analyze the pair correlations of all the quantities, and well confirm ${E_{{\rm{iso}}}}$-${E_{\rm p}}$, ${L_{{\rm{iso}}}}$-${E_{\rm p}}$, ${L_{{\rm{iso}}}}$-${\Gamma_0}$ and $E_{\rm{iso}}$-${\Gamma_0}$ relations, and find that the corresponding relations in the comoving frame do still exist, but have large dispersions. This suggests not only the well-known spectrum-energy relations are intrinsic correlations, but also the observed correlations are governed by the Doppler effect. In addition, the peak energies of long GRBs are independent of durations both in the rest frame and in the comoving frame. And there is a weak anticorrelation between the peak energy and Lorentz factor.

\end{abstract}


\section{Introduction} \label{sec:intro}

Gamma-ray bursts (GRBs) are the most luminous explosions in the universe \citep[see e.g.,][]{2004RvMP...76.1143P,2006RPPh...69.2259M,2007ChJAA...7....1Z}. According to the traditional classification schemes, GRBs can be divided into long bursts (longer than 2.0 seconds) and short ones (shorter than 2.0 seconds), based on the well-known bimodal distribution of their durations \citep{1993ApJ...413L.101K}. The observed spectra of GRBs are typically fitted by an empirical smoothly jointed broken power-law function, the so-called Band function \citep{1993ApJ...413..281B}, characterized by a peak energy ($E_{\rm p,obs}$) in the $\nu F_{\nu}$ spectrum. A variety of empirical relations involved the rest frame peak energy, $E_{\rm p}=E_{\rm p,obs} (1+z)$, have been discussed. For example, \citet{2002A&A...390...81A} found a tight correlation between the rest frame isotropic gamma-ray energy $E_{\rm iso}$ and $E_{\rm p}$ (so called Amati relation). A similar correlation was found between the isotropic luminosity $L_{\rm iso}$ and $E_{\rm p}$ (Wei \& Gao 2003; Yonetoku et al. 2004, so called Yonetoku relation). \citet{2004ApJ...616..331G} found a tighter relation by replacing $E_{\rm iso}$ with the geometrically corrected energy of the GRB jets. These correlations not only give important clues to understand GRB physics (even though some correlations still lack straightforward theoretical interpretations; see e.g., Fan et al. 2012), but also provide new ways to constrain cosmological parameters (e.g., Wang et al. 2015 for a recent review).

The intrinsic properties of GRBs with measured redshifts have been extensively investigated, but the additional correction of the initial Lorentz factor ($\Gamma_0$) is still to be considered. $\Gamma_0$ is a key parameter to better understand the physics of GRBs. Several methods have been developed to estimate the value of $\Gamma_0$ based on different hypotheses \citep{2001ApJ...555..540L,2007ApJ...664L...1P,2007A&A...469L..13M,2008MNRAS.384L..11G,2010ApJ...725.2209L,2010ApJ...709L.172R,2015ApJ...800L..23Z}. Although the number of GRBs with estimated $\Gamma_0$ is still limited, several interesting correlations involving $\Gamma_0$ have been reported in the literature. For example, Liang et al. (2010) discovered a correlation between $\Gamma_0$ and $E_{\rm iso}$ (see also Ghirlanda et al. 2012 ). \citet{2012ApJ...751...49L} and Fan et al. (2012) showed that $L_{\rm iso}$ is also correlated with $\Gamma_0$. \citet{2012MNRAS.420..483G} further analyzed the spectrum-energy correlation in the comoving frame and found that these correlations still exist (see also Ghirlanda et al. 2018). Recently, \citet{2015ApJ...813..116L} reported a {\bf strong} three-parameter correlation among $L_{\rm iso}(E_{\rm iso})$, $E_{\rm p}$ and $\Gamma_0$. This correlaion provides a simple way to estimate the initial Lorentz factor of GRBs, and gives us an opportunity to analyze the intrinsic properties of GRBs in the comoving frame with a reasonable large sample.

In this paper, we compile a long GRB sample with known redshift and well measured spectra observed by two satellites, Fermi \citep{2009ApJ...697.1071A} and Konus-Wind (hereafter KW, Aptekar et al. 1995), and obtain the pseudo values of $\Gamma_0$ for 225 long GRBs using the ${L_{{\rm{iso}}}}$-${E_{\rm p}}$-${\Gamma_0}$ correlation found by \citet{2015ApJ...813..116L}. We analyze the distributions of temporal and spectral characteristics of long GRBs in different frames, and study the pair correlations of all quantities. The data selection and sample are presented in section 2. The results are shown in section 3. Summary and discussion are given in section 4. The cosmological constants in this paper are $h=0.71$, $\Omega_M=0.27$, $\Omega _{\Lambda}=0.73$. The symbolic notation $Q_n=Q/10^n$ is adopted.

\section{Sample and Data Analysis} \label{sec:data}

In order to investigate the intrinsic properties of GRBs, a sample of GRBs with measured redshift $z$ and well constrained spectral parameters is needed. We concentrate our analysis on the GRBs detected by the Fermi and KW detectors. Some GRBs observed by other detectors such as BeppoSAX, HETE-2, Integral, are not considered in this paper to minimize the influence of different instruments (with different sensitivities and energy bands). The Fermi GRBs are selected from the Fermi catalog\footnote{https://heasarc.gsfc.nasa.gov/W3Browse/fermi/fermigbrst.html} till the end of December 2017, in total there are 2233 events. Firstly, we choose the bursts with well measured spectral parameters and known redshifts.\footnote{http://www.mpe.mpg.de/~jcg/grbgen.html} The traditional criteria of long GRBs ($T_{\rm 90} >$ 2 s) has been used and then we have 95 long GRBs. Among them, GRB 100816A and GRB 170817A with durations of 2.045 s and 2.048 s, have been classified as short bursts, which are thus excluded from our sample. Our Fermi sample thus contains 93 GRBs, as summarized in Table 1. Using the same selection criteria, we have a KW sample consisting of 131 long GRBs (see Table 2). The KW data are mainly taken from \citet{2017ApJ...850..161T} and the GCN reports (\emph{GCN Circular Archive}). In addition, we note that there are 46 overlapped GRBs in the two samples.

Below we estimate the rest frame parameters. In this paper, both $E_{\rm iso}$ and $L_{\rm iso}$ are corrected to the energy band of $1-10^{4}$ keV in the rest frame. The isotropic energy $E_{\rm iso}$ is estimated as

\begin{equation}
  {E_{{\rm{iso}}}} = 4\pi D_L^2{S_\gamma }k/(1 + z),
\end{equation}
where $S_{\rm \gamma}$ is the time integral fluence, in units of erg cm$^{-2}$, $D_{\rm L}$ is the luminosity distance, and $k$ is the $k$-correction factor defined as

\begin{equation}
  k = \frac{{\int_{1 keV/(1 + z)}^{{{10}^4 keV}/(1 + z)} {EN(E)dE} }}{{\int_{{e_{\min }}}^{{e_{\max }}} {EN(E)dE} }}.
\end{equation}
Here $e_{\rm min}$ and $e_{\rm max}$ are the observational energy band of fluence, $N(E)$ denotes the photon spectrum of GRBs. All the GRB spectra are assumed to be a Band function (Band et al. 1993), defined as

\begin{equation}
  N(E) = \left\{ {\begin{array}{*{20}{l}}
{A{{\left( {\frac{E}{{100keV}}} \right)}^\alpha }\exp \left( { - \frac{E}{{{E_0}}}} \right)}\\
{A{{\left( {\frac{E}{{100keV}}} \right)}^\beta }{{\left[ {\frac{{\left( {\alpha - \beta} \right){E_0}}}{{100keV}}} \right]}^{\alpha - \beta}}\exp \left({\beta-\alpha}\right)},
\end{array}} \right.
\end{equation}
where $\alpha$ is the low energy photon index, and $\beta$ is the high energy photon index. $E_{\rm 0}$ is the break energy, $E_{\rm 0} = E_{\rm p}/(2+\alpha)$. The isotropic luminosity $L_{\rm iso}$ is estimated as

\begin{equation}
  {L_{iso}} = 4\pi D_L^2{F_p}k,
\end{equation}
here $F_{\rm p}$ is the peak flux, in units of erg cm$^{-2}$ s$^{-1}$. The rest frame peak energy $E_{\rm p}$ and duration $T_{\rm 90}$ are estimated as

\begin{equation}
  {E_{\rm p}} = {E_{\rm p,obs}}(1 + z),
\end{equation}
and
\begin{equation}
  {T_{\rm 90}} = {T_{\rm 90,obs}}/(1 + z).
\end{equation}
All the errors of these quantities are calculated by the error propagation formula. The initial Lorentz factor $\Gamma_0$ can be derived from $L_{\rm iso}$ and $E_{\rm p}$ \citep{2015ApJ...813..116L} as

\begin{equation}
  {\Gamma _0} = {10^{3.33}}L_{\rm iso,52}^{0.46}{{E_{\rm p}}^{ - 0.43}}.
\end{equation}

Using the pseudo values of $\Gamma_0$, we can further estimate the comoving frame isotropic energy $E'_{\rm iso}$, isotropic luminosity $L'_{\rm iso}$, spectrum peak energy $E'_{\rm p}$ and duration $T'_{\rm 90}$ \citep{2012MNRAS.420..483G}, which are defied as

\begin{equation}
  {L'_{\rm iso}} = 3/4{L_{\rm iso}}/\Gamma _0^2,
\end{equation}

\begin{equation}
  {E'_{\rm iso}} = {E_{\rm iso}}/\Gamma _0,
\end{equation}

\begin{equation}
  {E'_{\rm p}} = {E_{\rm p}}/\Gamma _0,
\end{equation}
and
\begin{equation}
  {T'_{\rm 90}} = {T_{\rm 90}}\Gamma _0.
\end{equation}
The derived temporal and spectral results of Fermi sample and KW sample in the rest frame and comoving frame are listed in Table 3 and Table 4, respectively.

\section{Results} \label{sec:resu}

\subsection{Observer Frame} \label{subsec:obs}

We analyze the distributions of the observed duration ($T_{\rm 90,obs}$), spectrum peak energy ($E_{\rm p,obs}$), fluence ($S_{\rm \gamma}$), peak flux ($F_{\rm p}$) and redshift ($z$) (see Figure 1). The Fermi sample and KW sample are shown as red and green step lines, respectively. It is clear that all the parameters follow the logarithm normal distributions.\footnote {To confirm this result, we make the normal statistic test by using the software Mathematica tool called "DistributionFitTest", and report the p-value in Table 5. The lognormal distribution can be accepted.}  We find that the values of $E_{\rm p,obs}$, $S_{\rm \gamma}$ and $F_{\rm p}$ of Fermi GRBs are systematically smaller than those of KW GRBs. On the contrary, the observed durations of Fermi GRBs are longer. To give a quantitative result, we make a Gaussian fitting and list the results in Table 5. For Fermi (KW) sample, the median values of $T_{\rm 90,obs}$, $E_{\rm p,obs}$, $S_{\rm \gamma}$ and $F_{\rm p}$ are $\sim$41~(30) s, $\sim$162~(231) keV, $\sim1.6~(4.2)\times10^{-5}$ erg cm$^{-2}$ and $\sim1.6~(4.8)\times10^{-6}$ erg cm$^{-2}$ s$^{-1}$, respectively. In addition, the dispersions of $E_{\rm p,obs}$, $S_{\rm \gamma}$ and $F_{\rm p}$ distributions for Fermi GRBs are larger than those of KW GRBs. This indicates that the ranges of the parameter distributions of Fermi GRBs are wider than those of KW sample. We also present the distribution of $z$ (see the last panel in Figure 1). It shows that the distribution ranges of redshift in two samples are basically consistent. The median value of $z$ is $\sim$1.3. To check if the samples have selection effect, the distribution of $z$ for all 412 long GRBs with measured redshift is analyzed and showed in the last panel of Figure 1. The typical value of $z$ for all long GRBs is $\sim$1.48. From the figure, one can find that the $z$ distributions of our samples are basically consistent with all long GRBs.

The difference of observed properties for two samples may be due to the different energy band of instruments. The detected energy band of Fermi/GBM (8-1000 keV) is lower than KW (10-10000 keV), so that, on average, the values of $E_{\rm p}$, $S_{\rm \gamma}$ and $F_{\rm p}$ are smaller for Fermi GRBs. It is known that the durations of GRBs strongly depend on the energy band, with an index of -0.4 \citep{1995ApJ...448L.101F,1996AAS...189.2803N,2007PASJ...59..857Z,2011AIPC.1358...13B,2013ApJ...763...15Q}, and the instrument selection biases (e.g., Bromberg et al. 2013; Kocevski \& Petrosian 2013). The energy band to calculate the duration of Fermi/GBM (8-1000 keV) is lower than that of KW (80-1200 keV). Therefore, the durations of Fermi GRBs should be longer. To further confirm this result, we make the distribution analysis of 46 GRBs observed simultaneously by two detectors (see in Figure 2). From the Figure 2 one can find that the most GRB durations derived from Fermi detector are obviously longer than those calculated from KW. In addition, $S_{\rm \gamma}$ and $F_{\rm p}$ derived from Fermi detector are smaller.

\subsection{Rest Frame} \label{ssec:rest}

The statistical properties and correlations of GRBs in the rest frame have been widely studied. We firstly present the distributions of $E_{\rm iso}$, $L_{\rm iso}$, $E_{\rm p}$ and $T_{\rm 90}$ in Figure 3. It shows that these four quantities are basically logarithmic normal distributions and are similar with those in the observer frame. This indicates that the observed distribution profiles are not significantly affected by the redshift distribution. In the same way, we make a lognormal test and a Gaussian fit and show the results in Table 5. From the results, we have $E_{\rm iso} \sim0.99~(1.75)\times10^{53}$ erg, $\sigma \sim$0.86~(0.77), $L_{\rm iso}\sim$ 2.4~(4.9)$\times10^{52}$ erg s$^{-1}$, $\sigma \sim $0.87 (0.80), $E_{\rm p} \sim$406~(562) keV, $\sigma \sim$0.46~(0.36), and $T_{\rm 90}\sim$16.3~(12.3) s, $\sigma \sim$ 0.51~(0.51) for Fermi (KW) sample, respectively.

The spectrum-energy correlation of GRBs is a very important relation in GRB research field and has been discussed extensively.
The $E_{\rm p}$-$E_{\rm iso}$ and $E_{\rm p}$-$L_{\rm iso}$ relations are analyzed and shown in the top panel of Figure 4. From this figure, one can find the two samples almost have the same properties, so we combine two samples together to make a quantitative analysis and obtain $E_{\rm p} \propto {E_{\rm iso}}^{0.36 \pm 0.03}$, with the correlation coefficient $r=0.71$, the chance probability $p < 10^{-4}$ and the dispersion $\sigma= 0.58$, and $E_{\rm p} \propto {L_{\rm iso}}^{0.34 \pm 0.03}$ ($r=0.71$, $p < 10^{-4}$ and $\sigma= 0.57$), where the data of 46 overlapped bursts are taken from Fermi. Our results are consistent with the previous \textbf {finding} presented in the literature (e.g., Amati et al. 2008; Ghirlanda et al. 2009; Zhang et al. 2009; Zhang et al. 2012).

In addition, we analyze the relations of other quantities and also present the results in Figure 4. We find that $E_{\rm p}$ is not correlated with $T_{\rm 90}$, which indicate that the spectral hardness is independent of the duration for long GRBs. There exists a weak correlation between $E_{\rm iso}$ and $T_{\rm 90}$ with $r = 0.17$, but $L_{\rm iso}$ and $T_{\rm 90}$ are anti-correlated, with $r = -0.20$. There is a tight correlation between $E_{\rm iso}$ and $L_{\rm iso}$ with $r = 0.9$, but GRB 130925A deviates this relation.

\subsection{Properties of $\Gamma_0$} \label{ssec:Gamma}

The pseudo Lorentz factors have been estimated from the three-parameter correlation ${L_{{\rm{iso}}}}$-${E_{\rm p}}$-${\Gamma _0}$ (Liang et al. 2015), so we can study the distribution of $\Gamma_0$ and the relations between $\Gamma_0$ and other quantities. Figure 5 shows the distribution of $\Gamma_0$ for Fermi and KW sample. There is a typical value of $\Gamma_0$ as $\sim 240~(300)$ for Fermi (KW) sample, and a small dispersion $\sigma \sim0.3~(0.28)$ shown in Table 1. \citet{2012MNRAS.420..483G} discussed the distribution of $\Gamma_0$ in homogeneous ISM (H) and wind density profile (W) case and obtained $\Gamma_0$ (H, W) $\sim138~(66)$, $\sigma \sim0.17~(0.2)$. The reconstructed distribution of $\Gamma_0$ by \citet{2018A&A...609A.112G} has the median value $\sim300~(150)$. Our results are consistent with the previous findings and the theoretical predictions (see Piran 2004).

The values of $\Gamma_0$ are derived from $L_{\rm iso}$ and $E_{\rm p}$, so these three quantities should be correlated (see in Table 6). The relations between $\Gamma_0$ and $E_{\rm iso}$ ($T_{\rm 90}$) are analyzed and shown in Figure 6. From this figure we find that there is a tight correlation between $\Gamma_0$ and $E_{\rm iso}$, with a correlation coefficient $r = 0.77$. We parametrize the correlation and obtain $\log E_{iso,52} = (-4.10 \pm 0.33) + (2.15 \pm 0.13) \times \log \Gamma_0$, which is consistent with the result of \citet{2015ApJ...813..116L}. We also find there is a trend that the intrinsic duration is anticorrelated with the $\Gamma_0$. The parameterized result is $\log T_{90} = (2.20 \pm 0.30) + (-0.44 \pm 0.12) \times \log \Gamma_0$, where the correlation coefficient is $r = -0.26$.

\subsection{Comoving Frame} \label{ssec:cov}

The properties of GRBs in the rest frame can not fully reflect the real physical processes, so the intrinsic properties in the comoving frame need to be investigated. Using the pseudo Lorentz factor, we obtain the parameters in the comoving frame and analyze their properties. We show the distributions of the isotropic energy $E'_{\rm iso}$, isotropic luminosity $L'_{\rm iso}$, spectrum peak energy $E'_{\rm p}$ and duration $T'_{\rm 90}$ in the comoving frame (see in Figure 7). As shown in Figure 7, these four quantities basically follow the lognormal distributions. Fitting these distributions we have the median value of $E'_{\rm iso} \sim4~(6)\times10^{50}$ erg , with a dispersion of $\sigma \sim0.67~(0.58)$ and the median value of $L'_{\rm iso} \sim3~(4)\times10^{47}$ erg s$^{-1}$ , with a dispersion of $\sigma \sim0.45~(0.35)$ for Fermi (KW) sample. The typical values of the comoving frame isotropic energy and isotropic luminosity are consistent with the results of \citet{2018A&A...609A.112G}. We also find that the dispersions in the comoving frame are smaller than those in the rest frame. This indicates that the intrinsic energy and luminosity tend to be concentrated. In addition, we obtain the typical values of  $E'_{\rm p}$ and $T'_{\rm 90}$ are $\sim2 $ keV and $\sim4000$ s, respectively. The dispersions of these two quantities are large and similar to which in the rest frame, even a little bigger than those in the observer frame. This shows that the intrinsic peak energies and durations are not the same for the different long GRBs.

It is well known that long GRBs basically comply with $E_{\rm p}$-$E_{\rm iso}$ and $E_{\rm p}$-$L_{\rm iso}$ correlations. \citet{2012MNRAS.420..483G} found that these two correlations still exist in the comoving frame (see also Liang et al. 2015). The pair relations of $E'_{\rm iso}$, $L'_{\rm iso}$, $E'_{\rm p}$ and $T'_{\rm 90}$ are studied and shown in Figure 8. The results of the regression analysis are listed in Table 6. We have $E'_{\rm p} = (-0.79\pm 0.09){L'_{\rm iso,46}}^{0.67\pm 0.06}$ with a correlation coefficient of $r=0.66$ and a dispersion of $\sigma=0.62$, and $E'_{\rm p} = (0.09\pm 0.04){E'_{\rm iso,50}}^{0.25 \pm 0.05}$ with $r=0.38$ and $\sigma=0.76$. This results further confirm that the spectrum-energy relations do exist in the comoving frame, but have very large dispersions, suggesting that the observed strong spectrum-energy correlations in the rest frame are likely governed by the Doppler boosting effect. From Figure 8, we also find that the comoving peak energy $E'_{\rm p}$ is independent of the duration $T'_{\rm 90}$. This is consistent with the result in the rest frame. It further confirms that the hardness and duration of long GRBs are independent quantities and describe the different physical processes. In addition, we also find there exists a tight relation between the isotropic energy $E'_{\rm iso}$ and the duration $T'_{\rm 90}$, $T'_{\rm 90} = (3.19\pm 0.04) {E'_{\rm iso,50}}^{0.54 \pm 0.05}$ with $r = 0.65$ and $\sigma$ = 0.78. This shows the isotropic energy is strongly related to duration in the comoving frame. For the isotropic luminosity $L'_{\rm iso}$ and duration $T'_{\rm 90}$, there is a weak correlation as  $T'_{\rm 90} = (3.13\pm 0.15){L'_{\rm iso,46}}^{0.28 \pm 0.09}$ with $r =0.22$ and $\sigma =1.00$.

We also analyze the relations between $\Gamma_0$ and all the comoving frame quantities and report the results in Figure 9 and Table 6. It is found that $E'_{\rm iso}$, $L'_{\rm iso}$ and $T'_{\rm 90}$ are weakly correlated with $\Gamma_0$, but $E'_{\rm p}$ is anticorrelated with $\Gamma_0$. This indicate that the harder the spectra of long GRBs, the slower the bulk motion of fireball.

\subsection{The redshift dependence}

It is known that the distributions of the isotropic energy, luminosity and peak energy, and their statistical correlations in the rest frame are affected by the severe truncation mainly due to the detection flux and energy limit of the telescope. Figure 10. shows the relations between $T_{\rm 90}$, $E_{\rm p}$, $E_{\rm iso}$ and $L_{\rm iso}$ and $z$. From this figure we can find that $T_{\rm 90}$ does not significantly depend on $z$ \footnote{The previous studies found that the intrinsic duration of GRBs is independent with redshift (see, Sakamoto et al. 2011; Gruber et al. 2011; Zhang et al. 2013) or has a smaller redshift evolution (P${\'e}$langeon et al. 2008; Dainotti et al. 2015).}, but $E_{\rm iso}$, $L_{\rm iso}$ and $E_{\rm p}$ all depend on $z$, which are consistent with the previous results reported by many authors (e.g., Lloyd-Ronning et al. 2002; Wei \& Gao 2003; Yonetoku et al. 2004; Wu et al. 2012; Salvaterra et al. 2012; Zhang et al. 2014). To obtain the intrinsic distributions and correlations, the first step is to remove the effect of redshift evolution.  A nonparametric $\tau$ statistical technique was widely used to resolve this problem (e.g., Lloyd-Ronning et al. 2002; Yonetoku et al. 2004; Wu et al. 2012; Dainotti et al. 2013, 2015, 2017, Zhang et al. 2014; Petrosian et al. 2015; Tsvetkova et al. 2017). We firstly consider $E_{\rm iso}$ observed by Fermi as an example.

Consider a set of observable $E_{{\rm iso},i}$ and $z_{i}$, where $i$ indexes the $i$th burst and in our case $i$ runs from 1 to 93. For the $i$th sample of ($z_{i}, E_{{\rm iso},i}$), we consider an associated set of
\begin{equation}
J_{i}=\{j|E_{{\rm iso},j}>E_{{\rm iso},i},z_{j}<z_{i,\rm lim}\},~~~~~~~{\rm for}~1\leq i\leq 93.
\end{equation}
in which the number of samples in the $J_{i}$ set is $N_{i}$. The $z_{i,\rm lim}$ is the redshift of the crossing point between two lines of $E=E_{{\rm iso},i}$ and the fluence limit corresponding to its ``isotropic-equivalent-energy" limit. If $z_{i}$ and $E_{{\rm iso},i}$ are independent to each other, one would expect the number of the following sample
\begin{equation}
R_{i}={\rm Number}\{j\in J_{i}|z_{j}\leq z_{i}\}
\end{equation}
to be uniformly distributed between 1 and $N_{i}$. To estimate the correlation degree between $E_{\rm iso}$ and $z$, one may introduce the test statistic $\tau$ parameterized as
\begin{equation}
\tau=\frac{\sum_{i}(R_{i}-E_{i})}{\sqrt{\sum_{i}V_{i}}},
\end{equation}
where $E_{i}=(N_{i}+1)/2$ and $V_{i}=(N_{i}^2-1)/12$ are the expected mean and the variance of the uniform distribution, respectively.

If $R_{i}$ follows an ideal uniform distribution, then the samples of $R_{i}\leq E_{i}$ and $R_{i}\geq E_{i}$ should be equal, and thus the statistic parameter $\tau$ tends to be zero. Note that the $\tau$ value here has been normalized by the square root of variance, so the correlation degree $z$ and $E_{\rm iso}$ can be measured in units of standard deviation.

We take the simple form of $g(z)=(1+z)^{q}$ in order to separate the isotropic energy evolution. The value $E_{\rm iso,c}\equiv E_{\rm iso}/g(z)$ represents the isotropic energy after removing the evolution effect. When $\tau$ is not equal to zero, we change the $q$ values until $\tau=0$ with a proper $q$. Finally, we find that the null hypothesis of the evolution is rejected at about 4.1 $\sigma$ confidence level and there is a significant evolution for the isotropic energy, $q_{\rm E}=2.44^{+0.59}_{-0.47}$. The de-evolved isotropic energy $E_{\rm iso,c}$ for Fermi sample can be obtained from $E_{\rm iso,c}=E_{\rm iso}/(1+z)^{2.44}$.

Using the same $\tau$ statistical method, we obtain $q_{\rm L}=2.65^{+0.56}_{-0.69}$ (2.9$\sigma$, for $L_{\rm iso}$), and $q_{\rm Ep}=1.05^{+0.34}_{-0.29}$ (3.6$\sigma$, for $E_{\rm{p}}$ ) for Fermi data, and $q_{\rm E}=1.73^{+1.2}_{-0.77}$ (2.6$\sigma$), $q_{\rm L}=3.24^{+0.29}_{-0.67}$ (4.6$\sigma$), and $q_{\rm Ep}=1.07^{+0.26}_{-0.18}$ (4.1$\sigma$) for KW data. As shown in Figure 10, we choose $5\times 10^{-7}{\rm erg}\,{\rm cm}^{-2}$ as the fluence limit, $2\times 10^{-7}{\rm erg}\,{\rm cm}^{-2}\,{\rm s}^{-1}$ as the flux limit and 30 keV as the lower limit of the observed peak energy for Fermi data, and $4\times 10^{-6}{\rm erg}\,{\rm cm}^{-2}$ as the fluence limit, $6\times 10^{-7}{\rm erg}\,{\rm cm}^{-2}\,{\rm s}^{-1}$ as the flux limit and 60 keV as the lower limit of the observed peak energy for KW data.\footnote{The result is not significantly affected by the chosen value of flux/fluence and peak energy limit as long as it is around the instrumental sensitivity (see Wu et al. 2012, Zhang et al. 2014)} Meanwhile, we can obtain the de-evolved luminosity and peak energy from $L_{\rm iso,c}=L_{\rm iso}/(1+z)^{q_{\rm L}}$ and $E_{\rm p,c}=E_{\rm p}/(1+z)^{q_{\rm Ep}}$.

After eliminating the redshift evolution effect, the intrinsic distributions and correlations are analyzed and shown in Figures 11-14. The statistical analysis results are listed in Table 5 and Table 7. From these results, we find that the main conclusions are not significantly changed when the instrument biases are taken into account. But the pair correlations become slightly weaker in the rest frame and tighter in the comving frame. We also find that the isotropic energy, luminosity and peak energy both in the rest frame and comving frame are more concentrated. The median values are $E_{\rm iso,c} \sim1~(4)\times10^{52}$ erg, with a dispersion of $\sigma \sim0.75~(0.65)$, $L_{\rm iso,c} \sim2~(3)\times10^{51}$ erg s$^{-1}$, $\sigma \sim0.66~(0.65)$ and $E_{\rm p,c} \sim160~(220)$ keV, $\sigma \sim0.41~(0.30)$ for Fermi (KW) sample in the rest frame. In the comoving frame, we have $E'_{\rm iso,c} \sim4~(13)\times10^{49}$ erg, $\sigma \sim0.64~(0.52)$, $L'_{\rm iso,c} \sim2.7~(2.4)\times10^{46}$ erg s$^{-1}$, $\sigma \sim0.44~(0.4)$ and $E'_{\rm p,c} \sim0.6~(0.7)$ keV, $\sigma \sim0.49~(0.4)$ for Fermi (KW) sample.

\section{Summary and Discussion} \label{sec:sum}

We compile a long GRB sample with the measured redshifts and well determined spectra observed by Fermi and Konus-Wind instruments. Using a tight ${L_{{\rm{iso}}}}$-${E_{\rm p}}$-${\Gamma_0}$ three-parameter correlation, we estimate the GRB initial Lorentz factors in our samples. The properties and correlations in the observer frame, rest frame and comoving frame are studied. We find that the distributions of the isotropic energy, luminosity, peak energy and duration in the different frame are basically lognormal and the comoving frame quantities are more concentrated.
When the redshift evolution effect has been taken into account, we obtain the median values are $E'_{\rm iso,c} \sim 8\times10^{49}$ erg, $L'_{\rm iso,c}\sim 2.5\times10^{46}$ erg s$^{-1}$, $E'_{\rm p,c}\sim 0.7$ KeV. The central value of ${\Gamma_0}$ is  $\sim 270$, which is consistent with the previous finding and the theoretical prediction. We further confirm the Amati relation, Yonetoku relation, and ${L_{{\rm{iso}}}}$ ($E_{\rm{iso}}$)-${\Gamma_0}$ relation, and find that the corresponding relations in the comoving frame still exist with large dispersions. This indicates these correlations are not a good standard candle and we need caution to make cosmological studies (e.g., Dainotti et al. 2013; Dainotti \& Amati 2018). In addition, the peak energies of long GRBs are independent of durations both in the rest frame and in the comoving frame. And there is a weak anticorrelation between the peak energy and Lorentz factor.

It is known that short GRBs are also comply with the similar ${E_{{\rm{iso}}}}$-${E_{\rm p}}$ and ${L_{{\rm{iso}}}}$-${E_{\rm p}}$ correlations (e.g., Zhang et al. 2012), but both correlations for short GRBs are dimmer than those of Long GRBs (e.g., Tsutsui et al. 2013). We wonder if the spectrum-energy correlation in the comoving frame also exist for short GRBs. However the Lorentz factors of short GRBs are not derived now, we cannot analyze the intrinsic properties of short GRBs in the comoving frame and this question need to be investigated in the future.

\section*{Acknowledgments}

We thank the anonymous referee for constructive comments. We also thank Dr. Yi-Zhong Fan for stimulating suggestions/discussions. We acknowledgment the use of public data from Fermi and Konus-Wind. This work was supported in part by NSFC under grants 11763003, and by the Guangxi Natural Science Foundation (No. 2017GXNSFAA198094).



\clearpage

\begin{figure*}
\centering
\label{figure1}
\includegraphics[angle=0,scale=0.58]{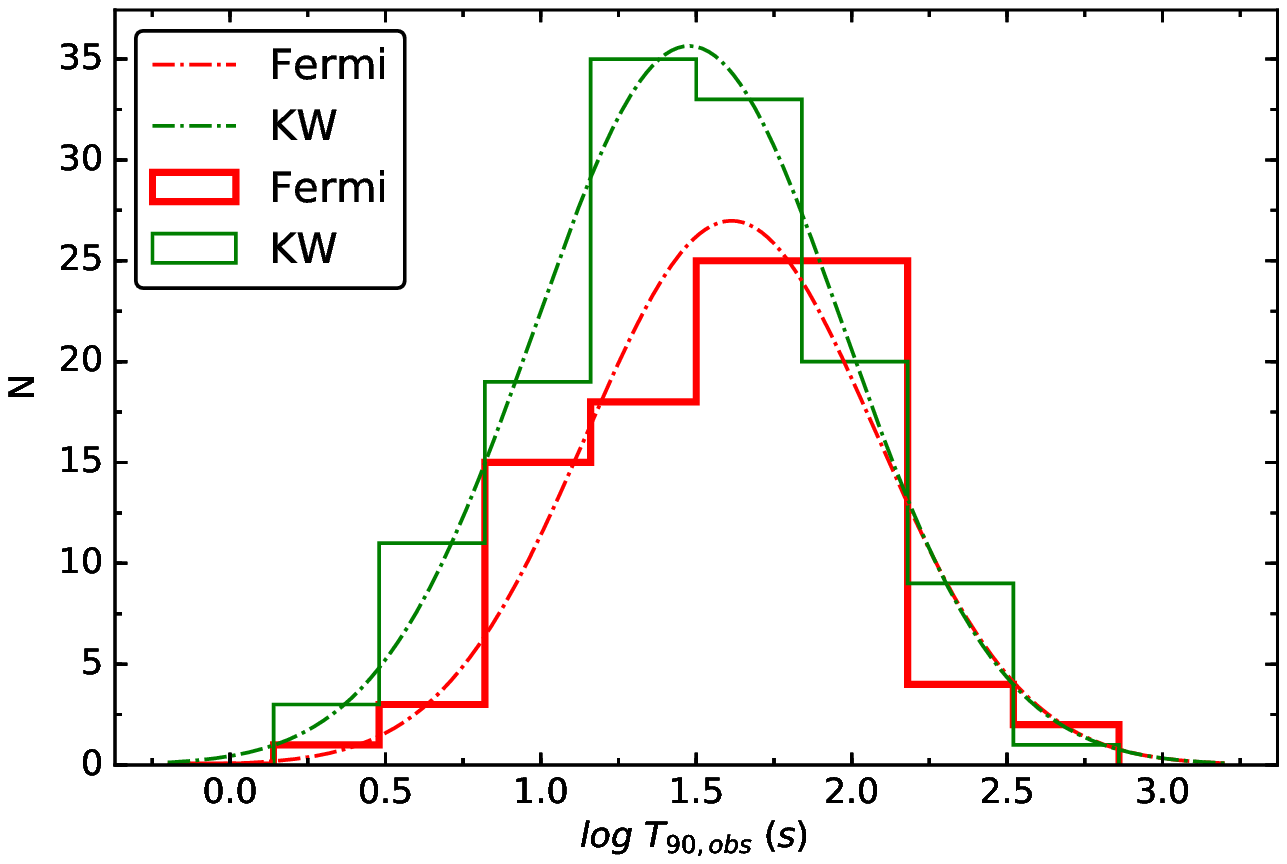}
\includegraphics[angle=0,scale=0.58]{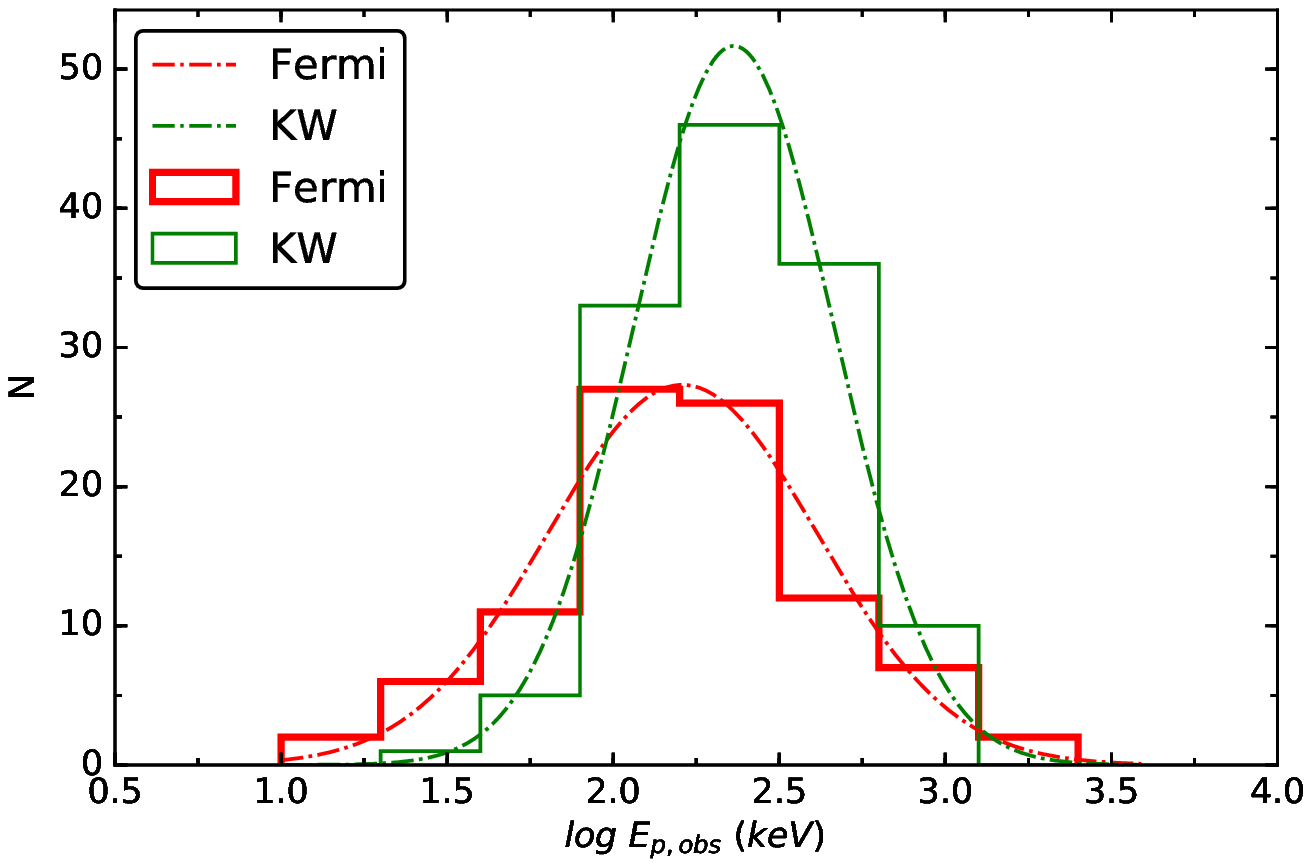}
\includegraphics[angle=0,scale=0.58]{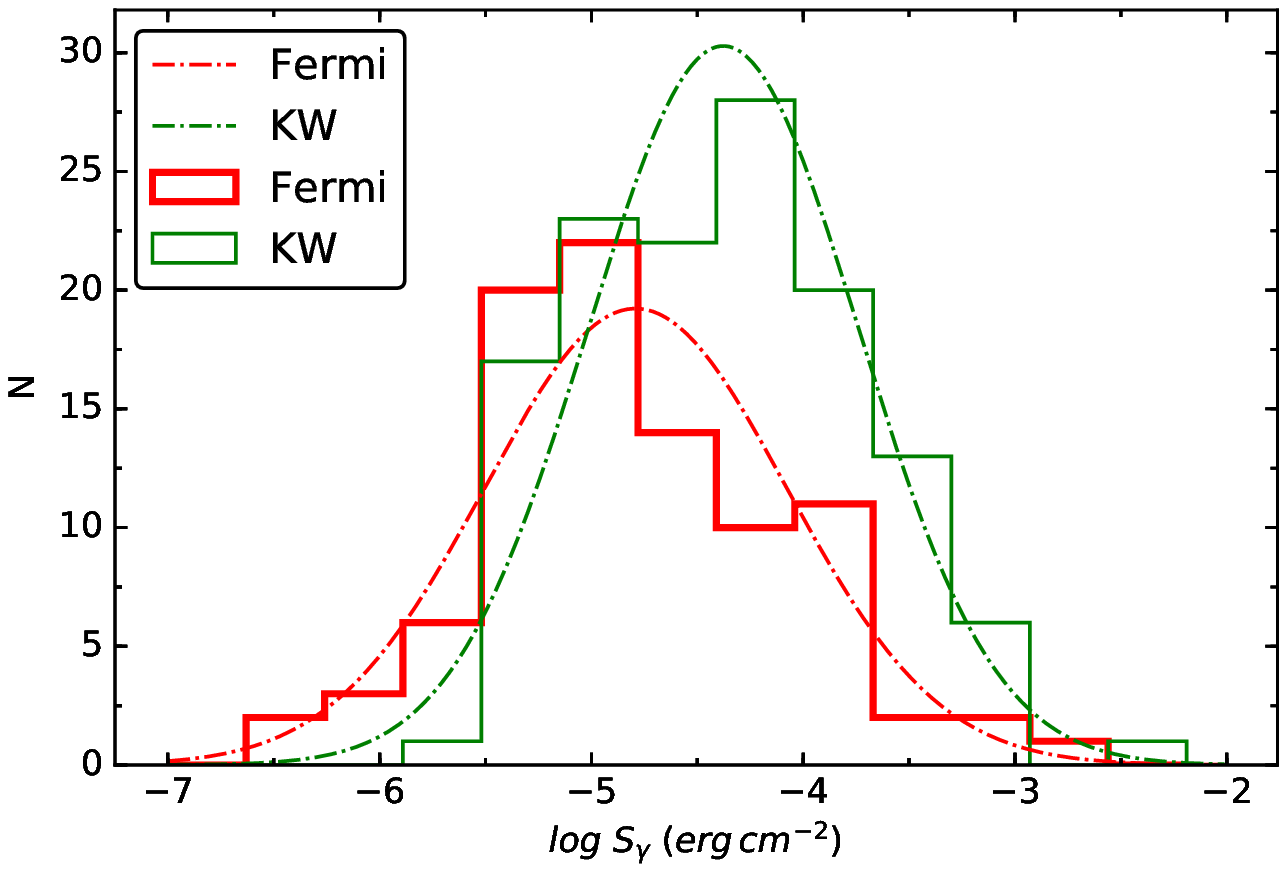}
\includegraphics[angle=0,scale=0.58]{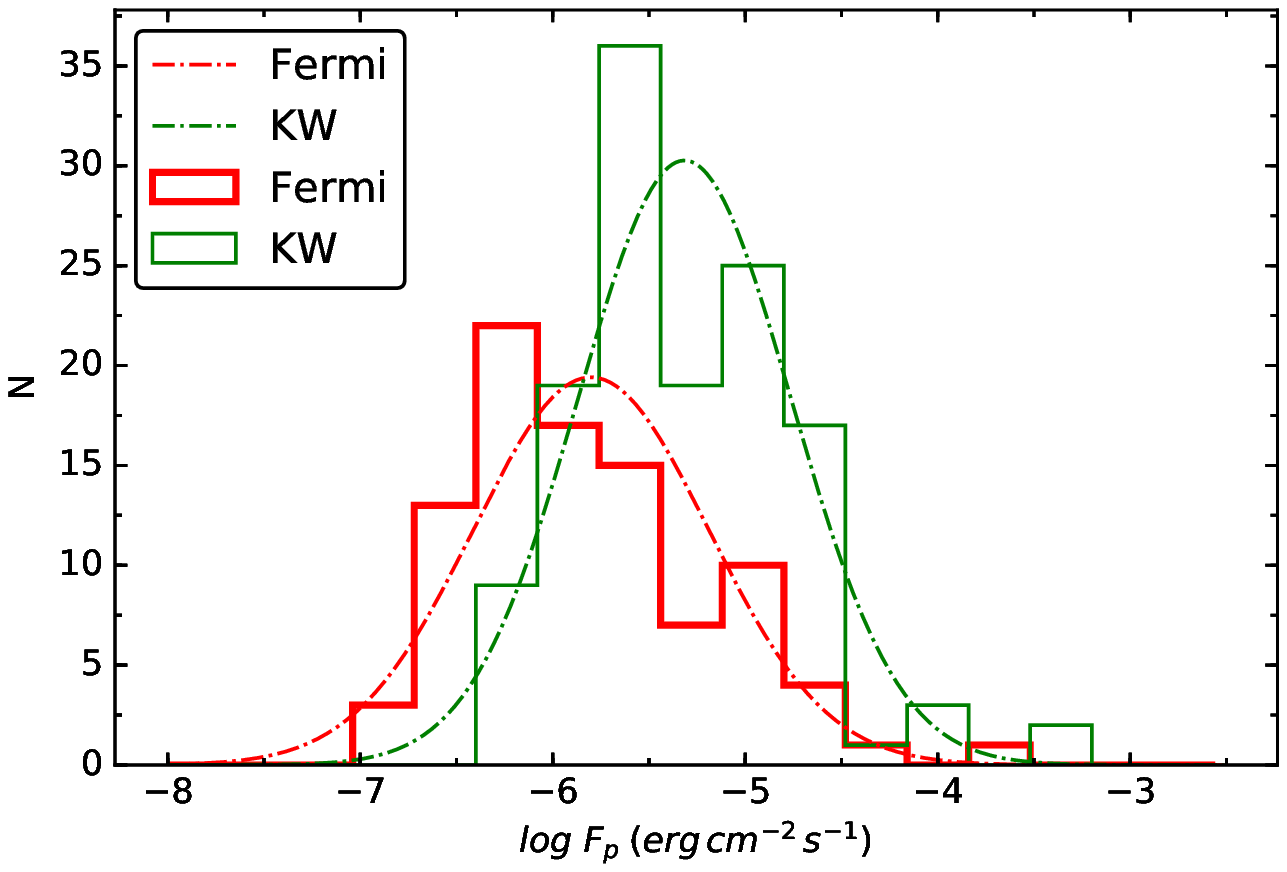}
\includegraphics[angle=0,scale=0.58]{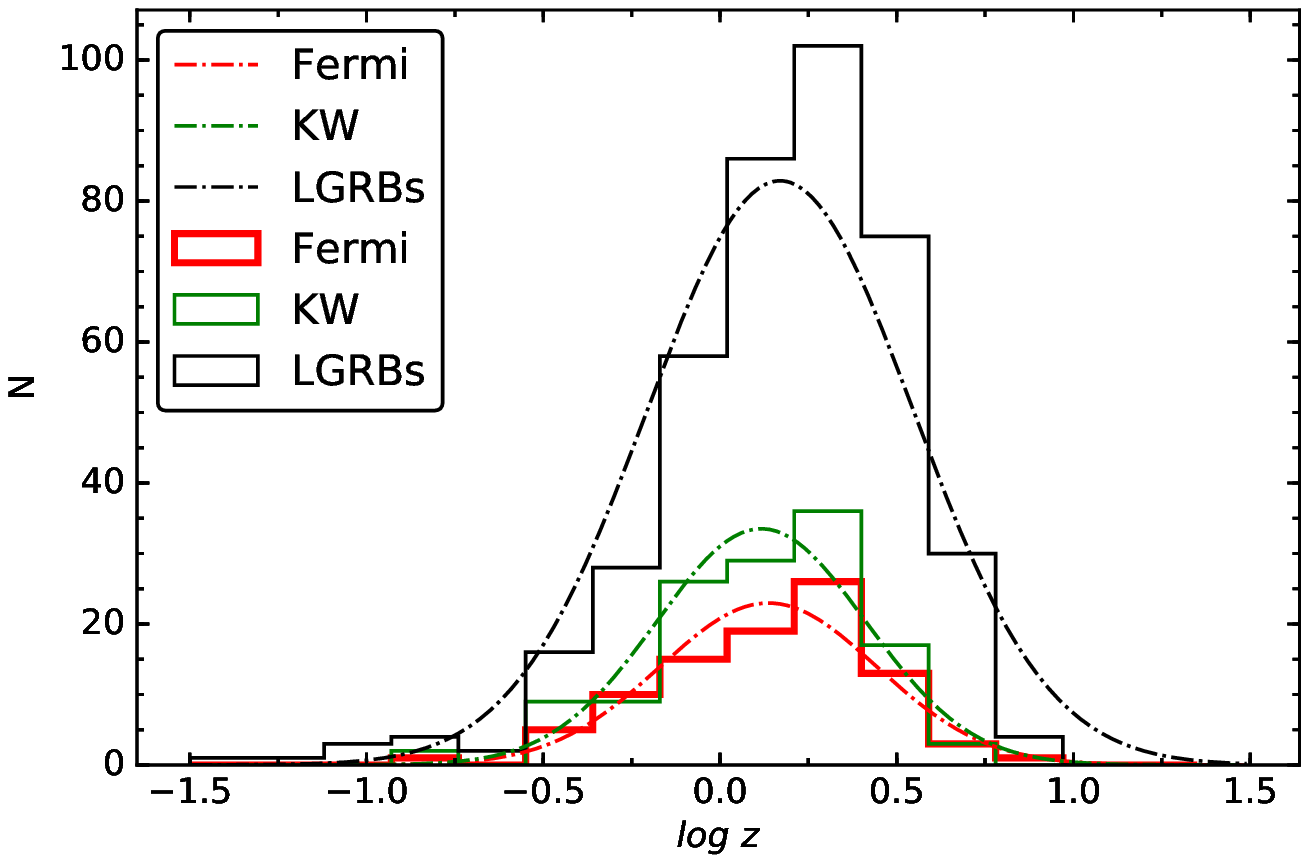}
\caption{Distributions of the observed duration $T_{\rm 90,obs}$, spectrum peak energy $E_{\rm p,obs}$, fluence $S_{\rm \gamma}$, peak flux $F_{\rm p}$ and redshift $z$. The red and green lines denote Fermi and KW sample, respectively. The dotted lines are best fitting curves. The black line in the last panel denotes all 412 long GRBs with measured redshift.}
\end{figure*}

\begin{figure*}
\centering
\label{figure2}
\includegraphics[angle=0,scale=0.5]{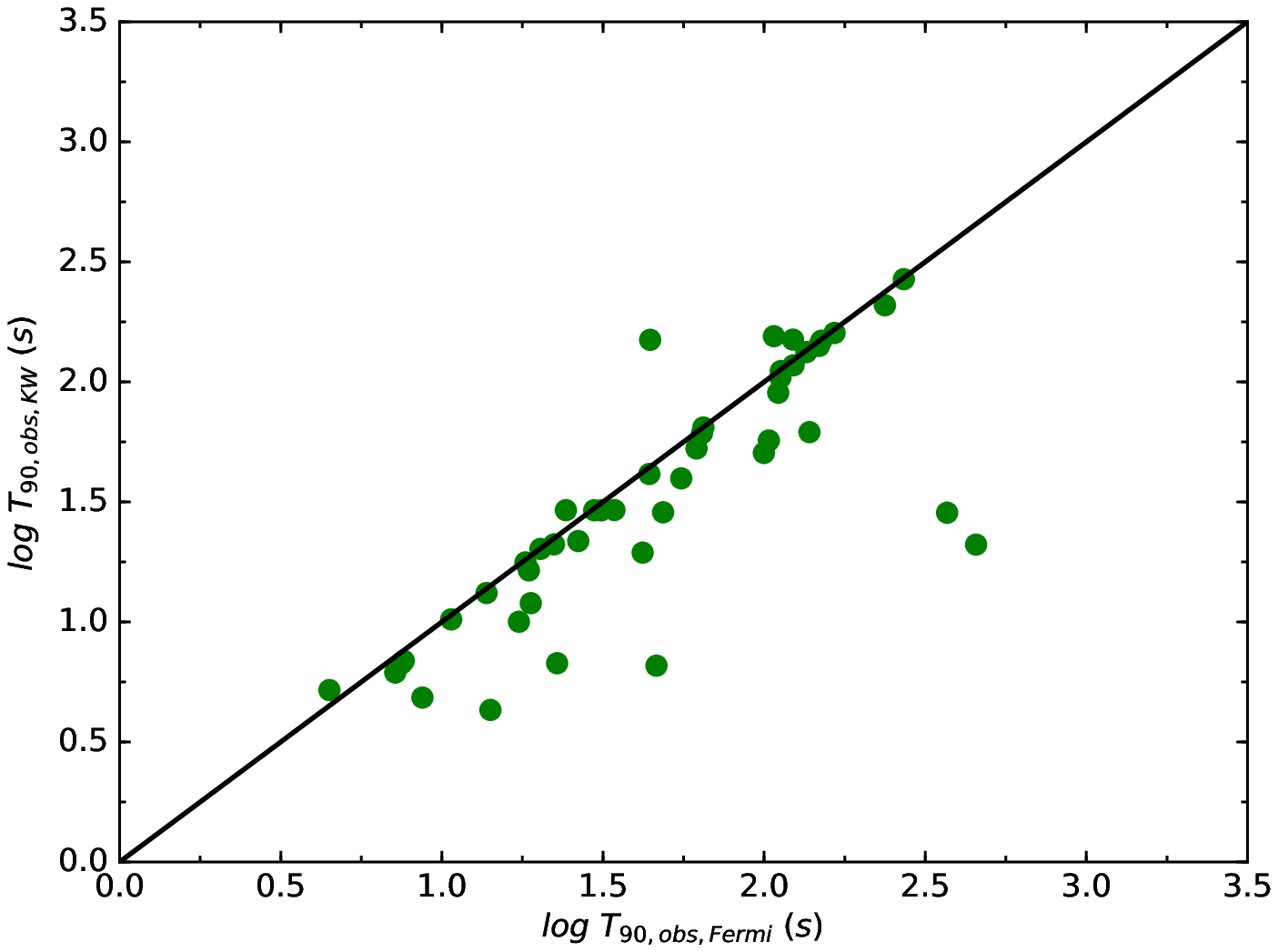}
\includegraphics[angle=0,scale=0.5]{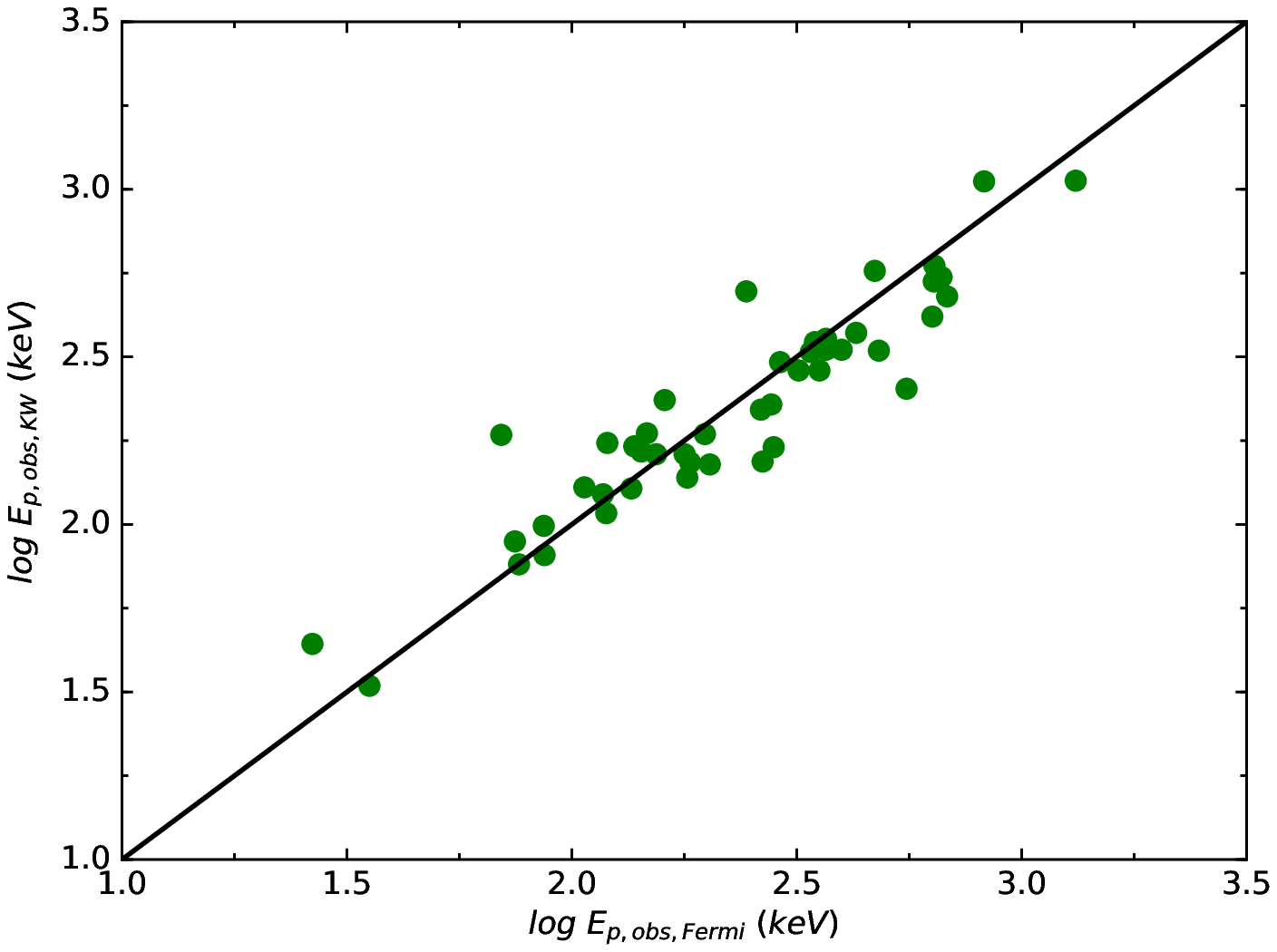}
\includegraphics[angle=0,scale=0.5]{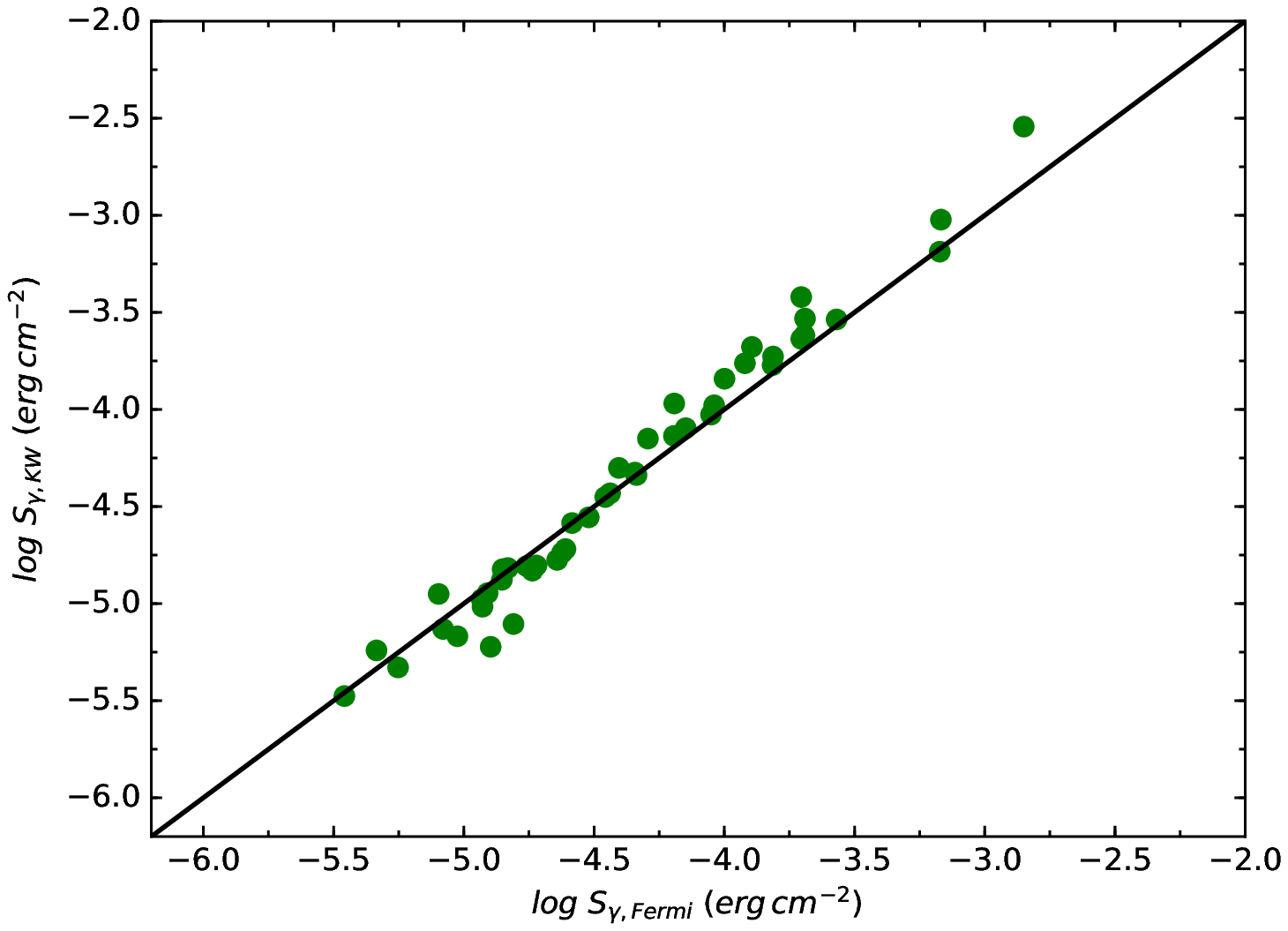}
\includegraphics[angle=0,scale=0.5]{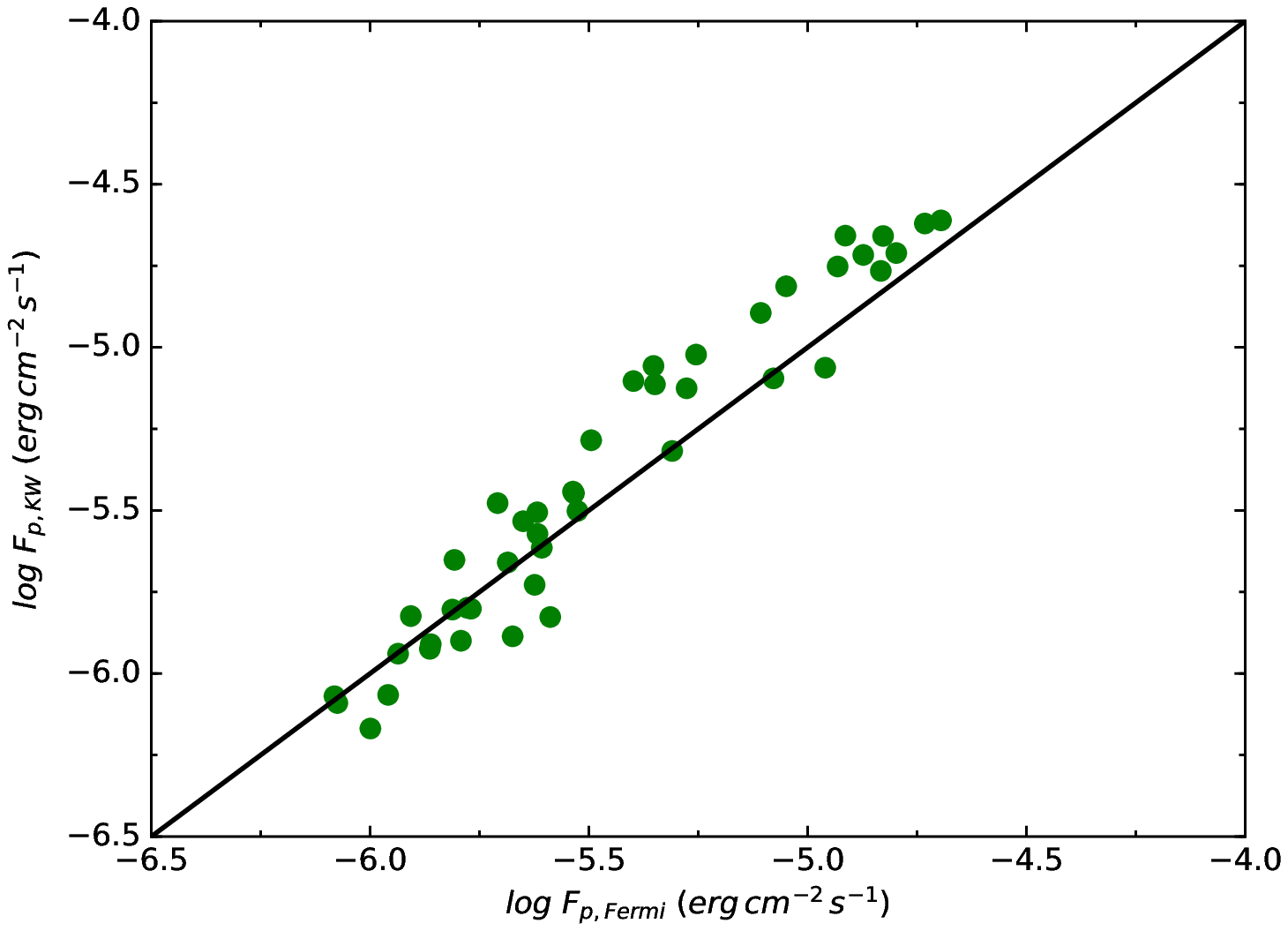}
\caption{Distributions of $T_{\rm 90,obs}$, $E_{\rm p,obs}$, $S_{\rm \gamma}$ and $F_{\rm p}$ for 46 GRBs observed simultaneously by Fermi and KW instruments. The lines superimposed is the line with slope 1.}
\end{figure*}

\begin{figure*}
\centering
\label{figure3}
\includegraphics[angle=0,scale=0.53]{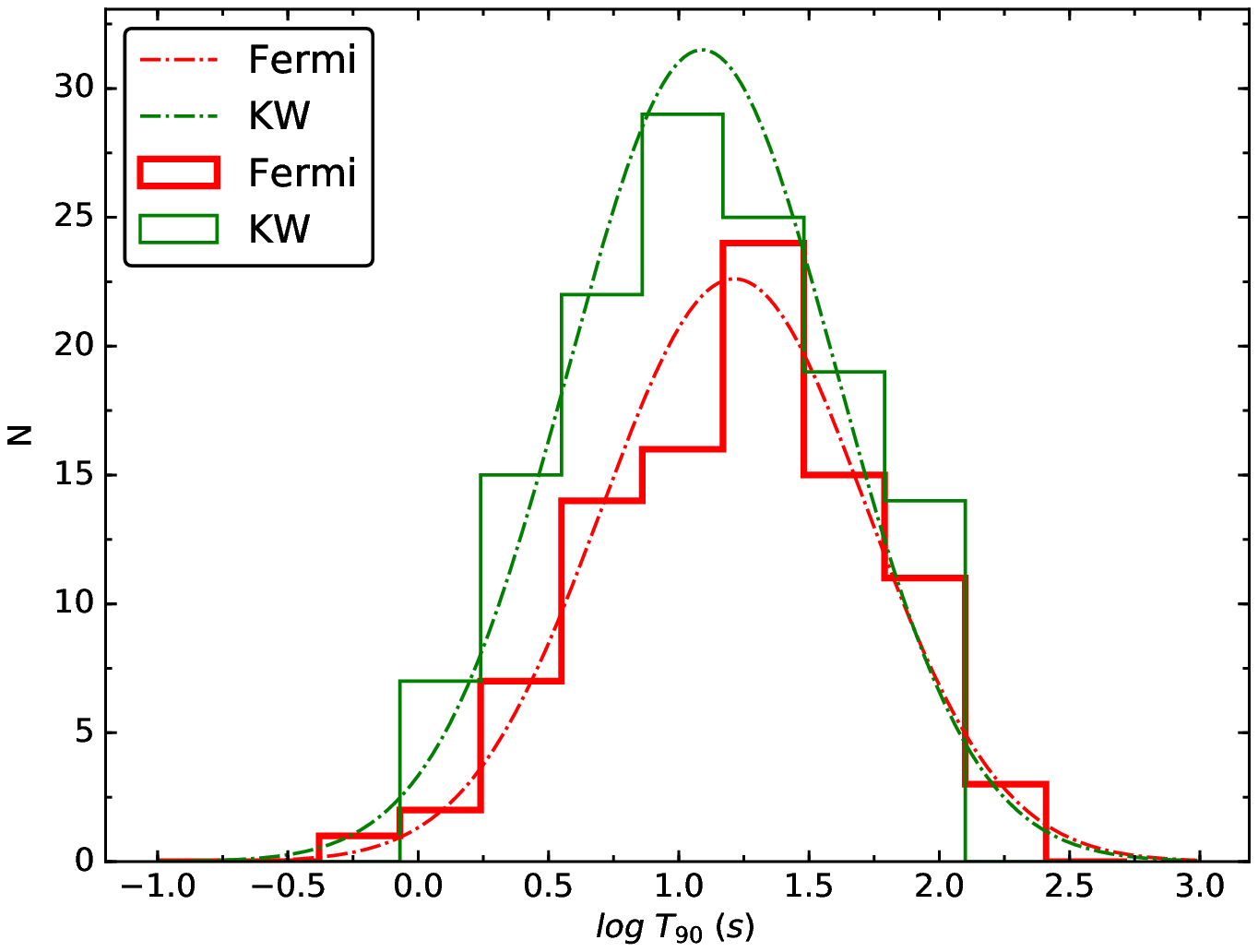}
\includegraphics[angle=0,scale=0.53]{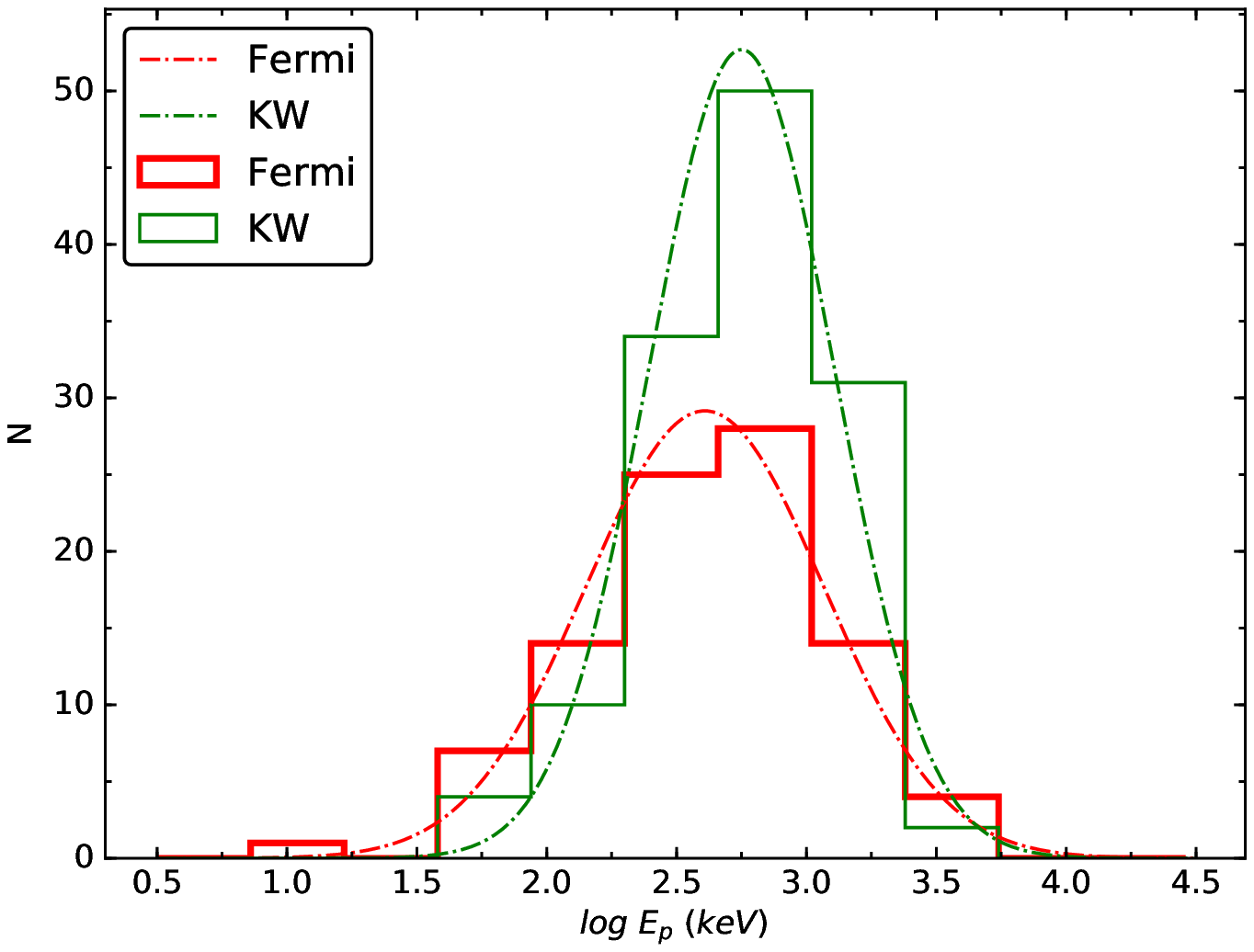}
\includegraphics[angle=0,scale=0.53]{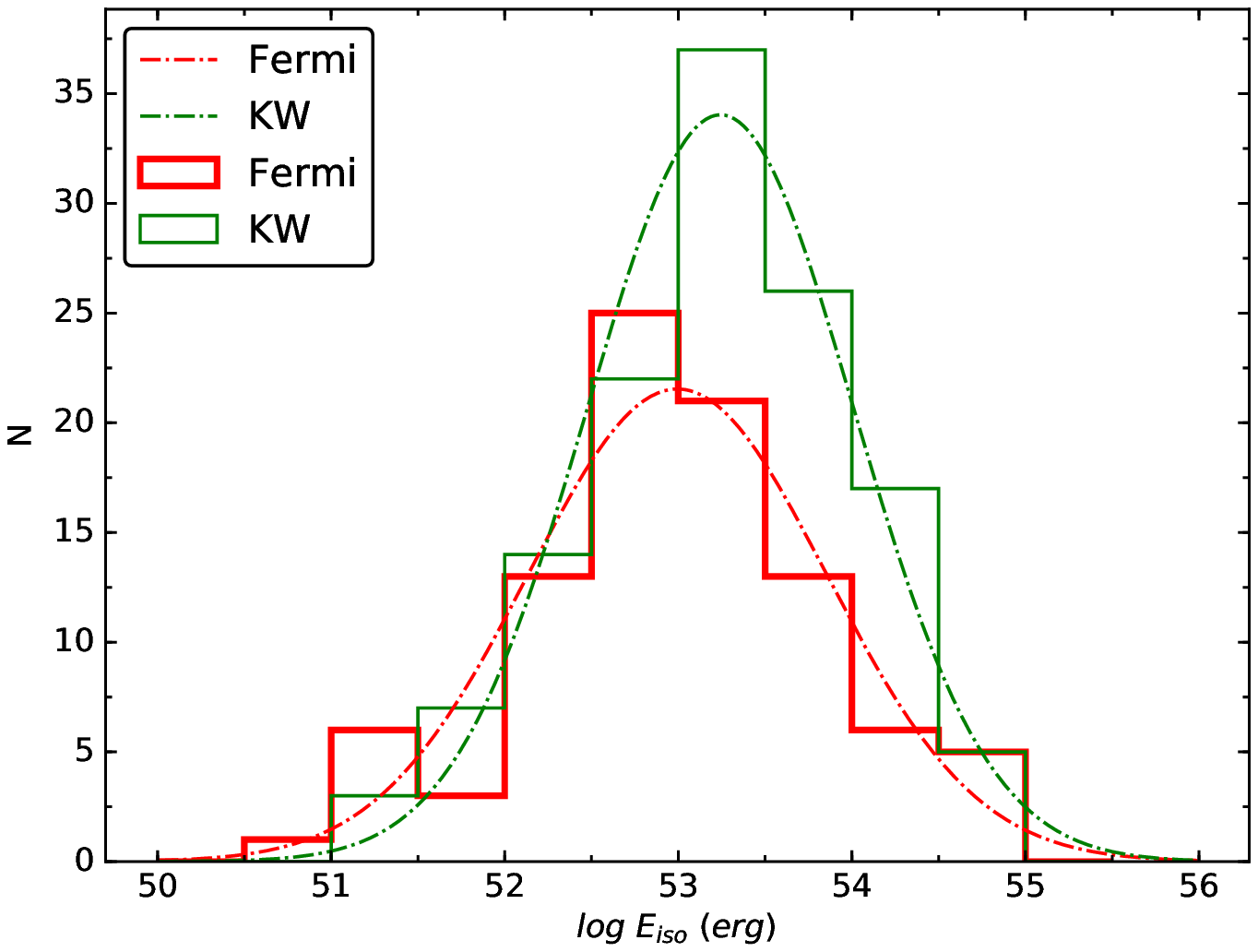}
\includegraphics[angle=0,scale=0.53]{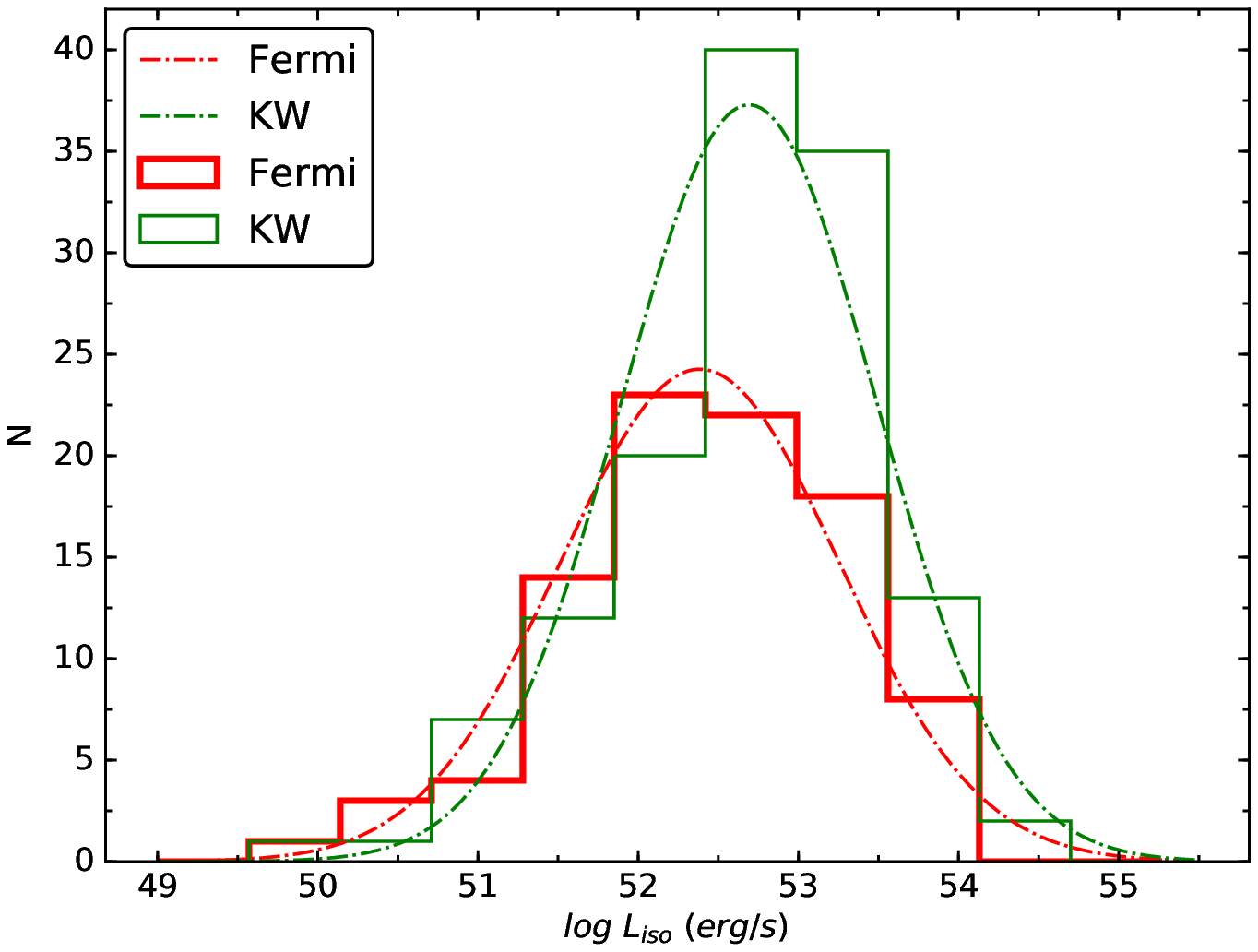}
\caption{Distributions of the duration $T_{\rm 90}$, spectrum peak energy $E_{\rm p}$, isotropic energy $E_{\rm iso}$ and luminosity $L_{\rm iso}$ in the rest frame. The red and green lines denote Fermi and KW sample, respectively. The dotted lines are Gaussian fitting curves.}
\end{figure*}

\begin{figure*}
\centering
\label{figure4}
\includegraphics[angle=0,scale=0.52]{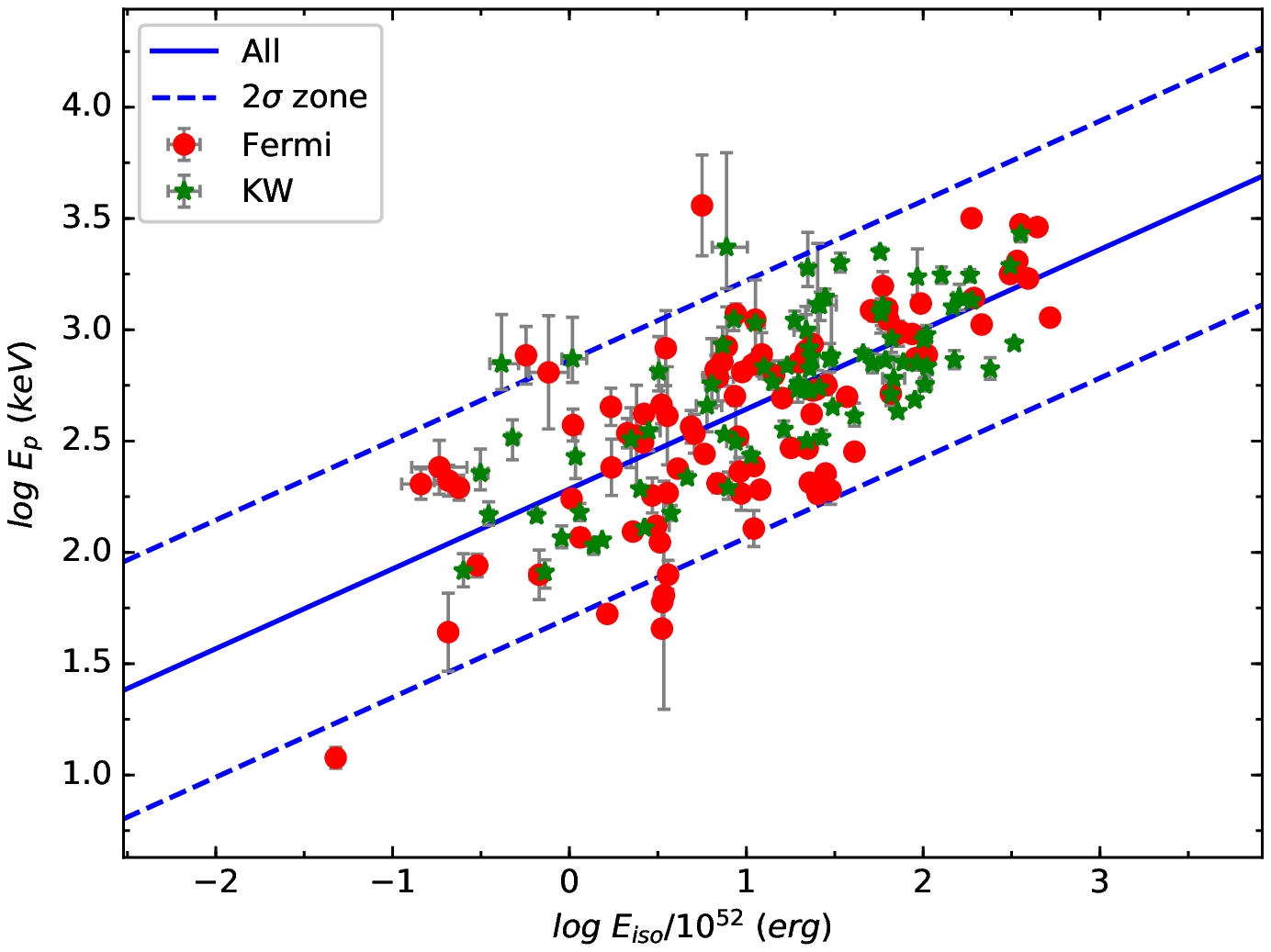}
\includegraphics[angle=0,scale=0.52]{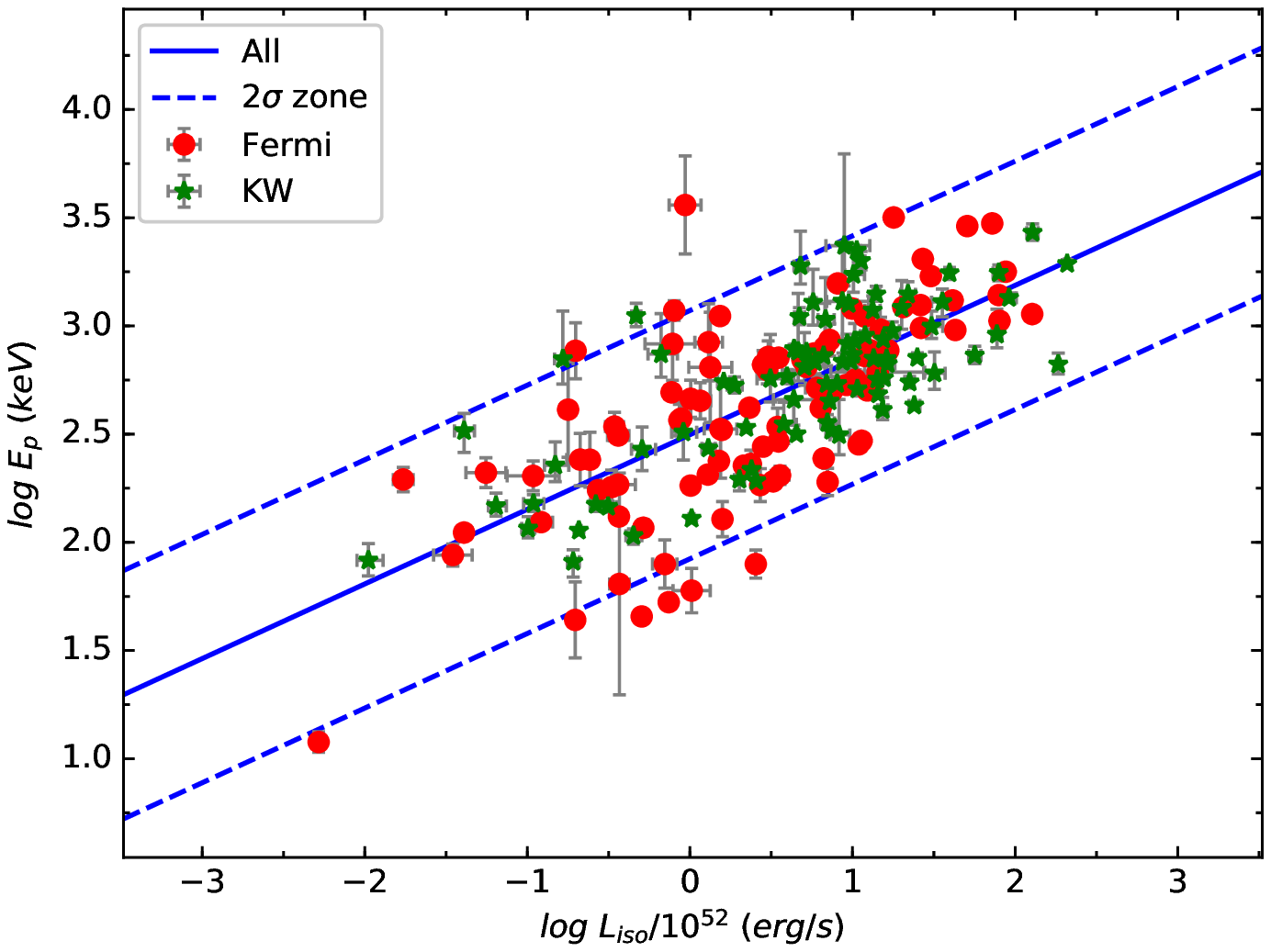}
\includegraphics[angle=0,scale=0.52]{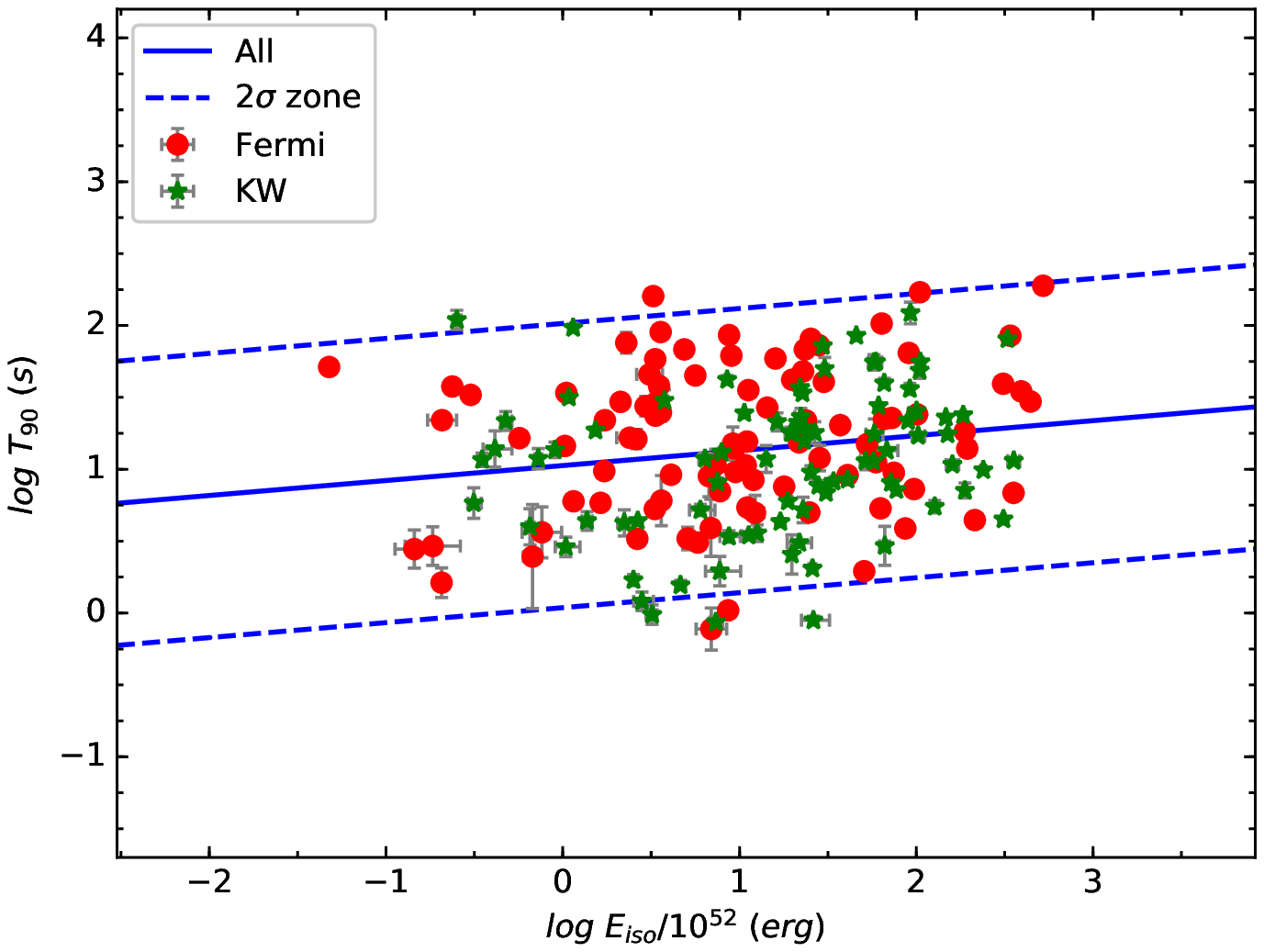}
\includegraphics[angle=0,scale=0.52]{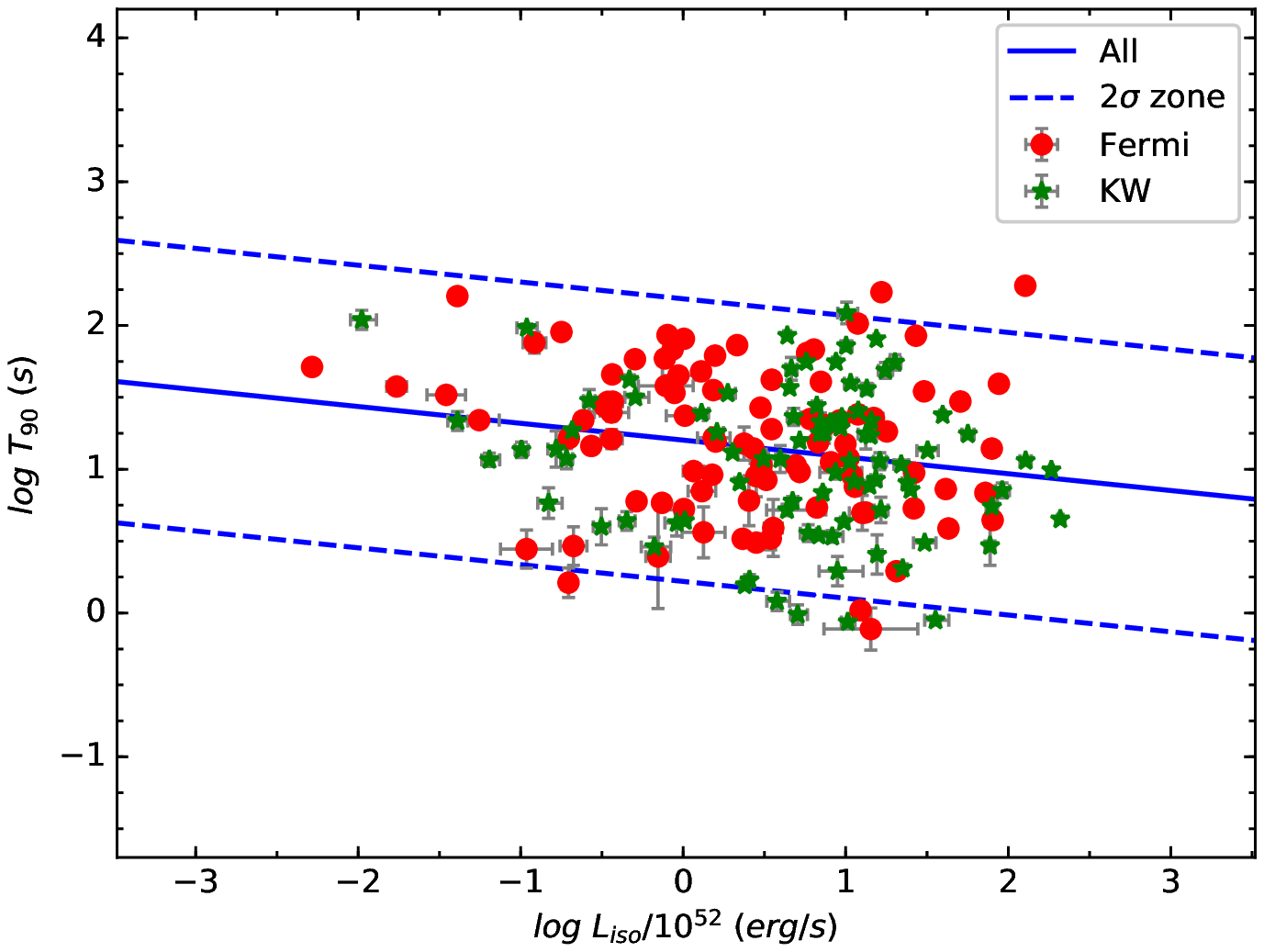}
\includegraphics[angle=0,scale=0.52]{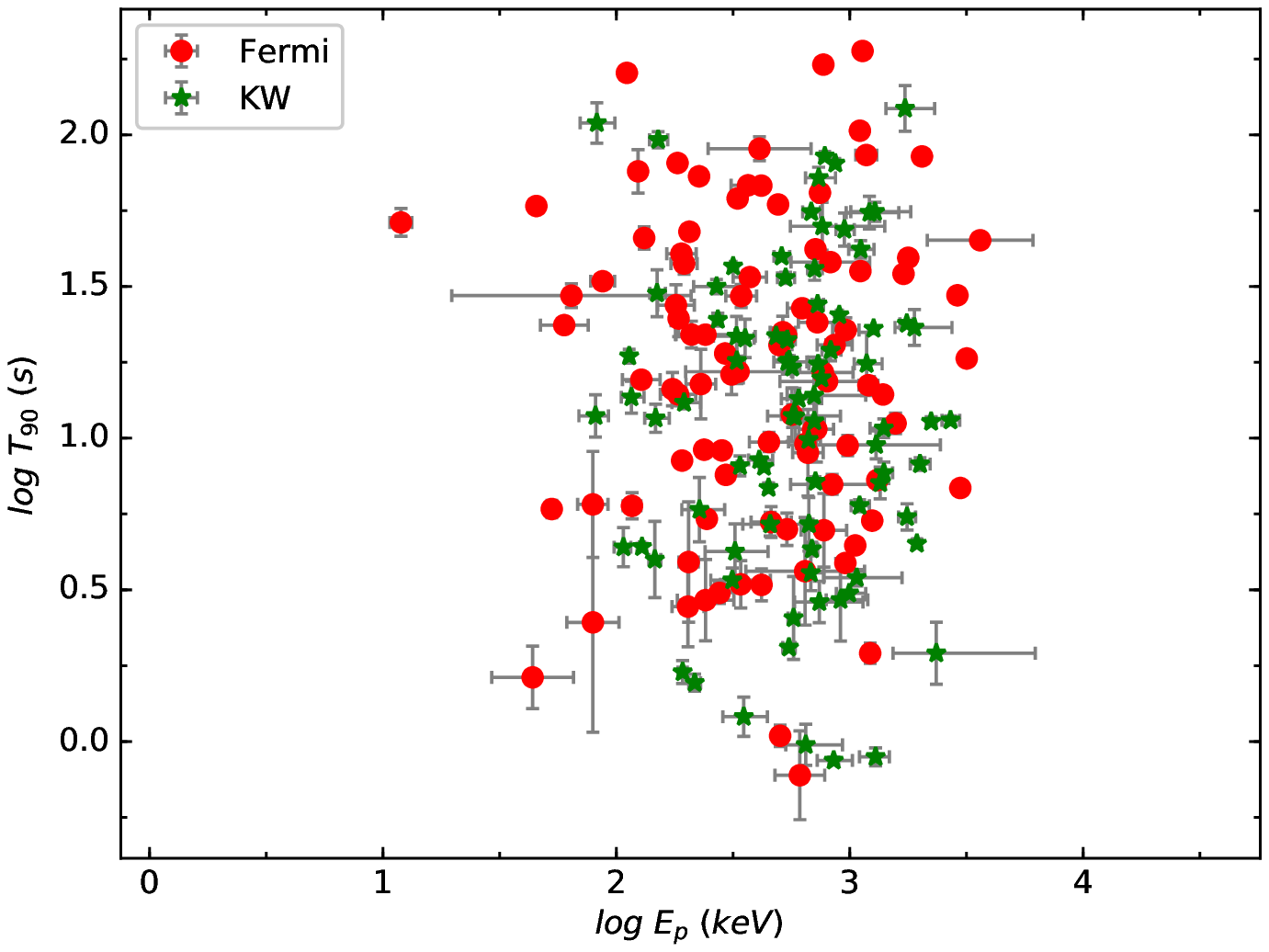}
\includegraphics[angle=0,scale=0.52]{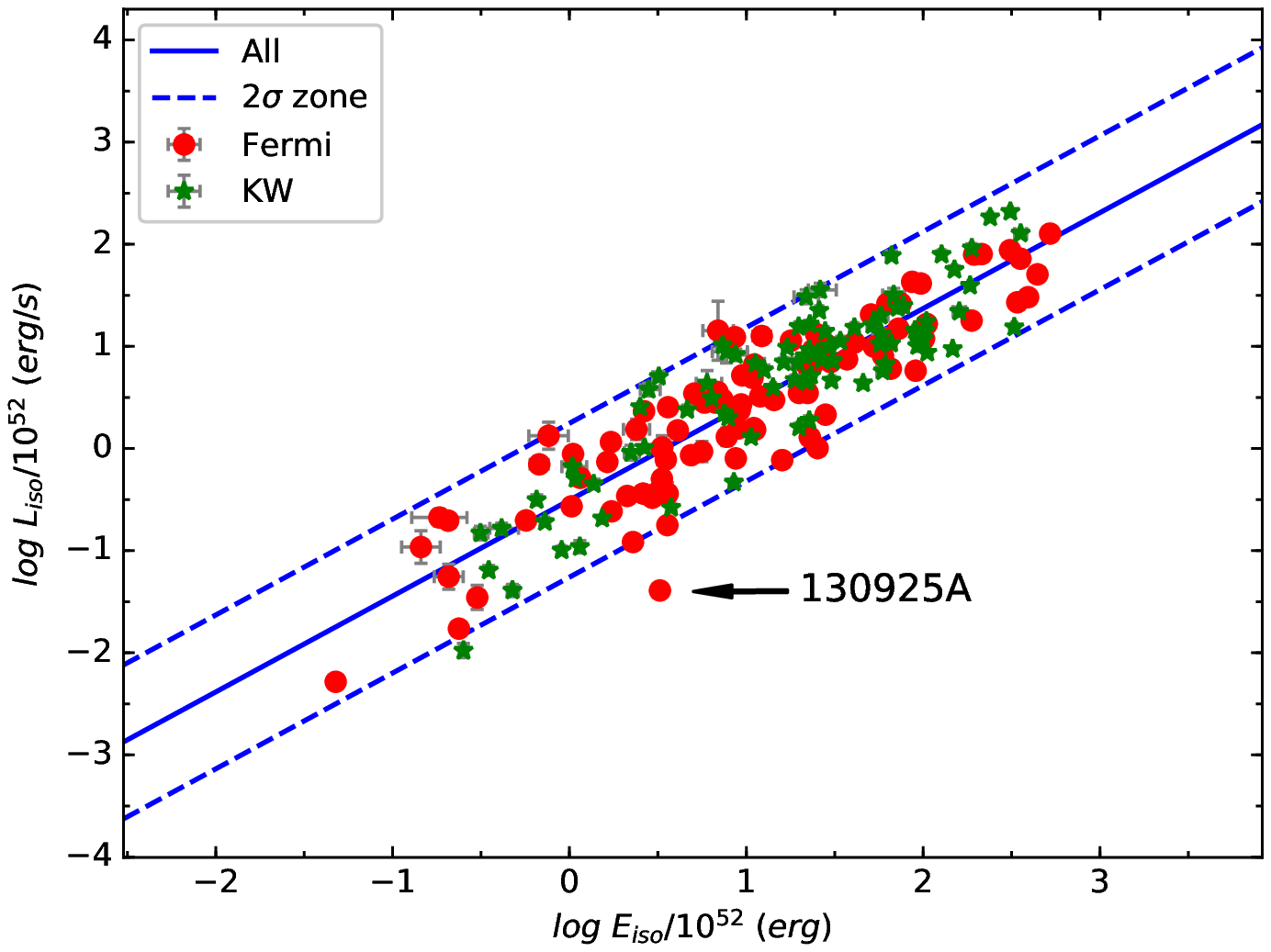}
\caption{Pair correlations among $E_{\rm p}$, $T_{\rm 90}$, $E_{\rm iso}$ and $L_{\rm iso}$ in the rest frame. Red solid circles represent Fermi sample and green asterisks represent KW sample. The solid and dashed lines represent the best-fit curves together with the 2$\sigma$ dispersion regions. The data of 46 overlapped GRBs are taken from the Fermi instrument.}
\end{figure*}

\begin{figure}
\centering
\label{figure5}
\includegraphics[angle=0,scale=0.58]{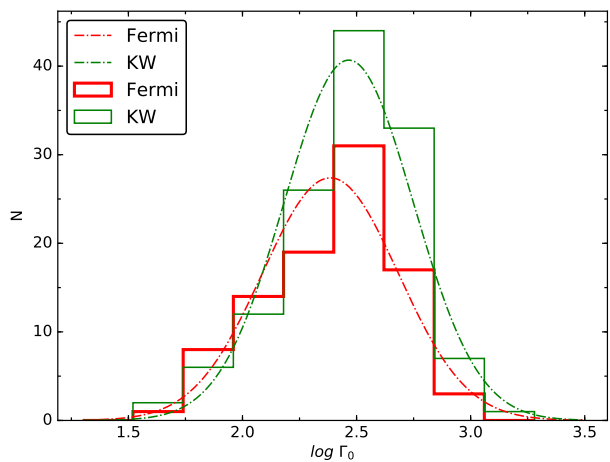}
\caption{Distribution of $\Gamma_0$, where the dashed lines are the best fitting curves.}
\end{figure}

\begin{figure*}
\centering
\label{figure6}
\includegraphics[angle=0,scale=0.58]{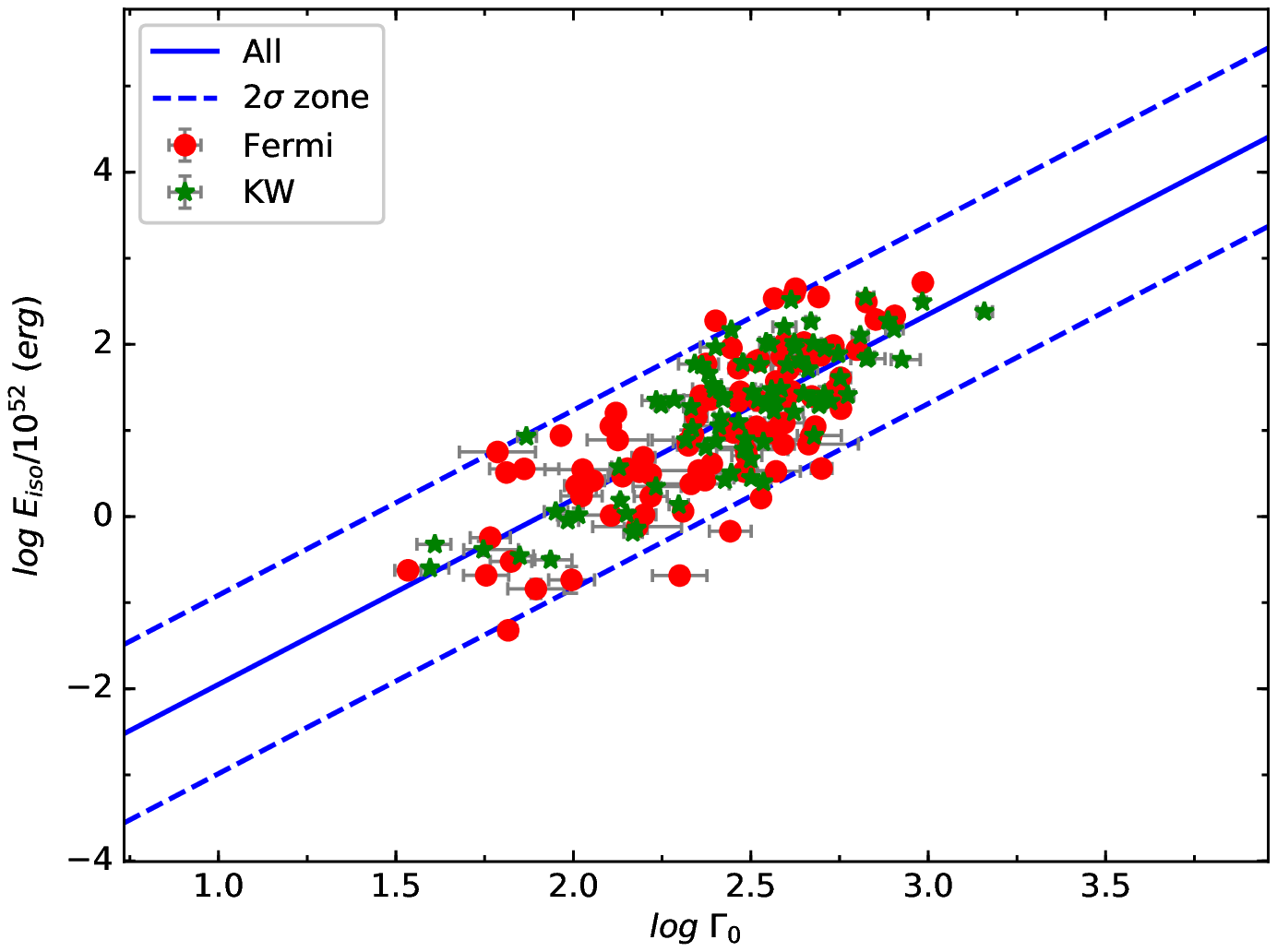}
\includegraphics[angle=0,scale=0.58]{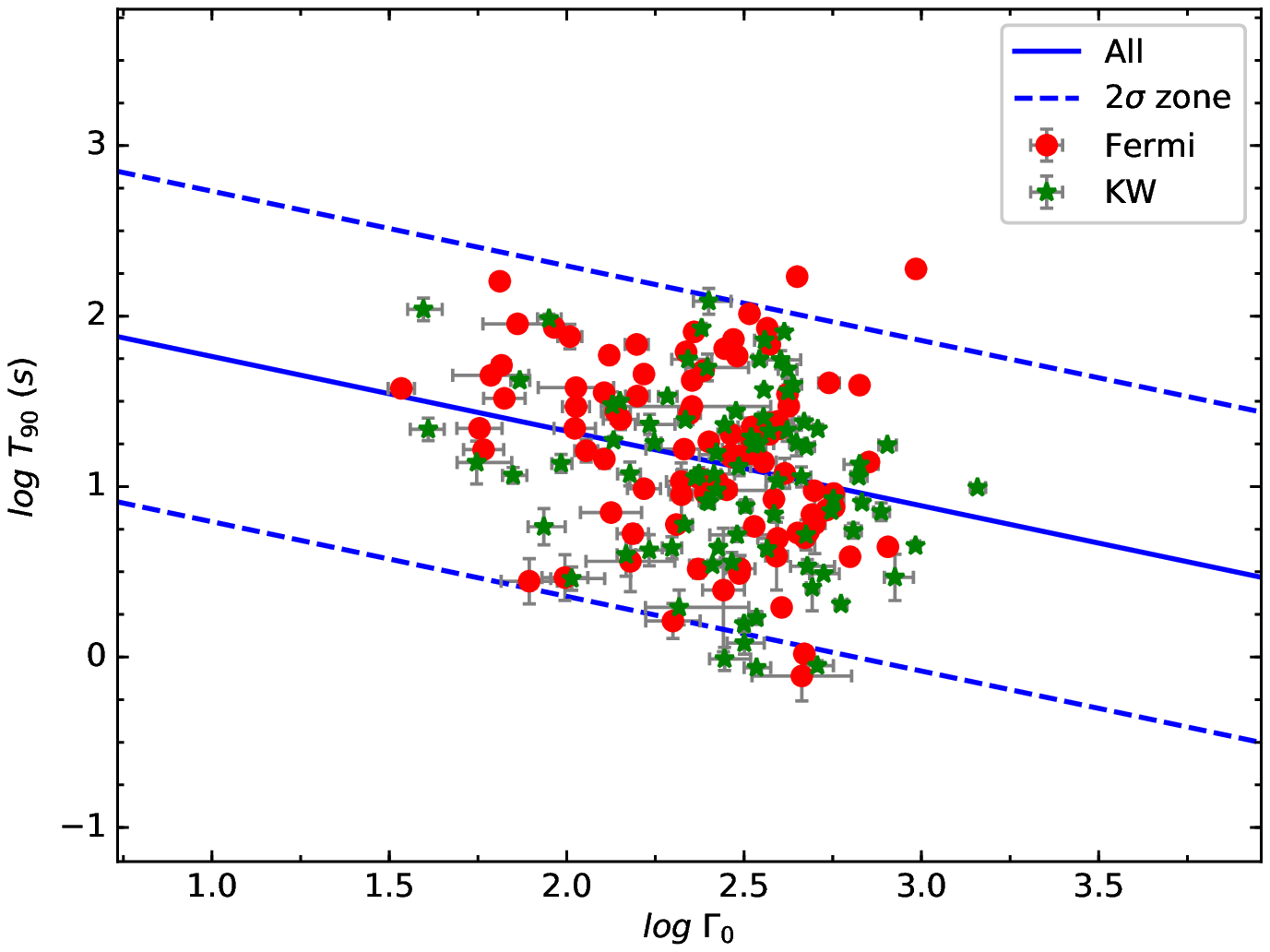}
\caption{Relations between $\Gamma_0$ and $E_{\rm iso}$($T_{\rm 90}$). The other symbols are the same as Figure 4.}
\end{figure*}

\begin{figure*}
\centering
\label{figure7}
\includegraphics[angle=0,scale=0.5]{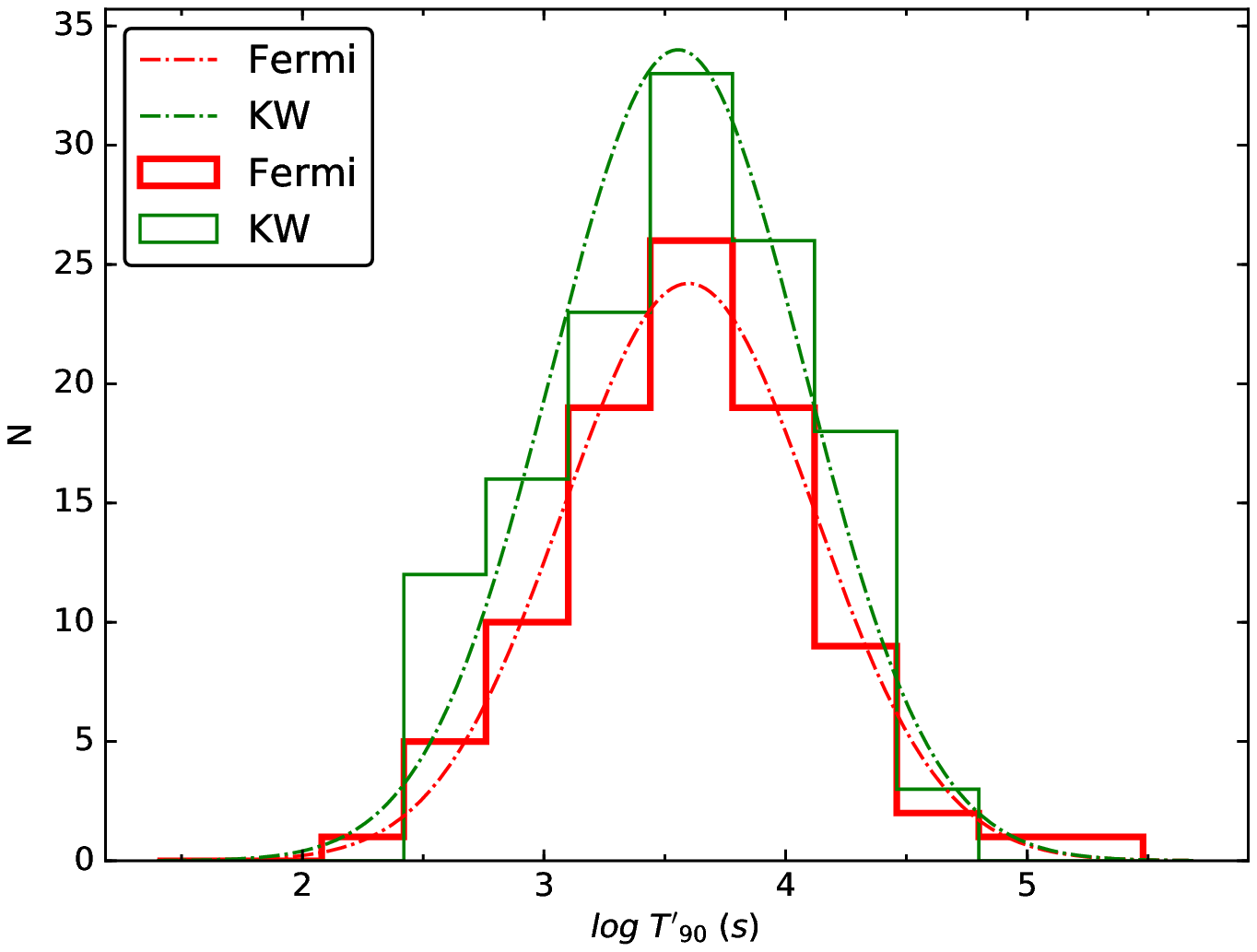}
\includegraphics[angle=0,scale=0.5]{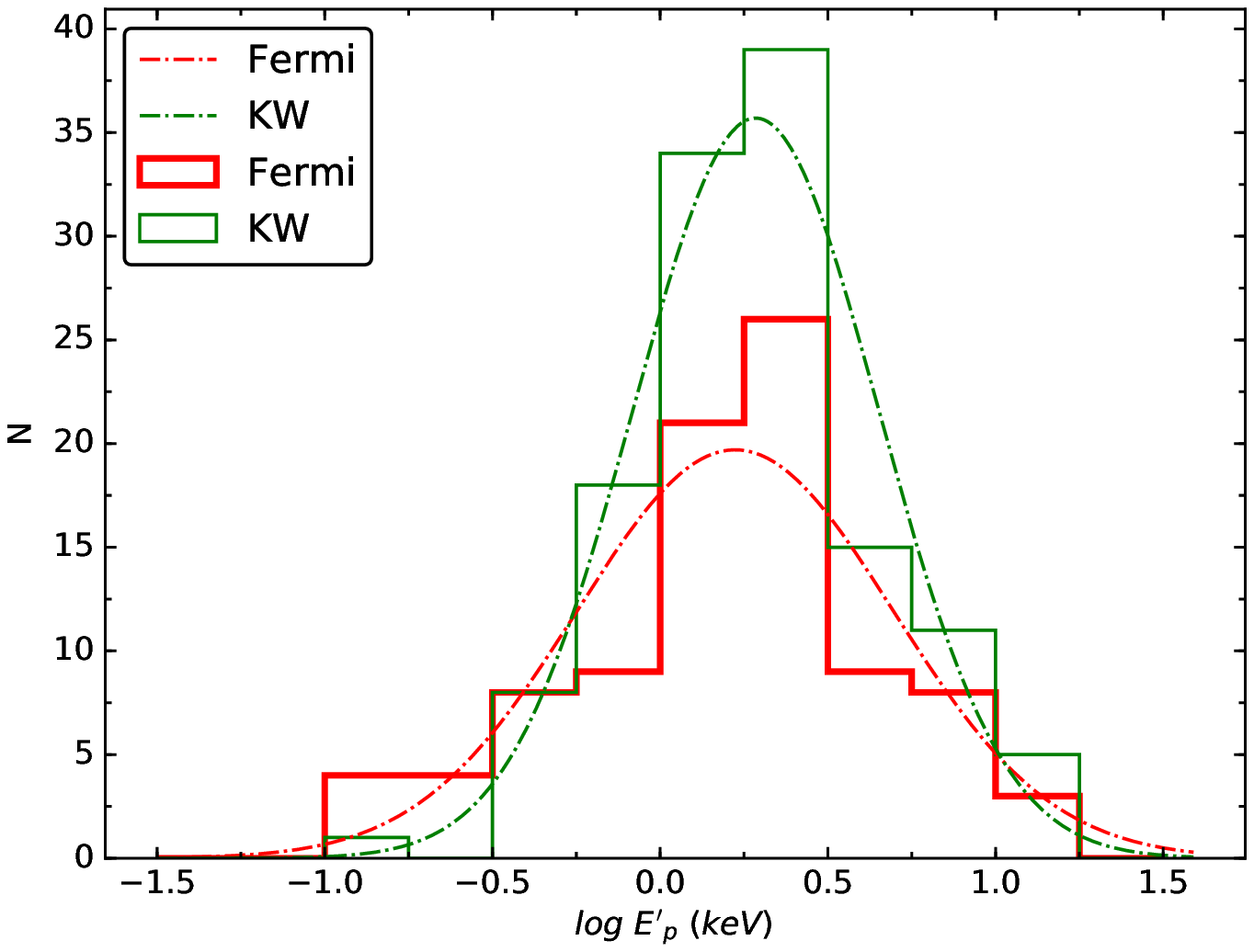}
\includegraphics[angle=0,scale=0.5]{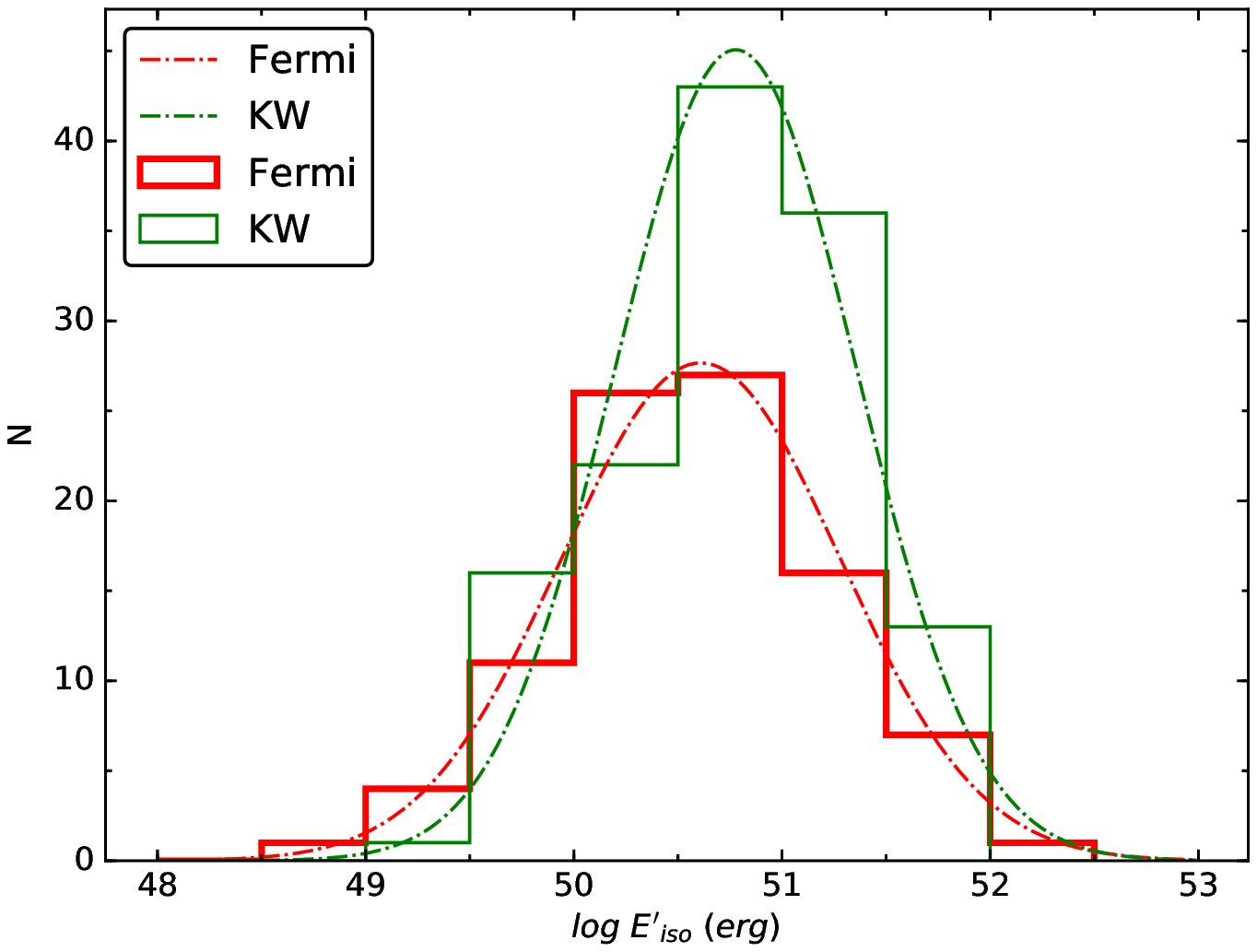}
\includegraphics[angle=0,scale=0.5]{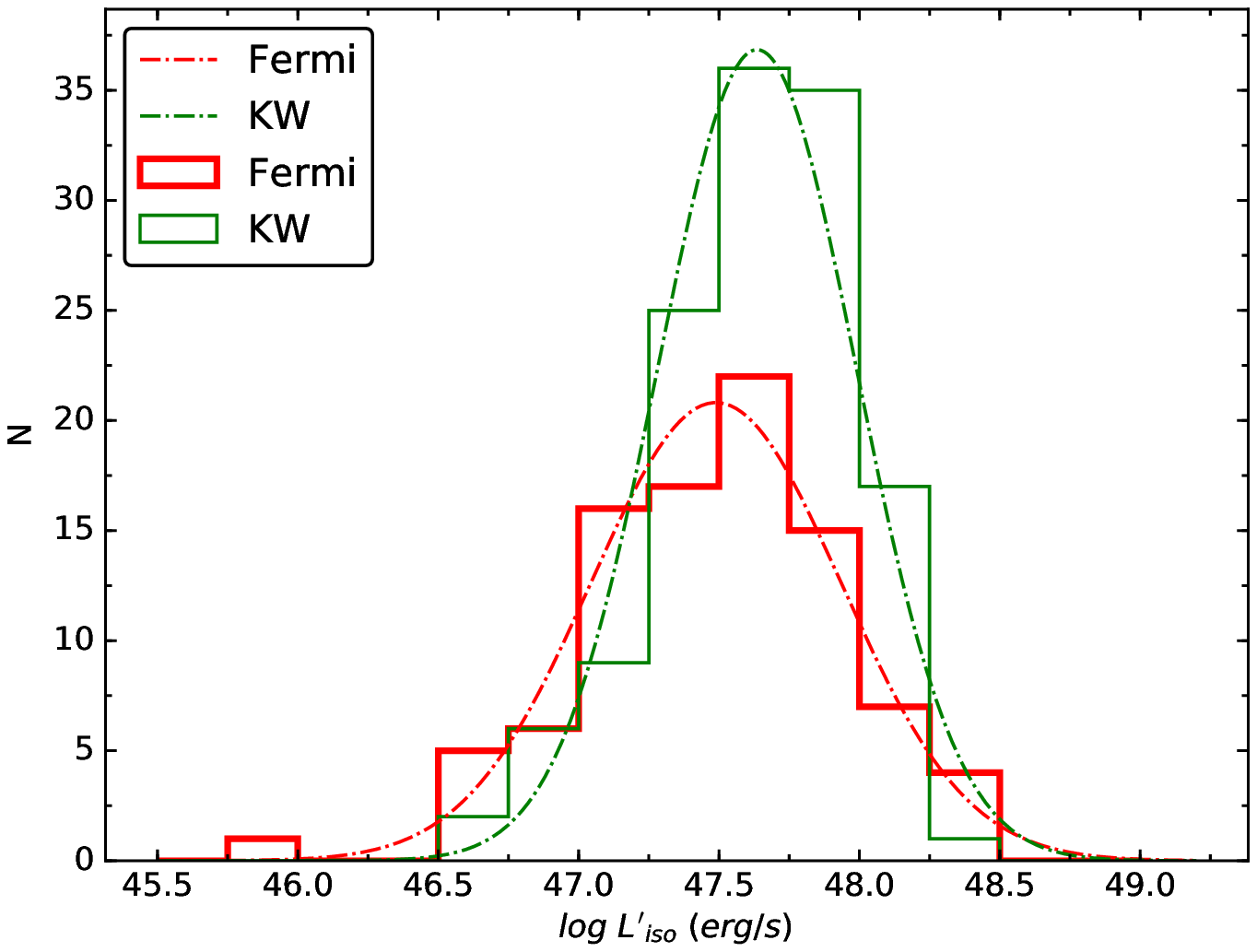}
\caption{Distributions of the comoving frame duration $T'_{\rm 90}$, spectrum peak energy $E'_{\rm p}$, isotropic energy $E'_{\rm iso}$ and isotropic luminosity $L'_{\rm iso}$. The other symbols are the same as Figure 3.}
\end{figure*}

\begin{figure*}
\centering
\label{figure8}
\includegraphics[angle=0,scale=0.52]{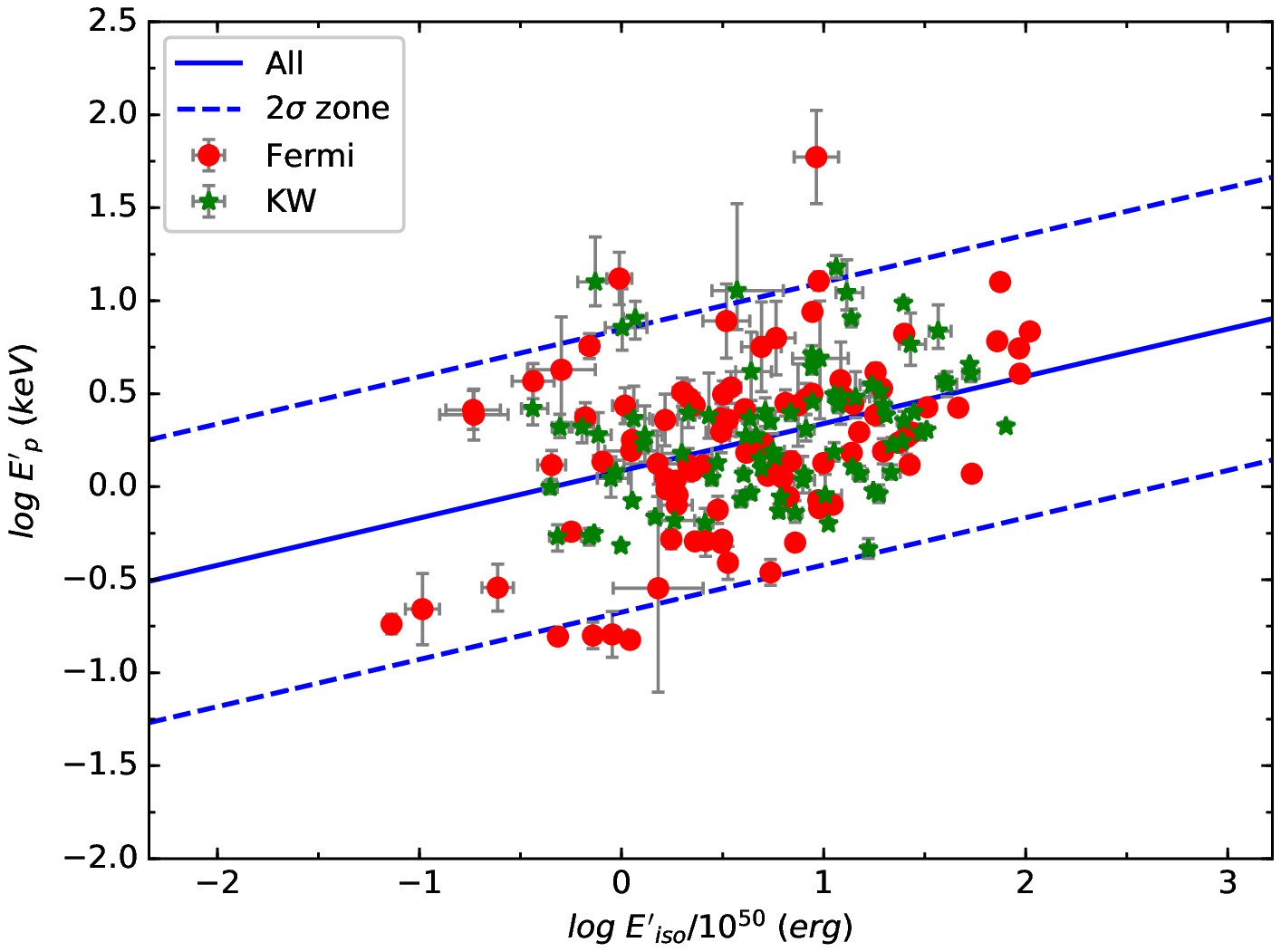}
\includegraphics[angle=0,scale=0.52]{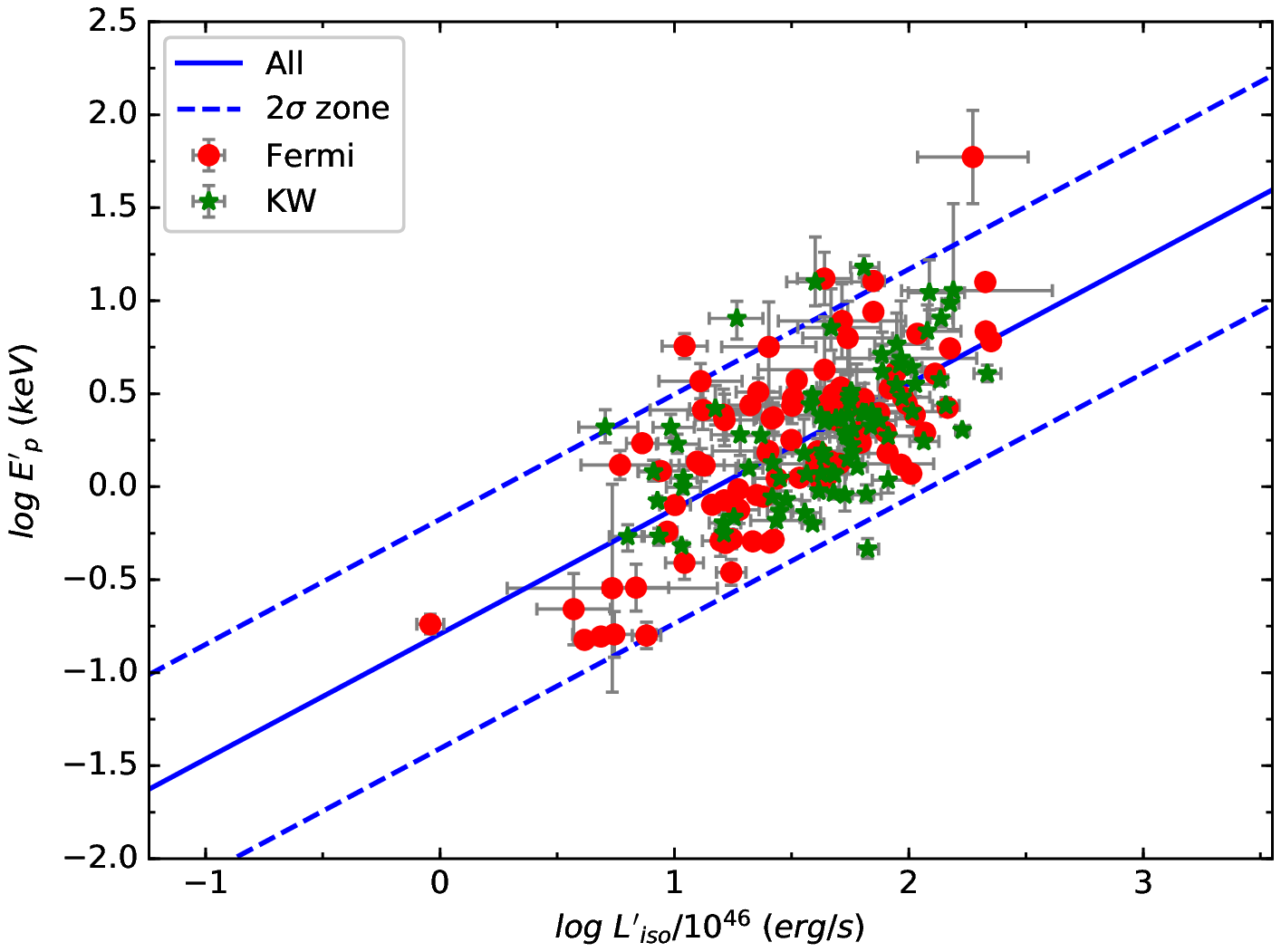}
\includegraphics[angle=0,scale=0.52]{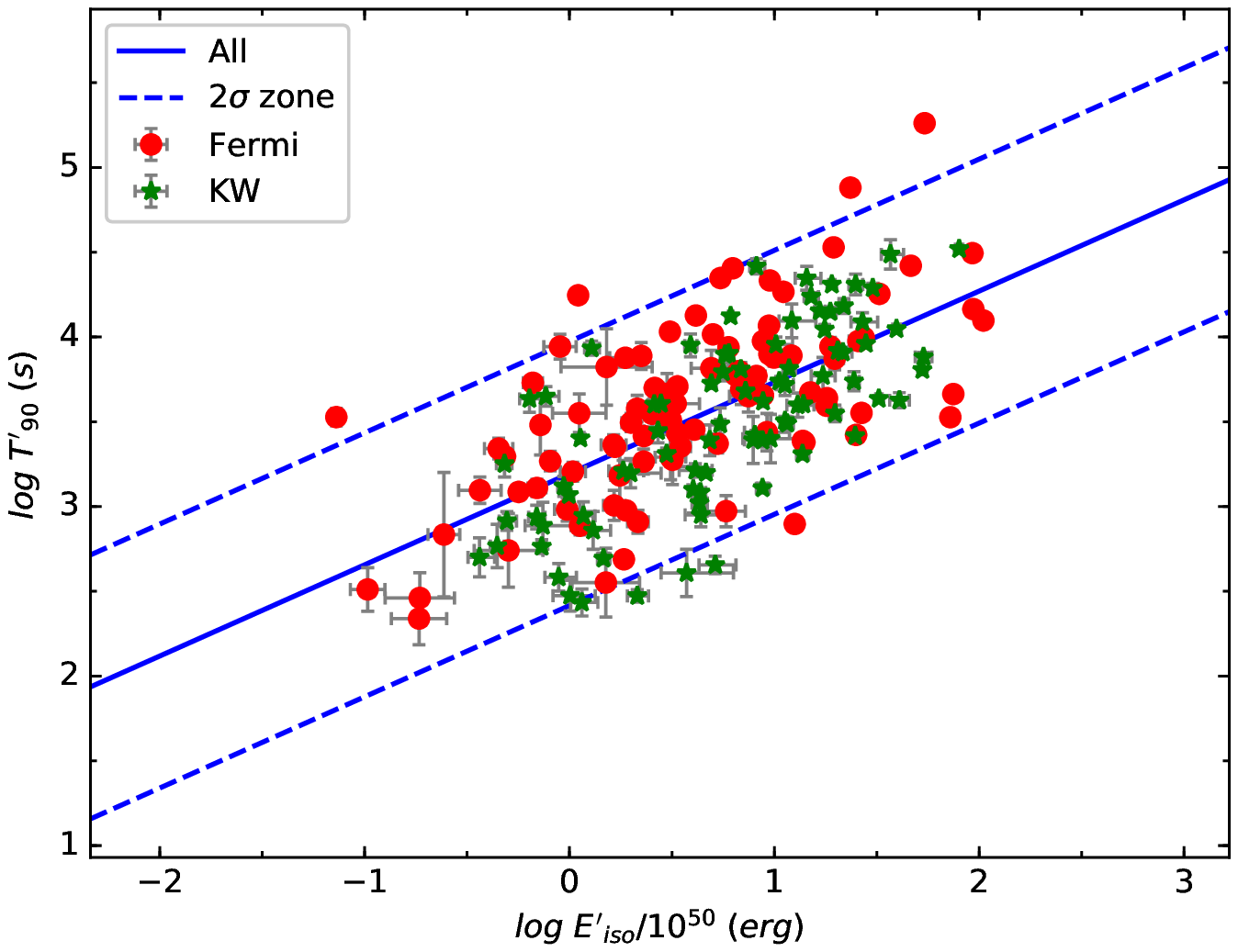}
\includegraphics[angle=0,scale=0.52]{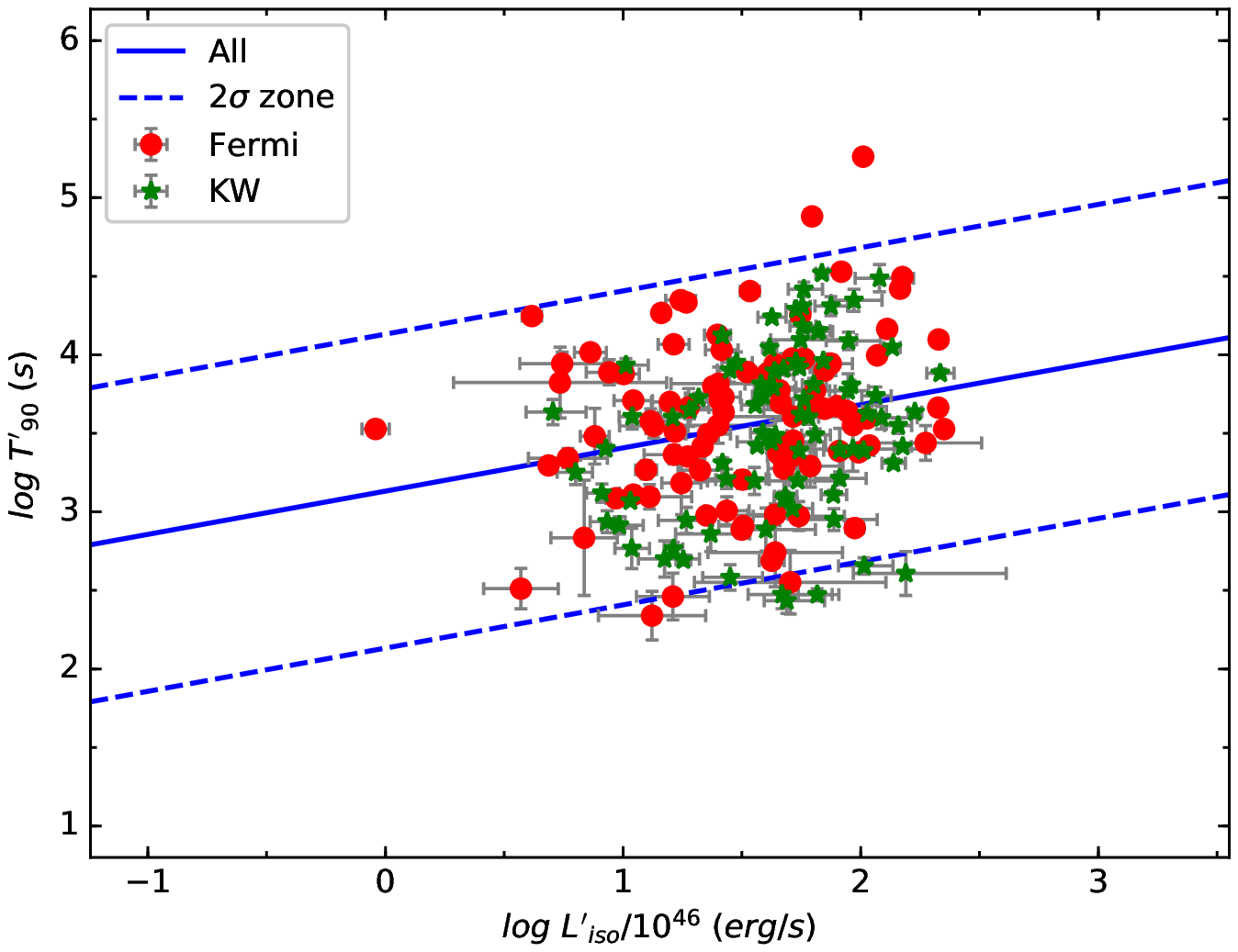}
\includegraphics[angle=0,scale=0.52]{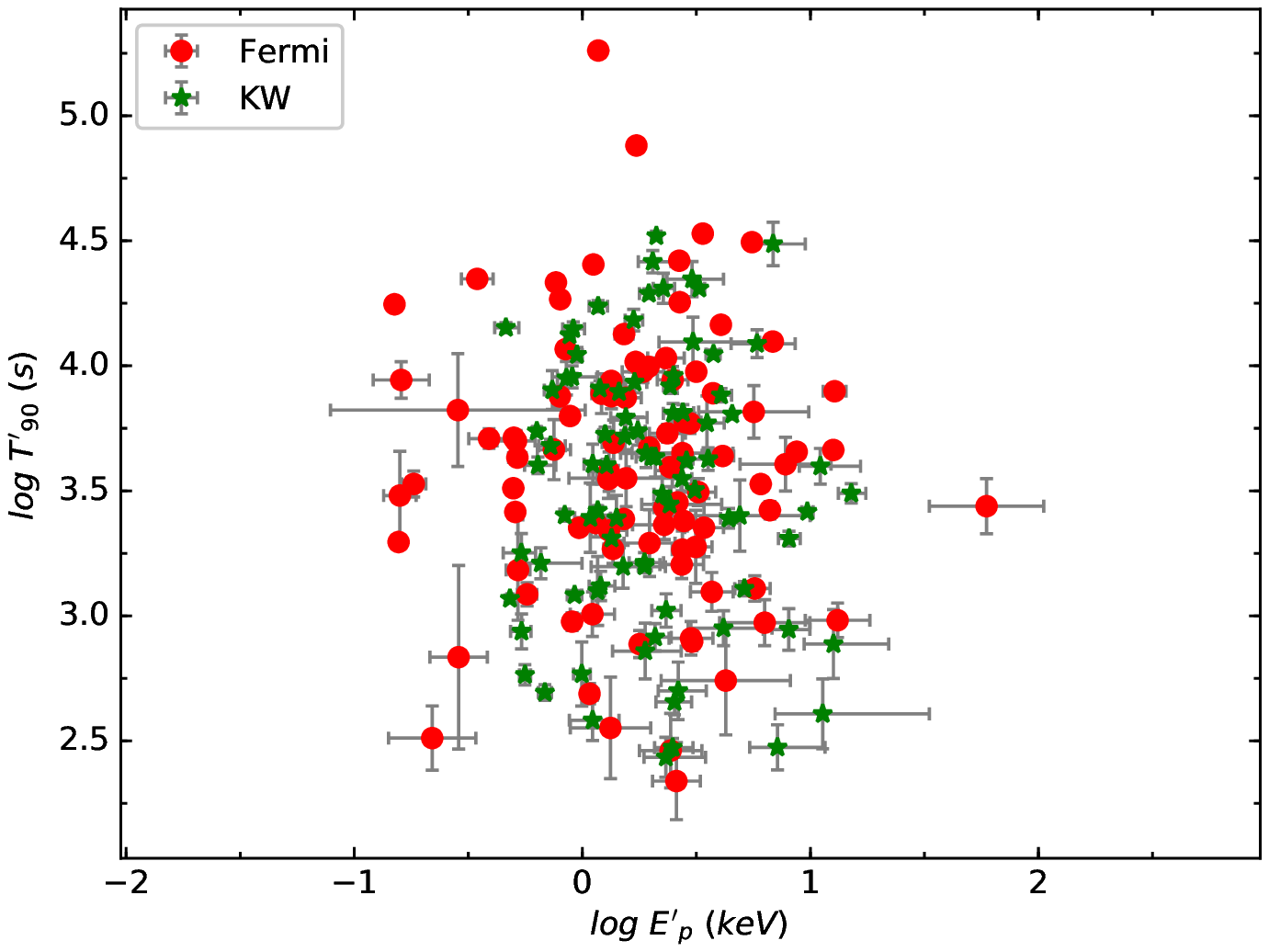}
\includegraphics[angle=0,scale=0.52]{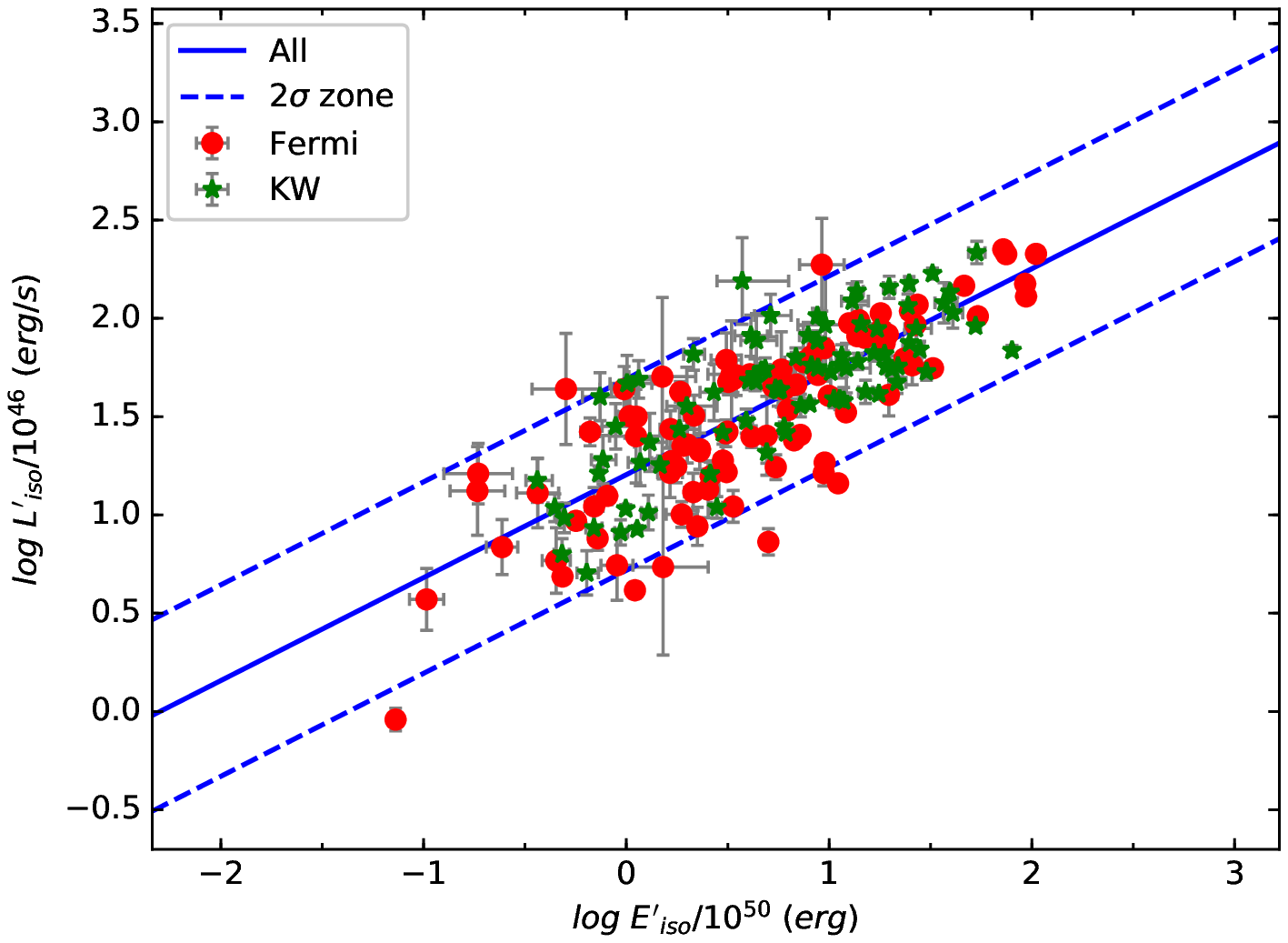}
\caption{Pair correlations among $E'_{p}$, $T'_{90}$, $E'_{iso}$ and $L'_{iso}$. The other symbols are the same as Figure 4.}
\end{figure*}

\begin{figure*}
\label{figure9}
\includegraphics[angle=0,scale=0.58]{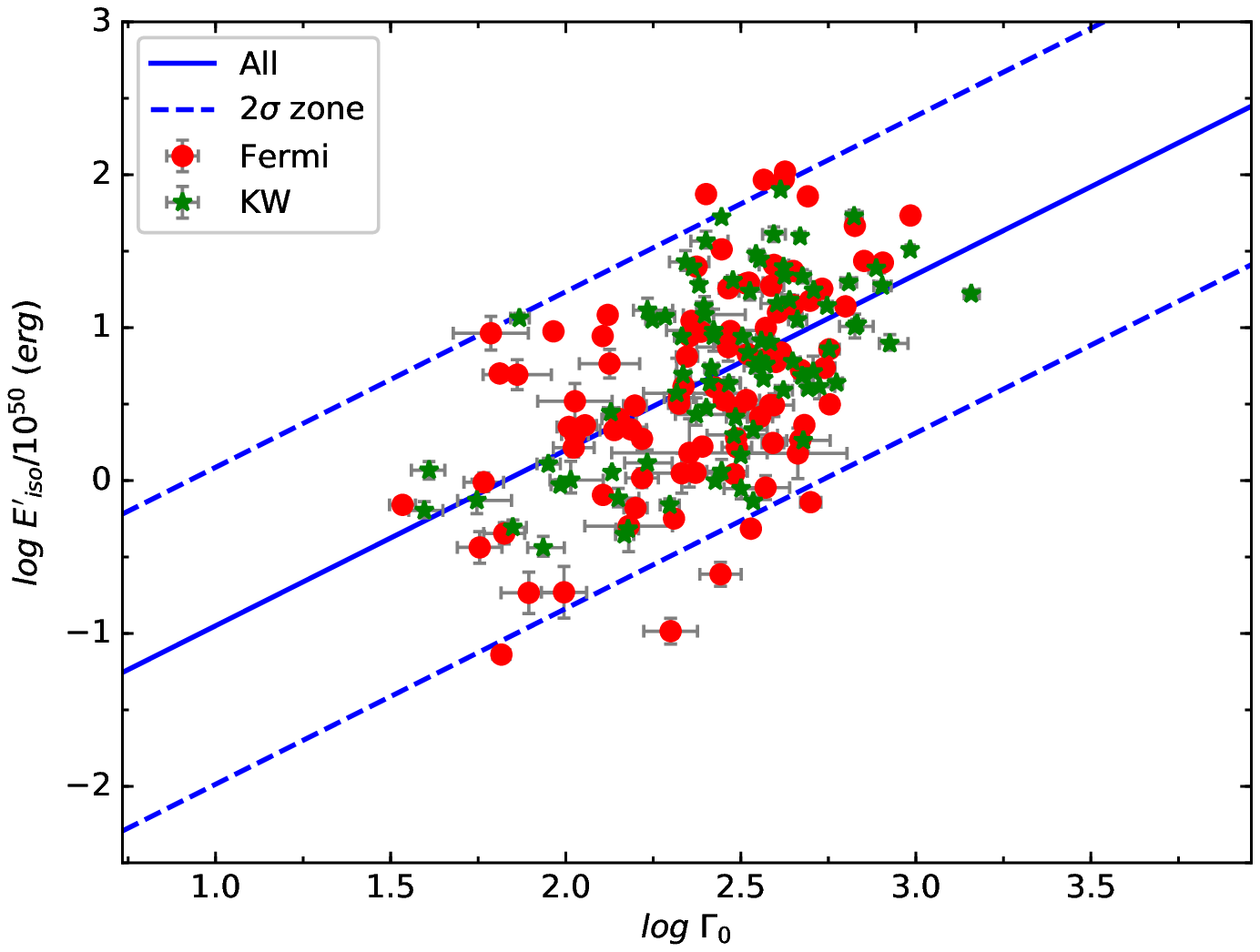}
\includegraphics[angle=0,scale=0.58]{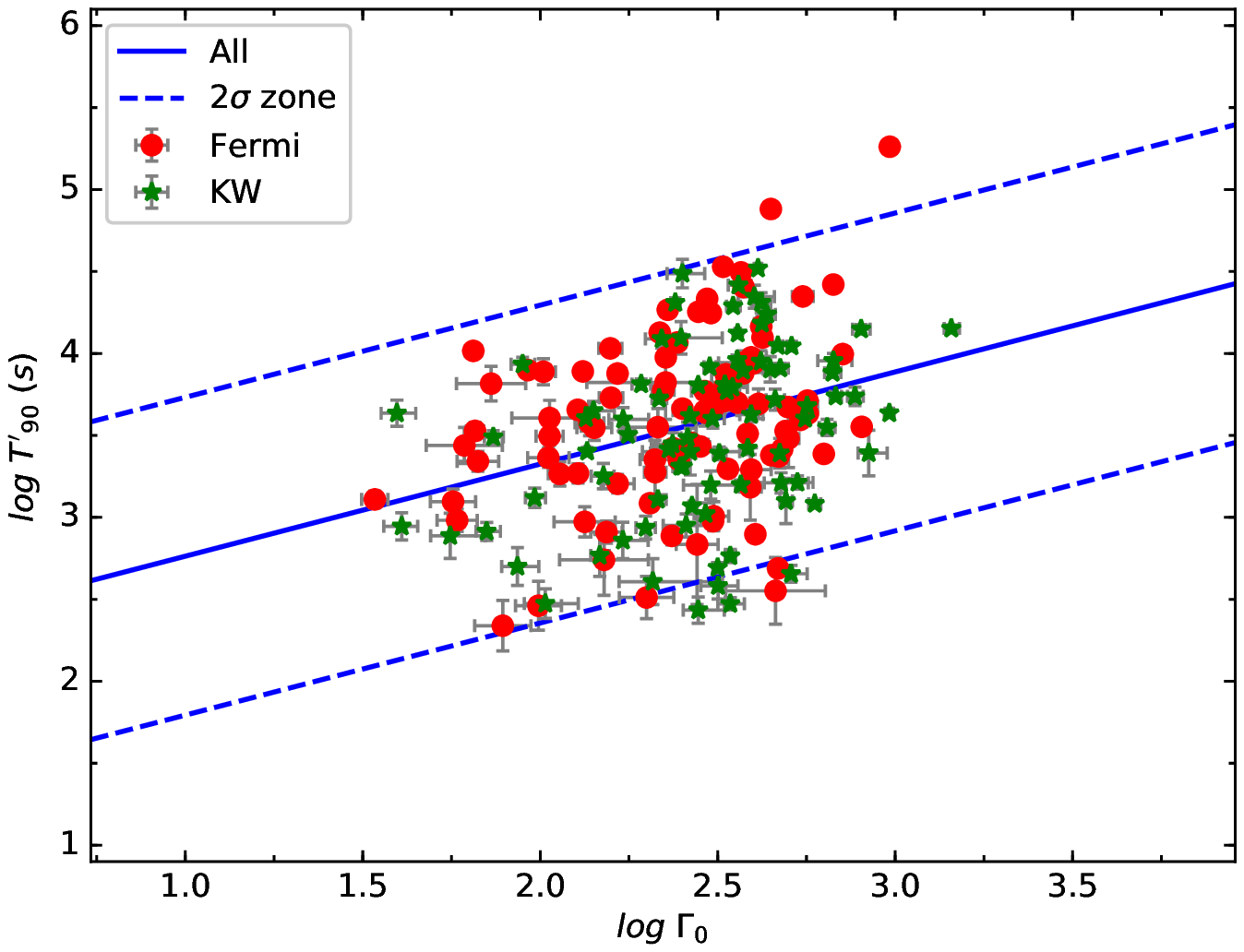}
\includegraphics[angle=0,scale=0.58]{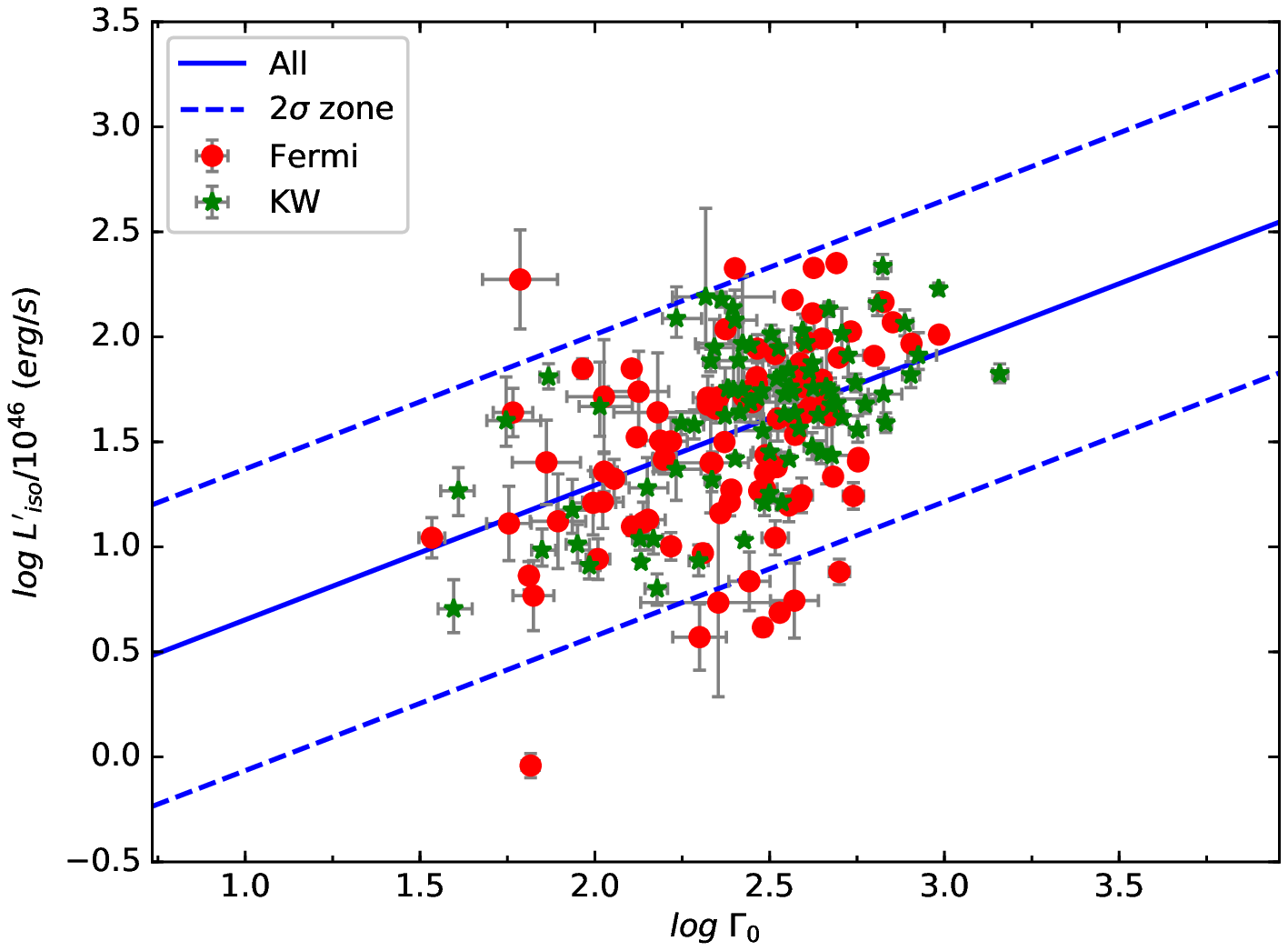}
\includegraphics[angle=0,scale=0.58]{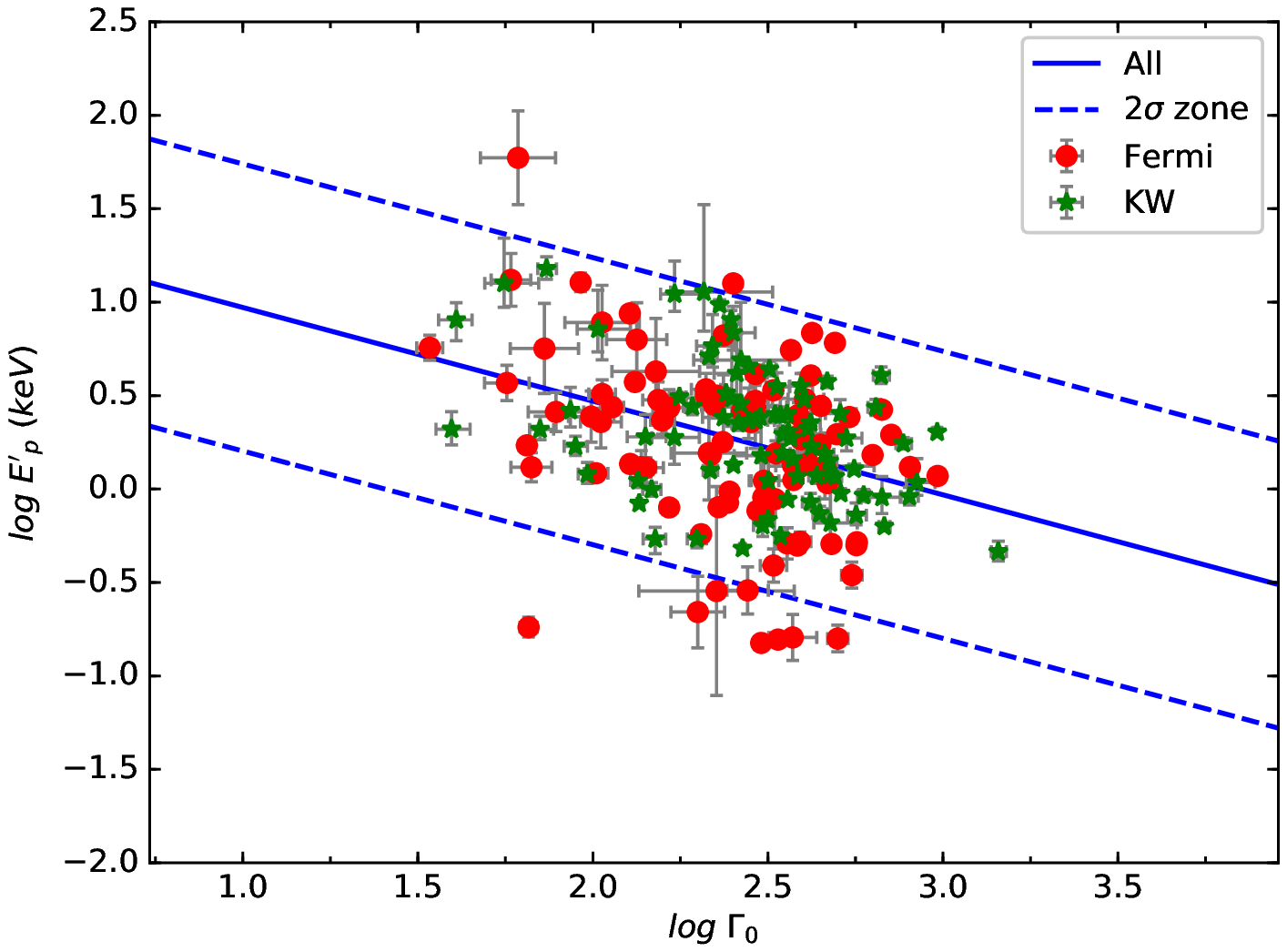}
\caption{Relations between $\Gamma_0$ and $E'_{\rm iso}$($T'_{\rm 90}$, $L'_{\rm iso}$, $E'_{p}$). The other symbols are the same as Figure 4.}
\end{figure*}

\begin{figure*}
\centering
\label{figure10}
\includegraphics[angle=0,scale=0.52]{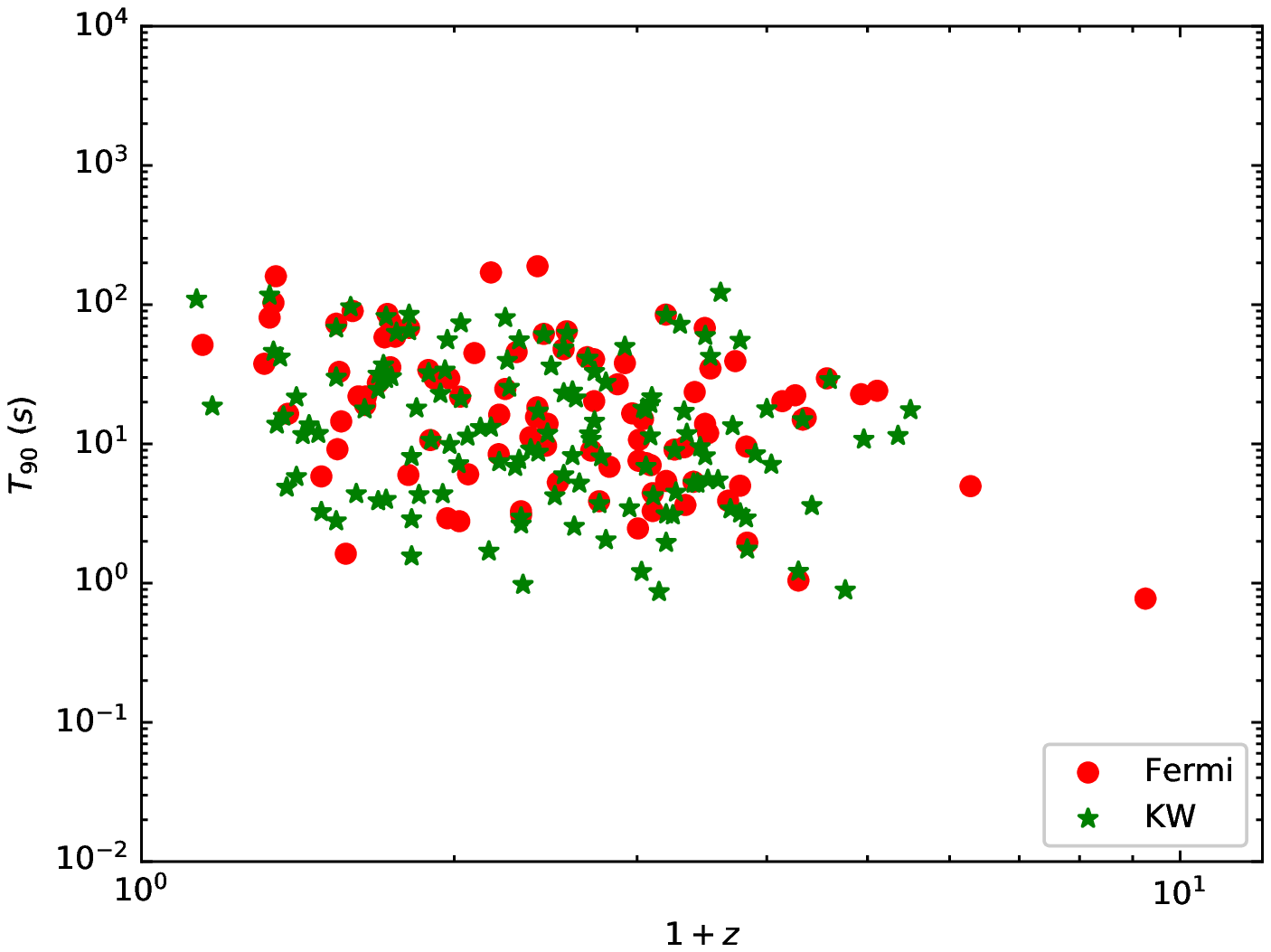}
\includegraphics[angle=0,scale=0.52]{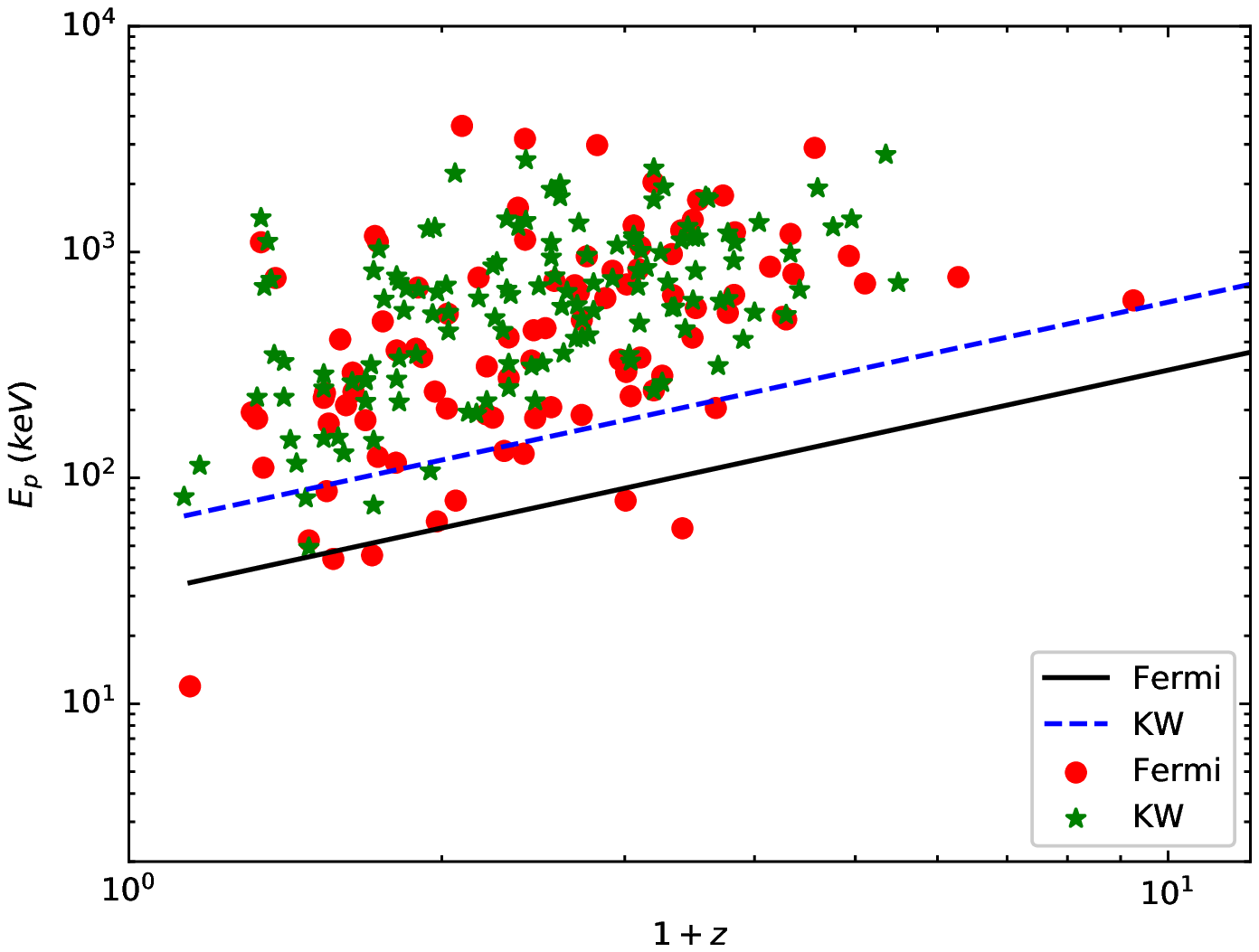}
\includegraphics[angle=0,scale=0.52]{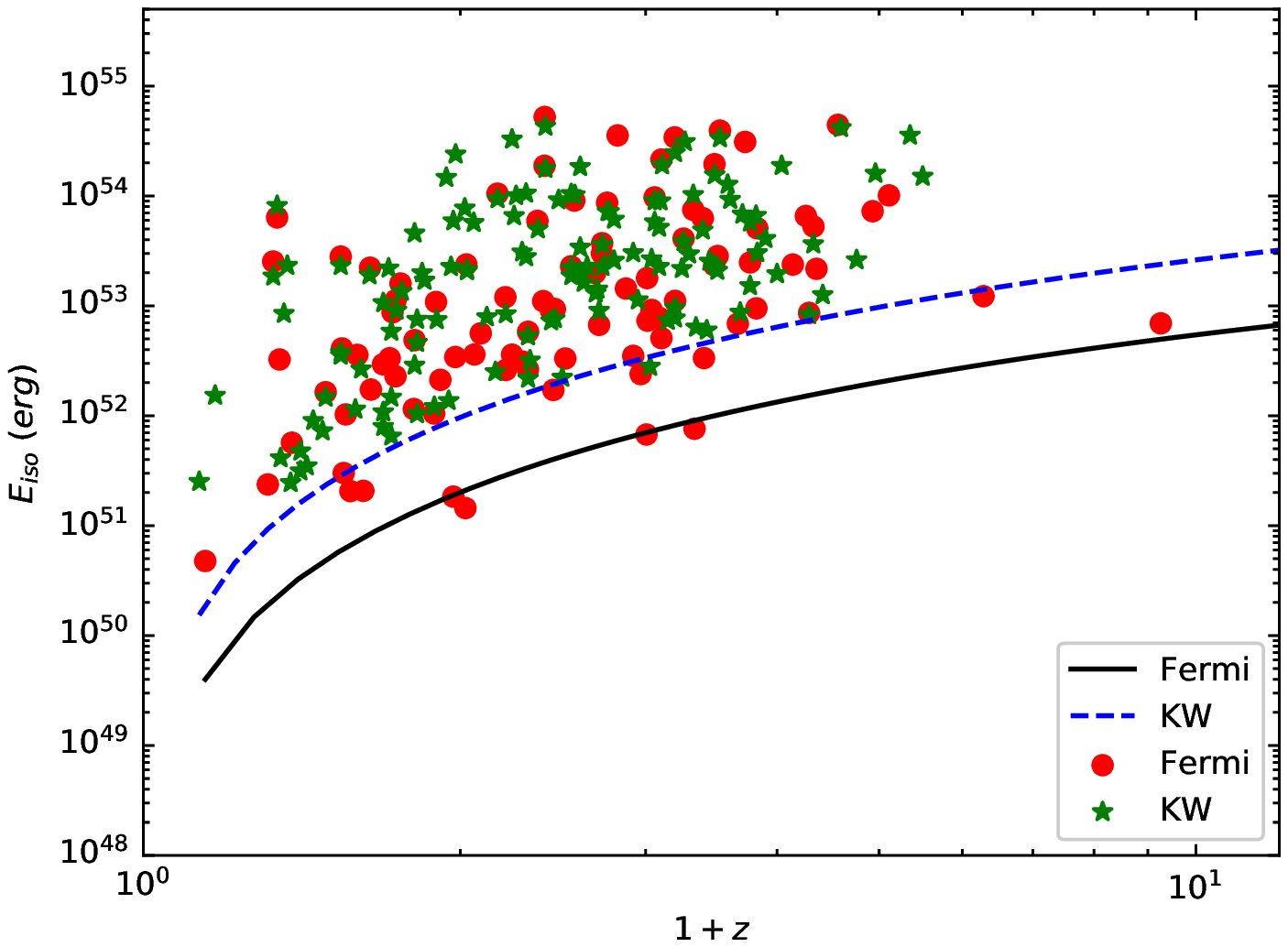}
\includegraphics[angle=0,scale=0.52]{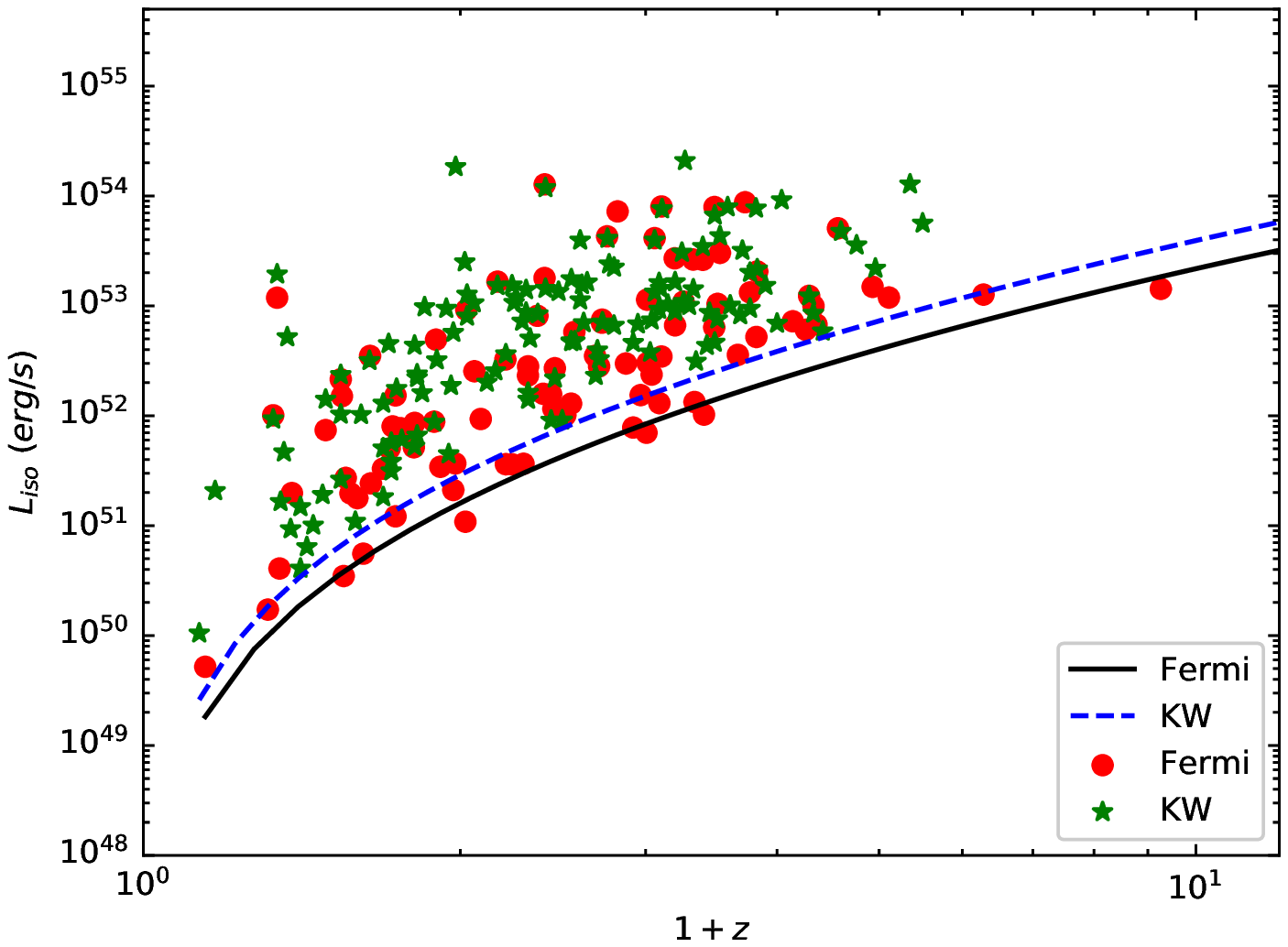}
\caption{ $T_{\rm 90}$, $E_{\rm p}$, $E_{\rm iso}$ and  $L_{\rm iso}$ vs. redshift in the rest frame. The observer frame limits (Section 3.5) are shown with the solid lines for Fermi data and dashed lines for KW data.}
\end{figure*}

\begin{figure*}
\centering
\label{figure11}
\includegraphics[angle=0,scale=0.52]{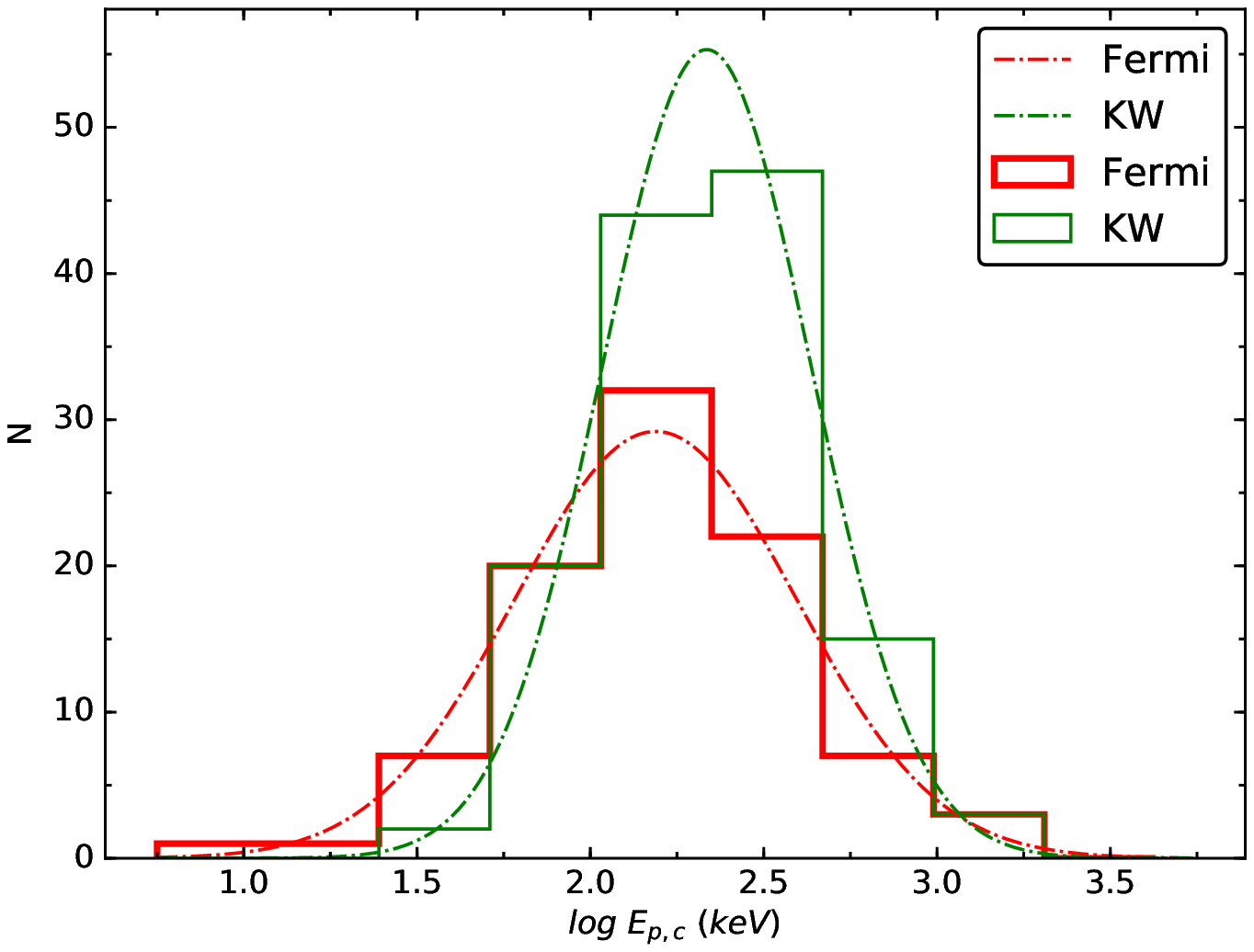}
\includegraphics[angle=0,scale=0.52]{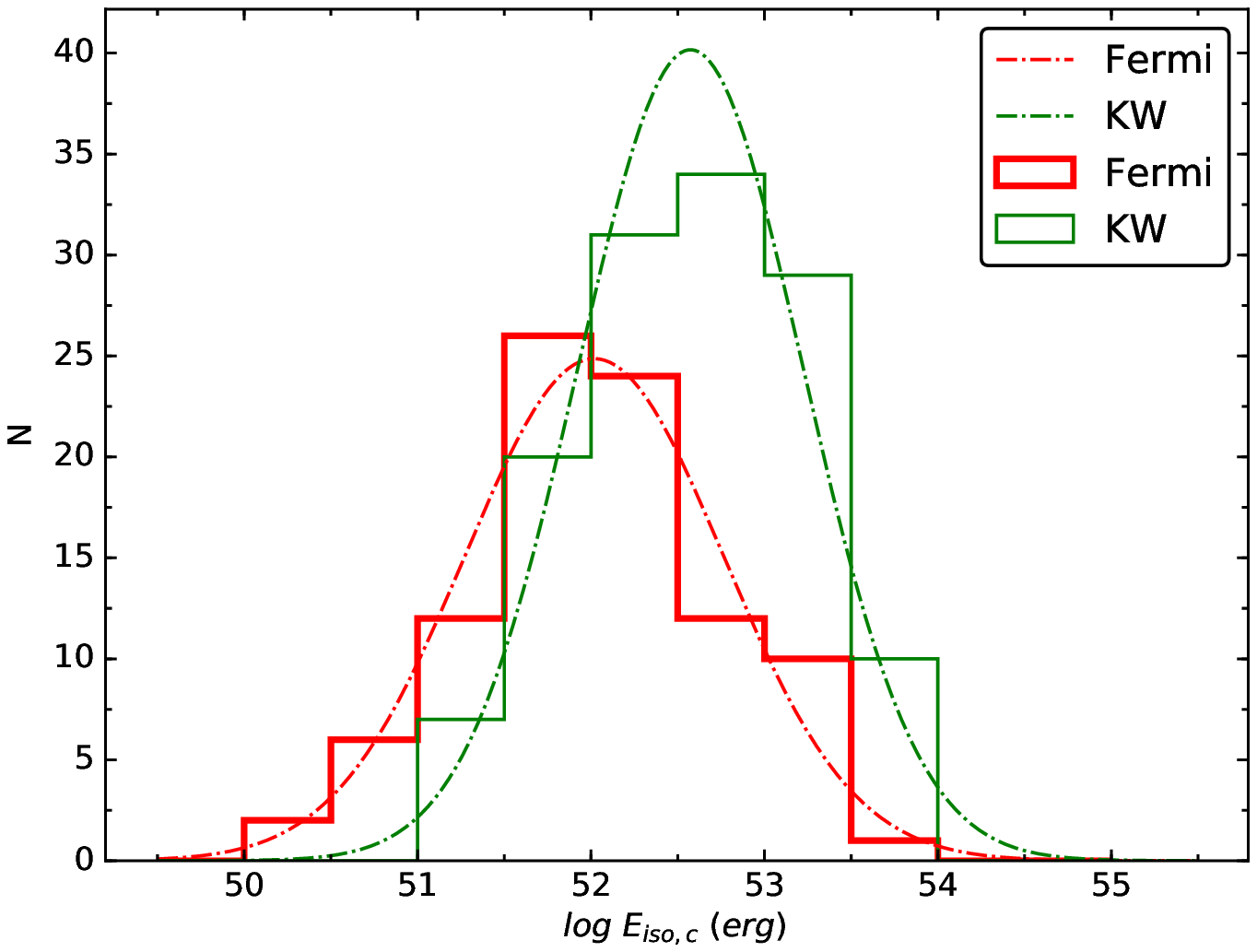}
\includegraphics[angle=0,scale=0.52]{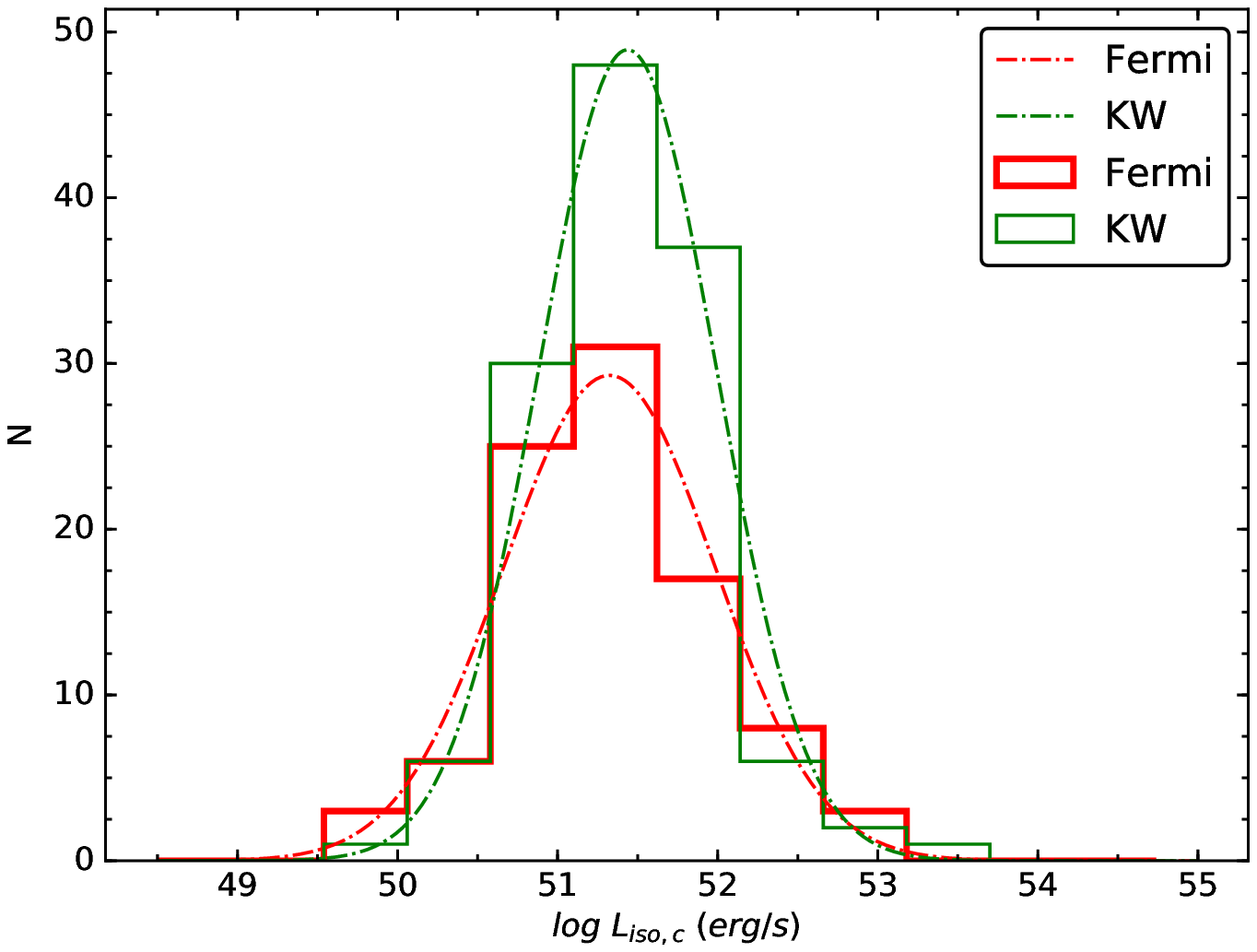}
\includegraphics[angle=0,scale=0.52]{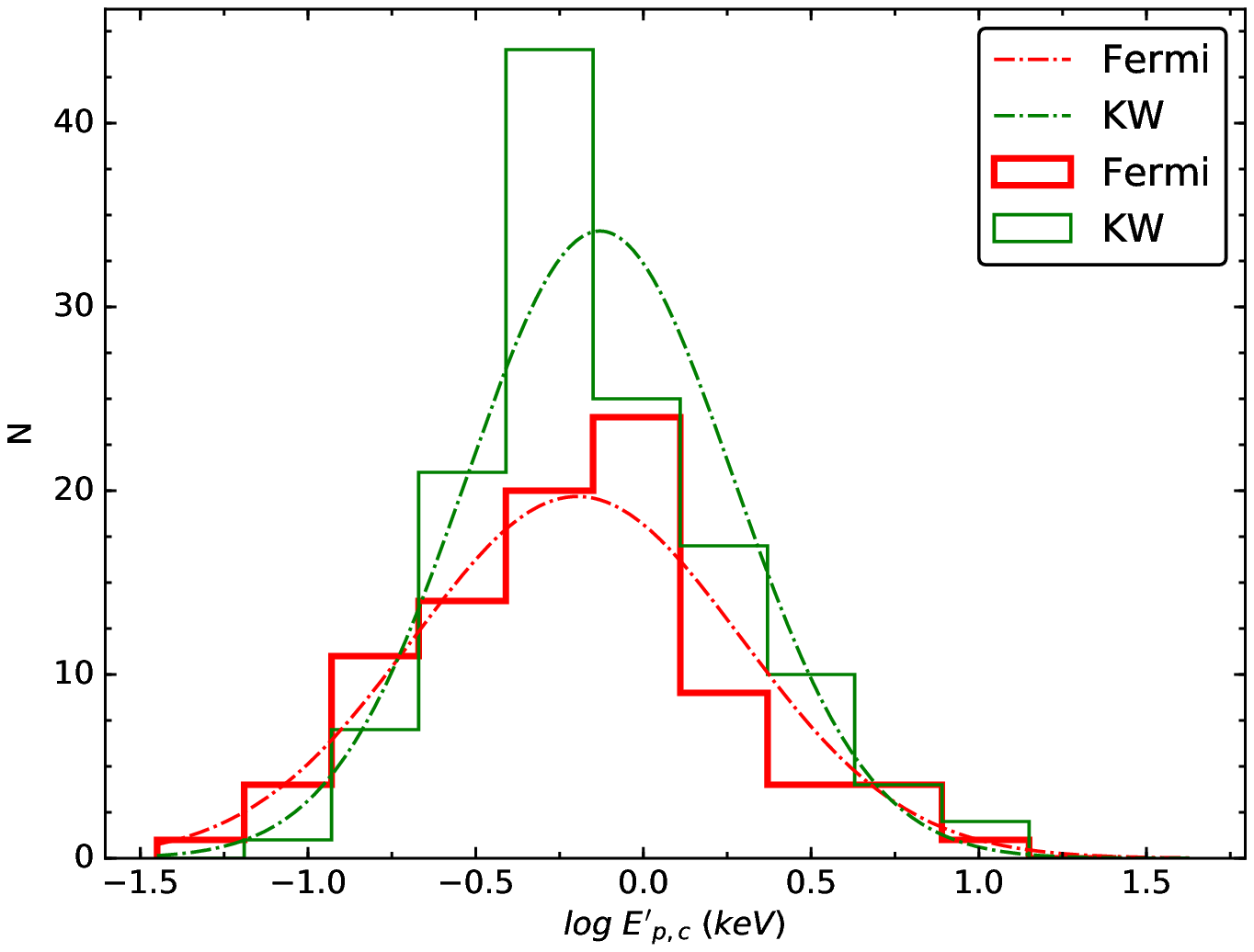}
\includegraphics[angle=0,scale=0.52]{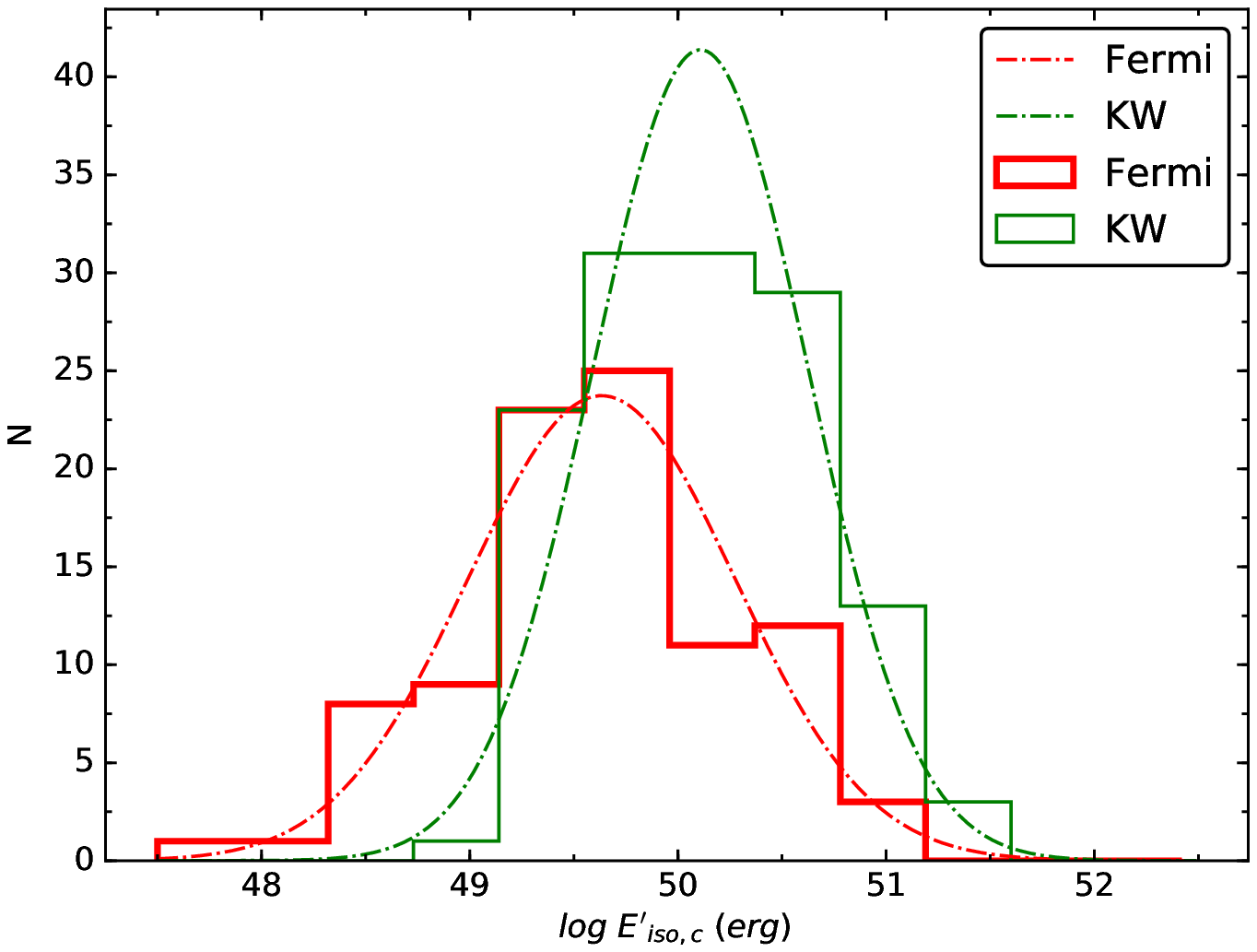}
\includegraphics[angle=0,scale=0.52]{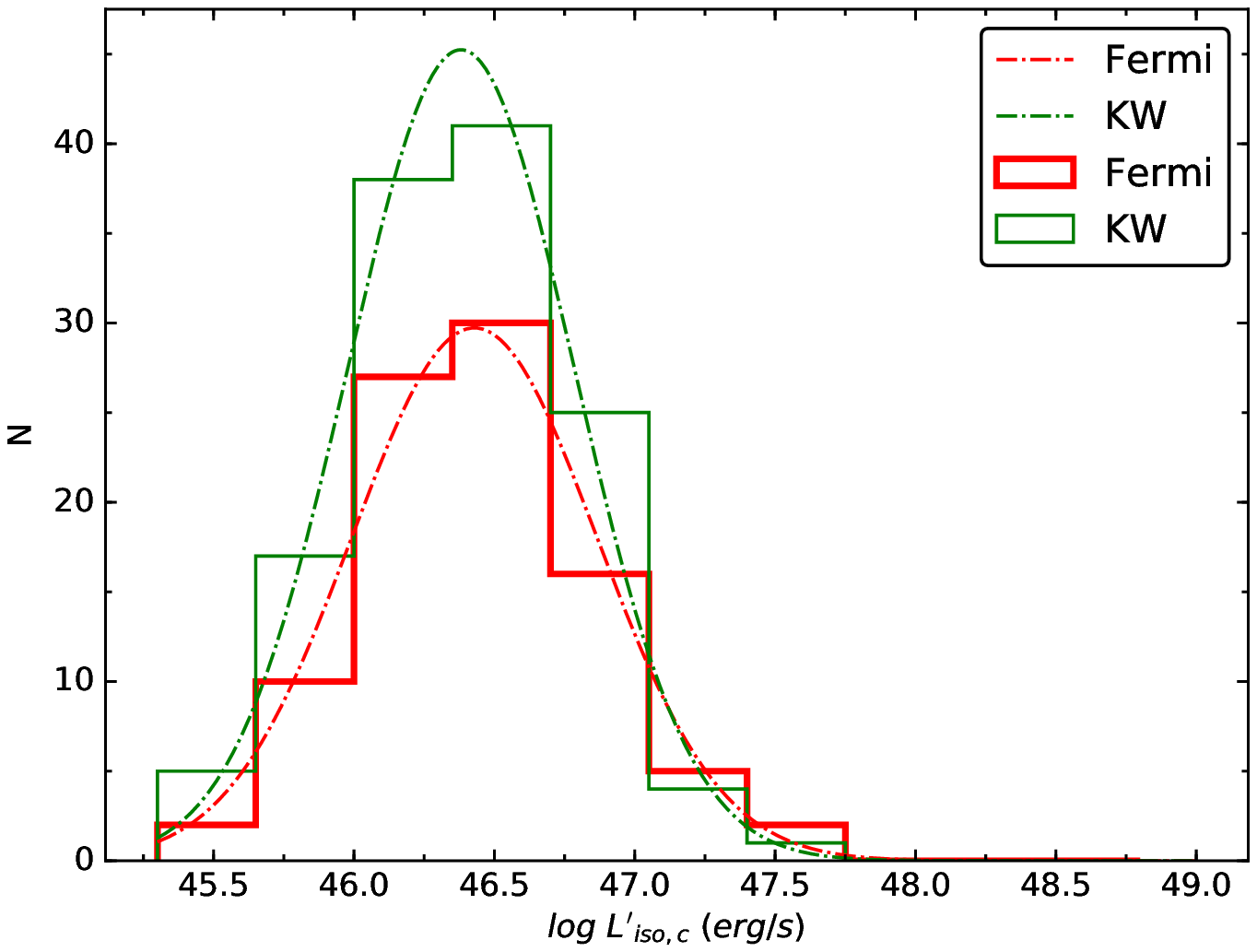}
\caption{Distributions of the de-evolved spectrum peak energy, isotropic energy  and luminosity in the rest frame and comving frame. The other symbols are the same as Figure 3.}
\end{figure*}

\begin{figure*}
\centering
\label{figure12}
\includegraphics[angle=0,scale=0.52]{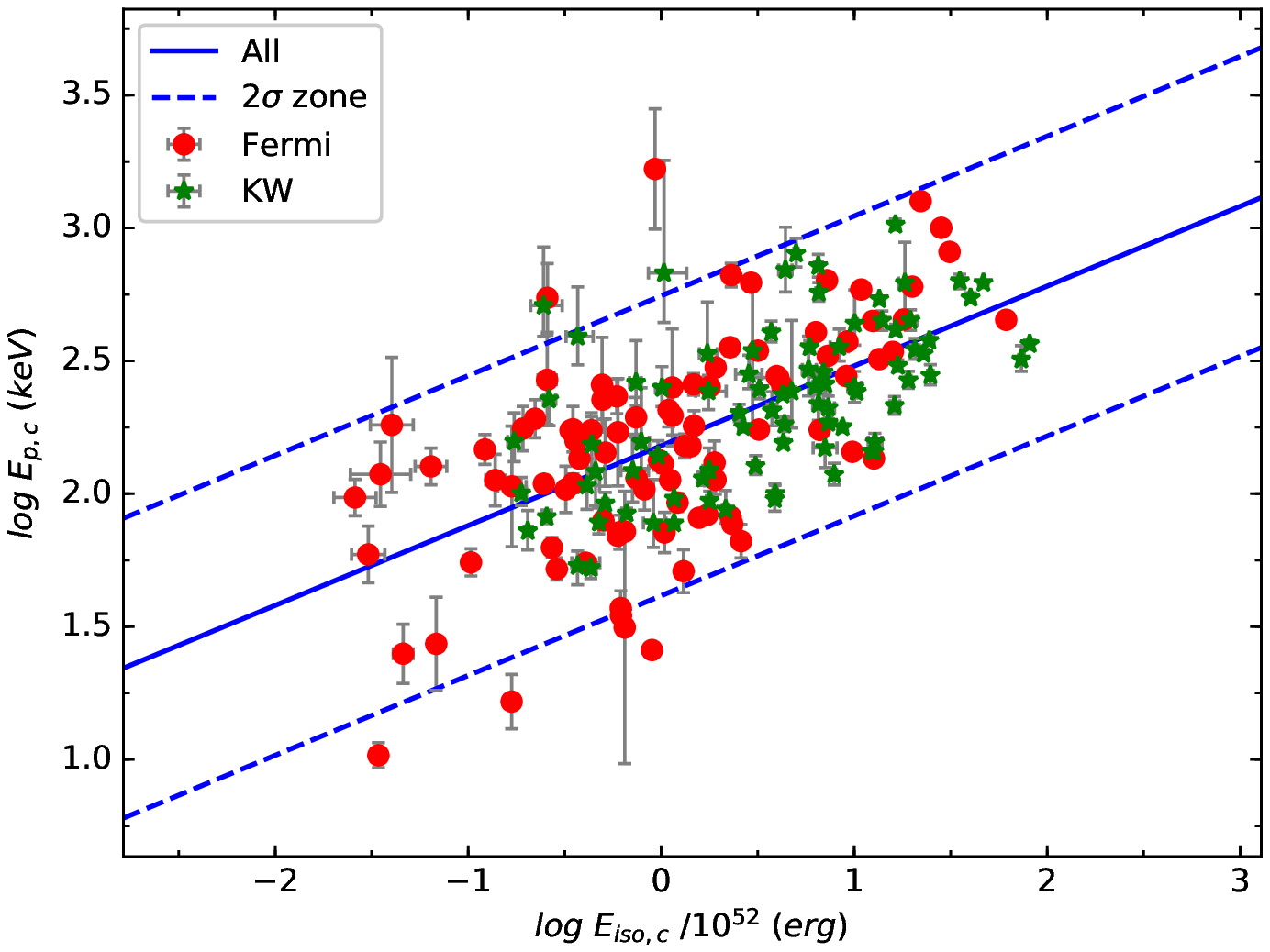}
\includegraphics[angle=0,scale=0.52]{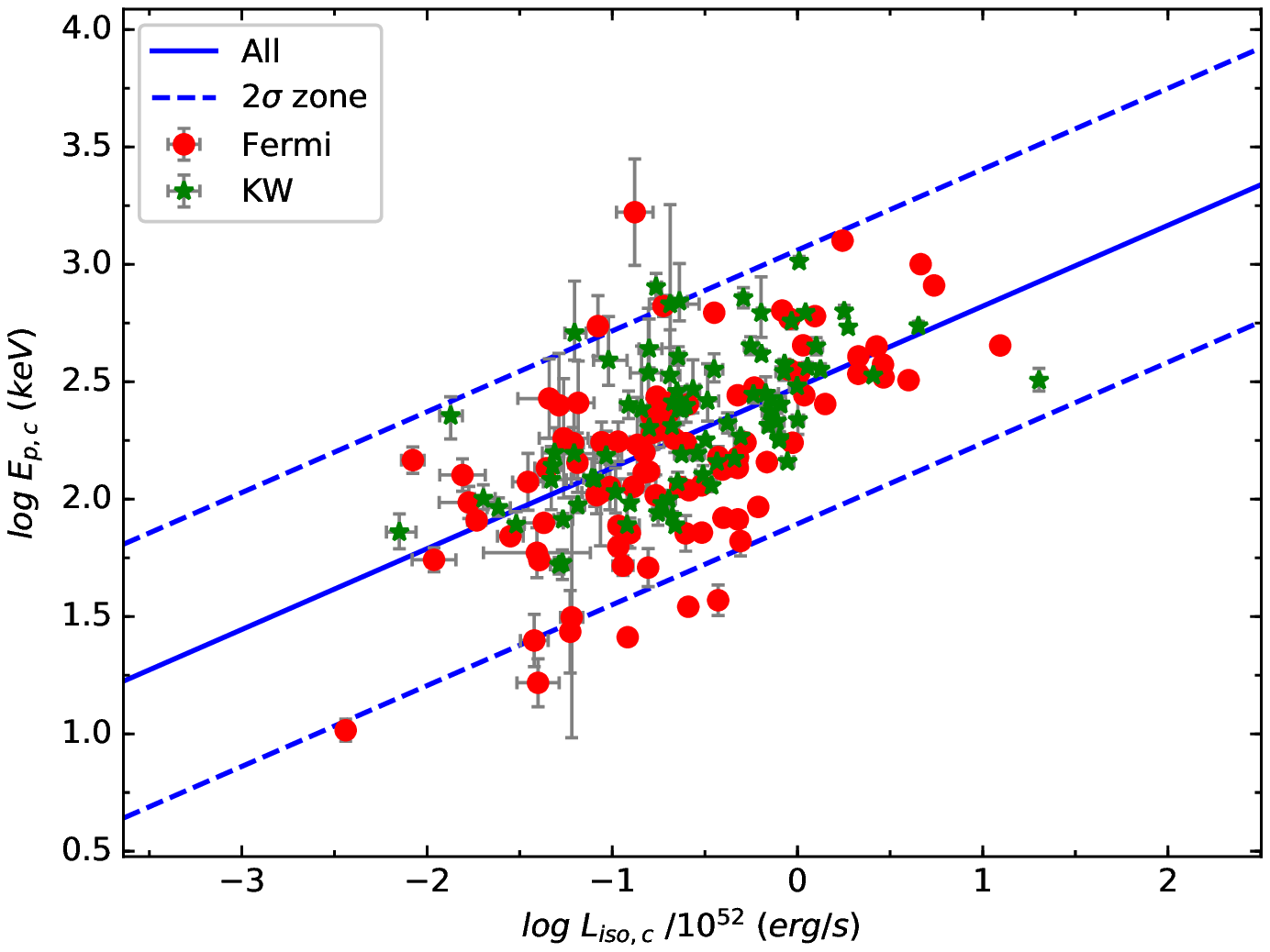}
\includegraphics[angle=0,scale=0.52]{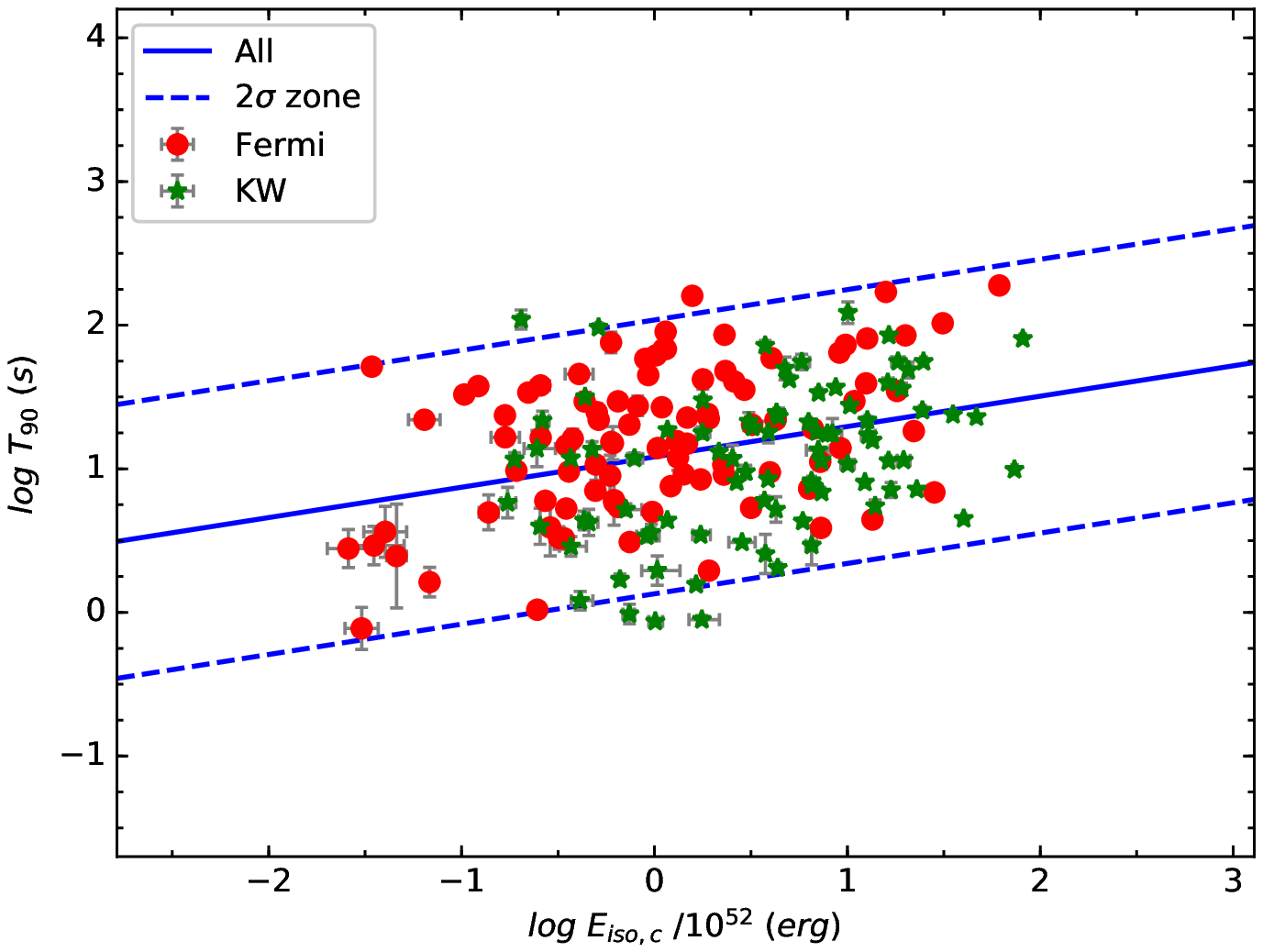}
\includegraphics[angle=0,scale=0.52]{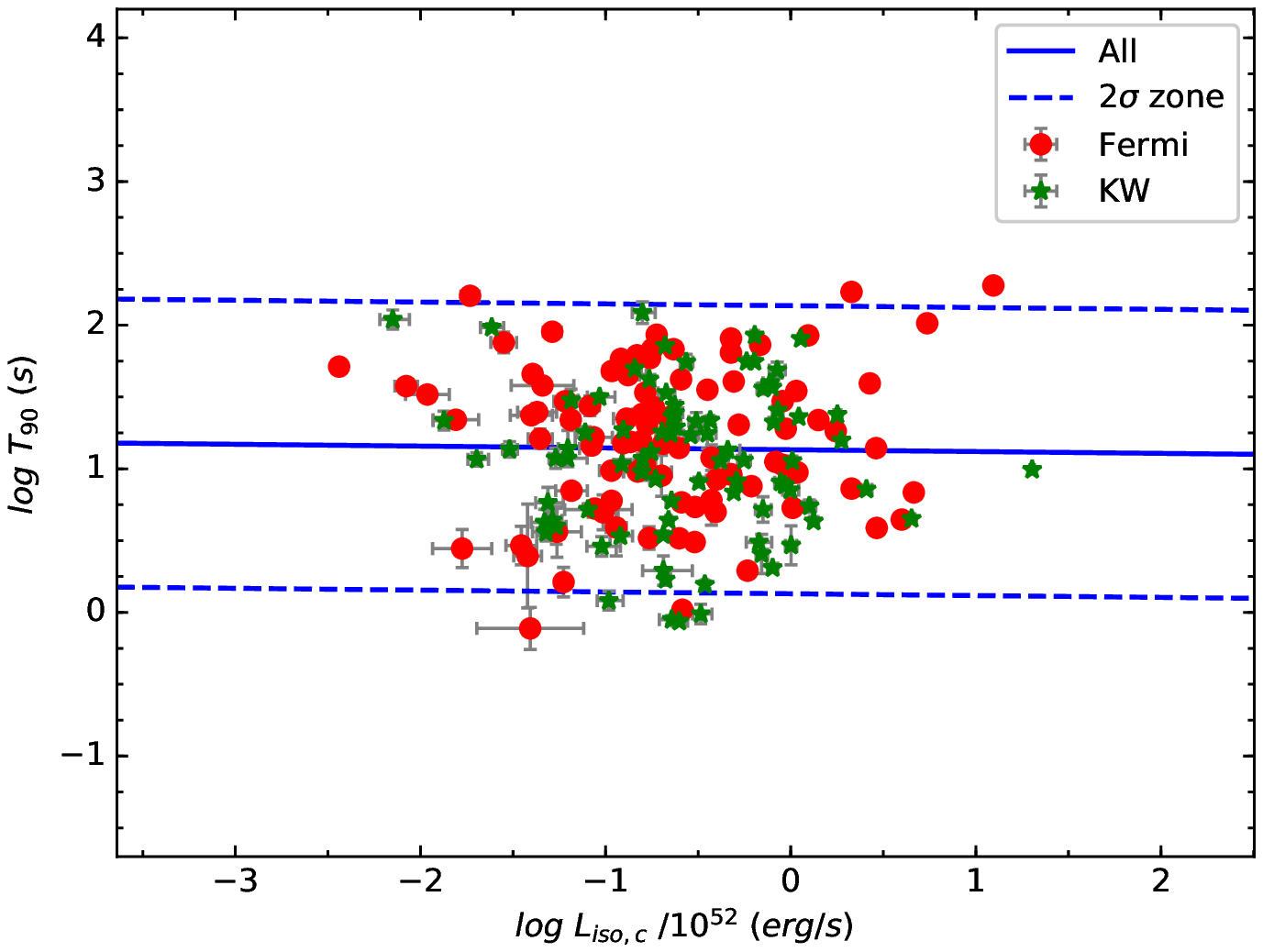}
\includegraphics[angle=0,scale=0.52]{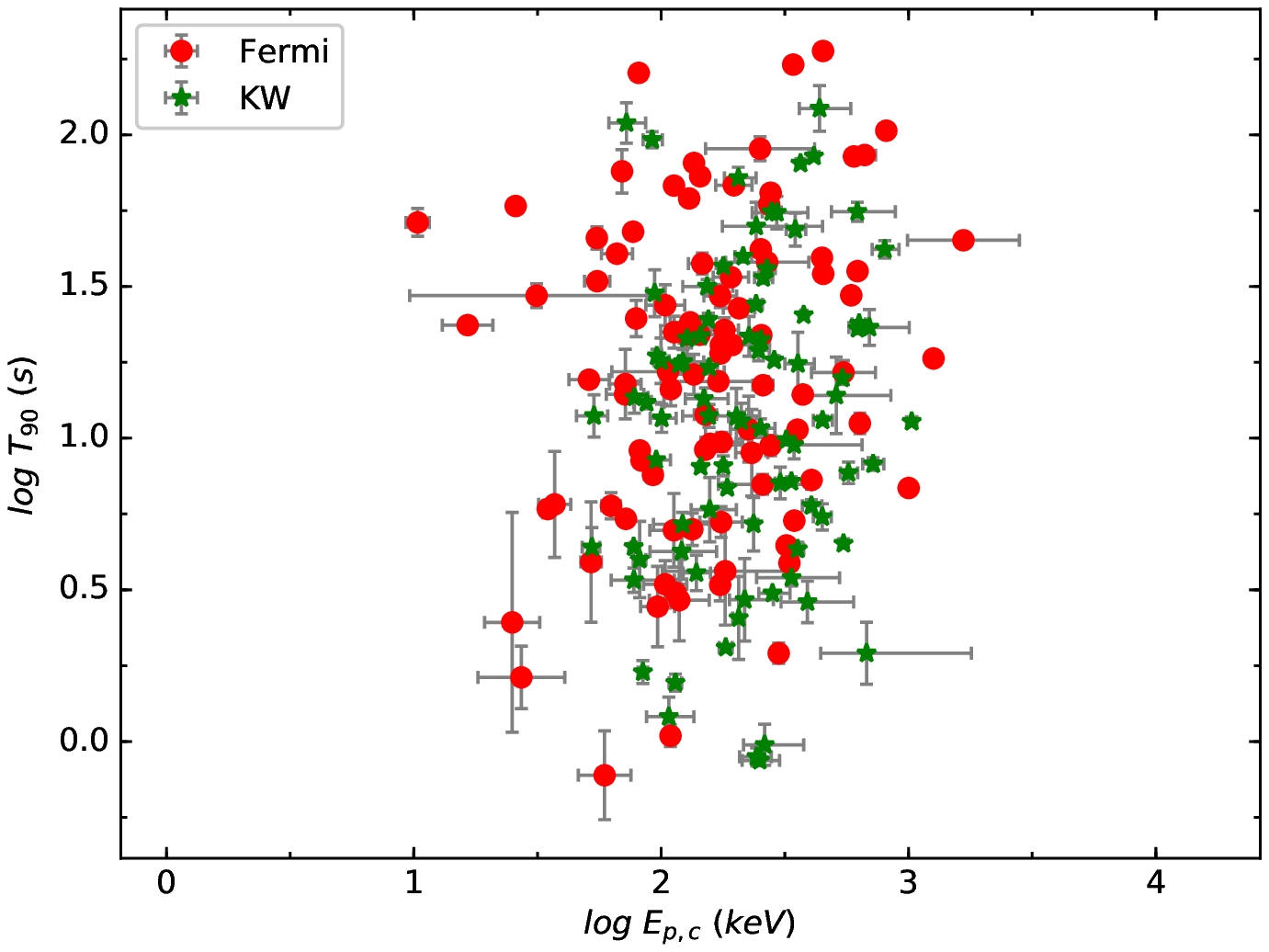}
\includegraphics[angle=0,scale=0.52]{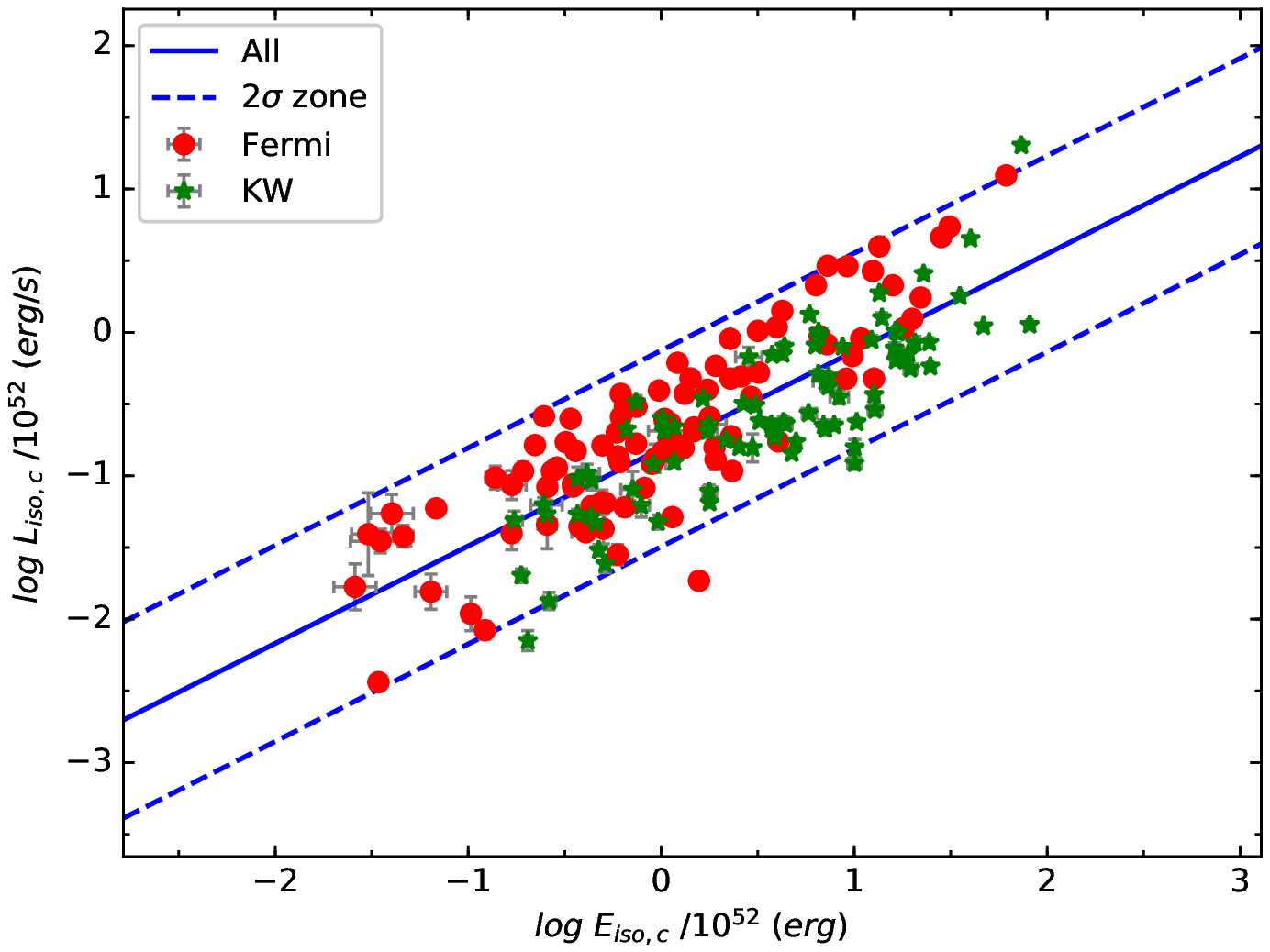}
\caption{Pair correlations among the rest frame quantities $E_{\rm p,c}$, $T_{\rm 90}$, $E_{\rm iso,c}$ and $L_{\rm iso,c}$, where the redshift evolution are corrected. The other symbols are the same as Figure 4.}
\end{figure*}

\begin{figure*}
\centering
\label{figure13}
\includegraphics[angle=0,scale=0.52]{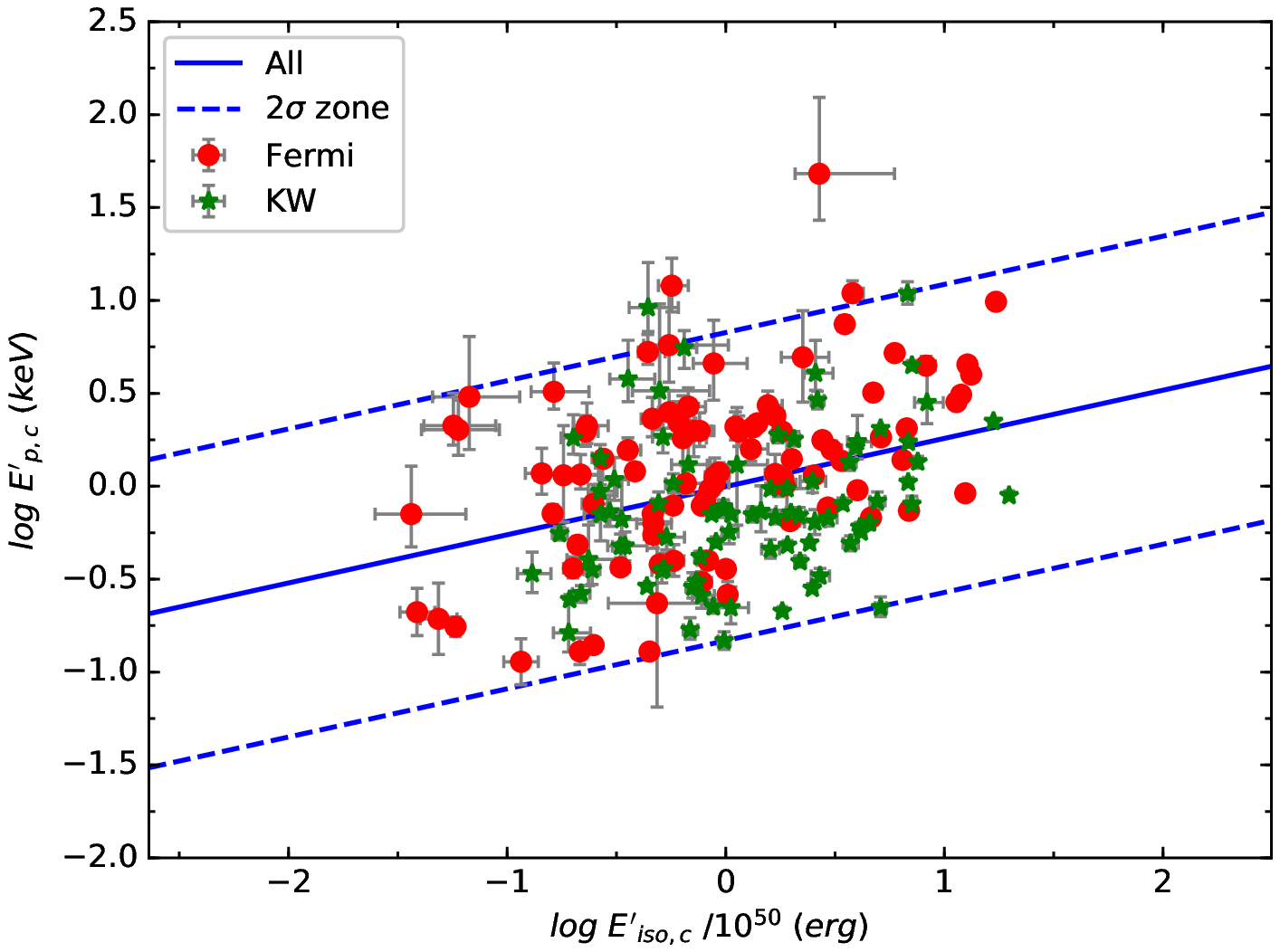}
\includegraphics[angle=0,scale=0.52]{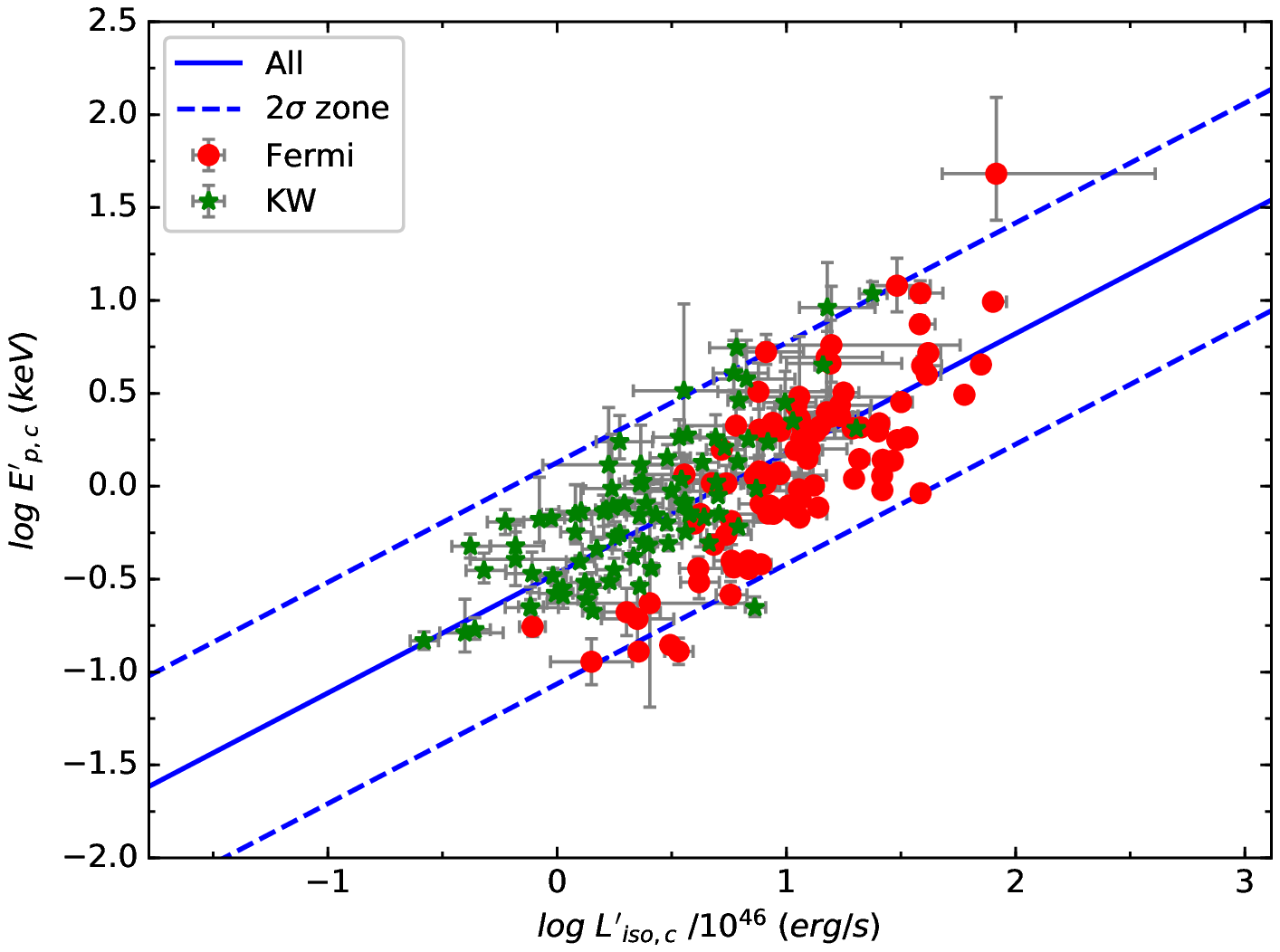}
\includegraphics[angle=0,scale=0.52]{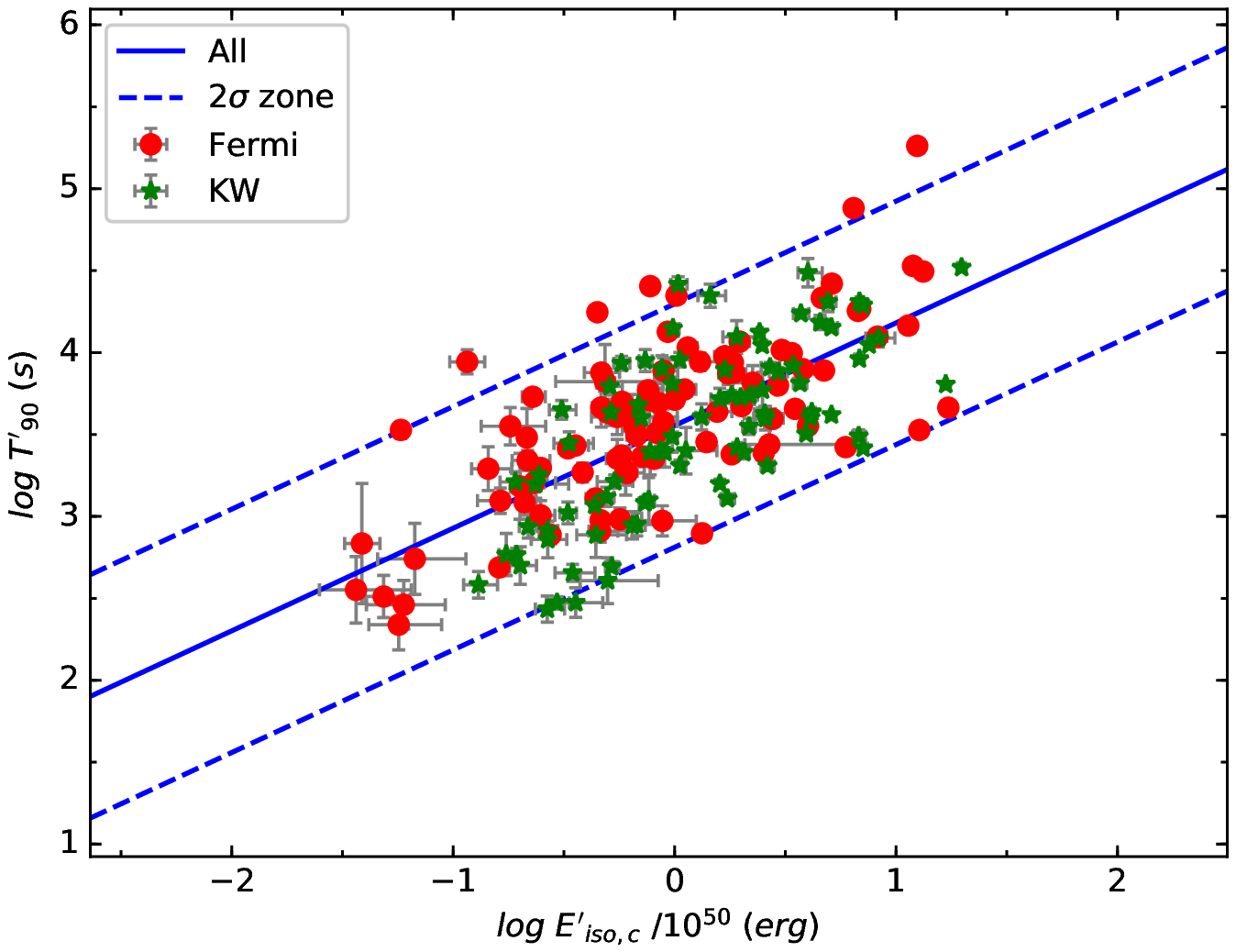}
\includegraphics[angle=0,scale=0.52]{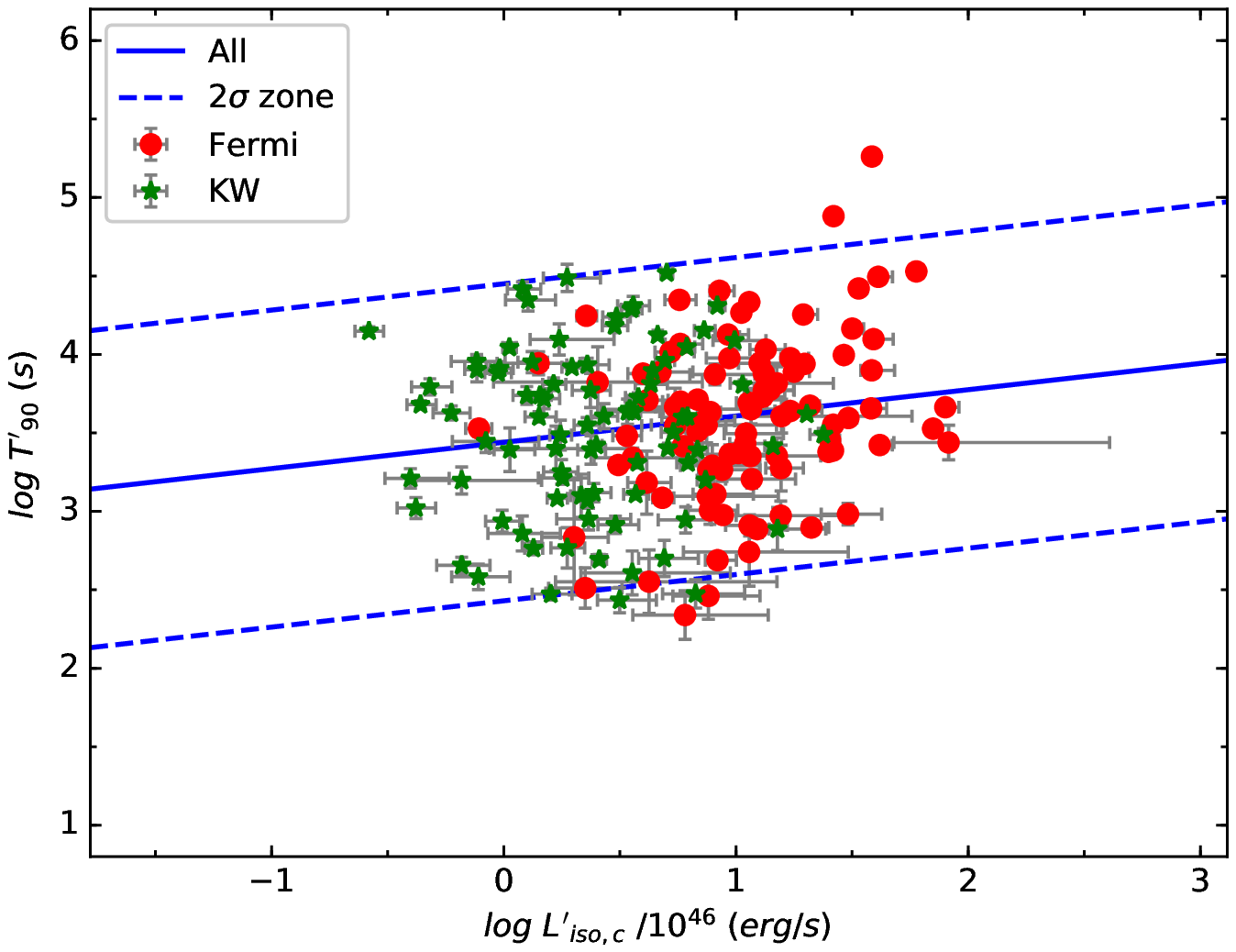}
\includegraphics[angle=0,scale=0.52]{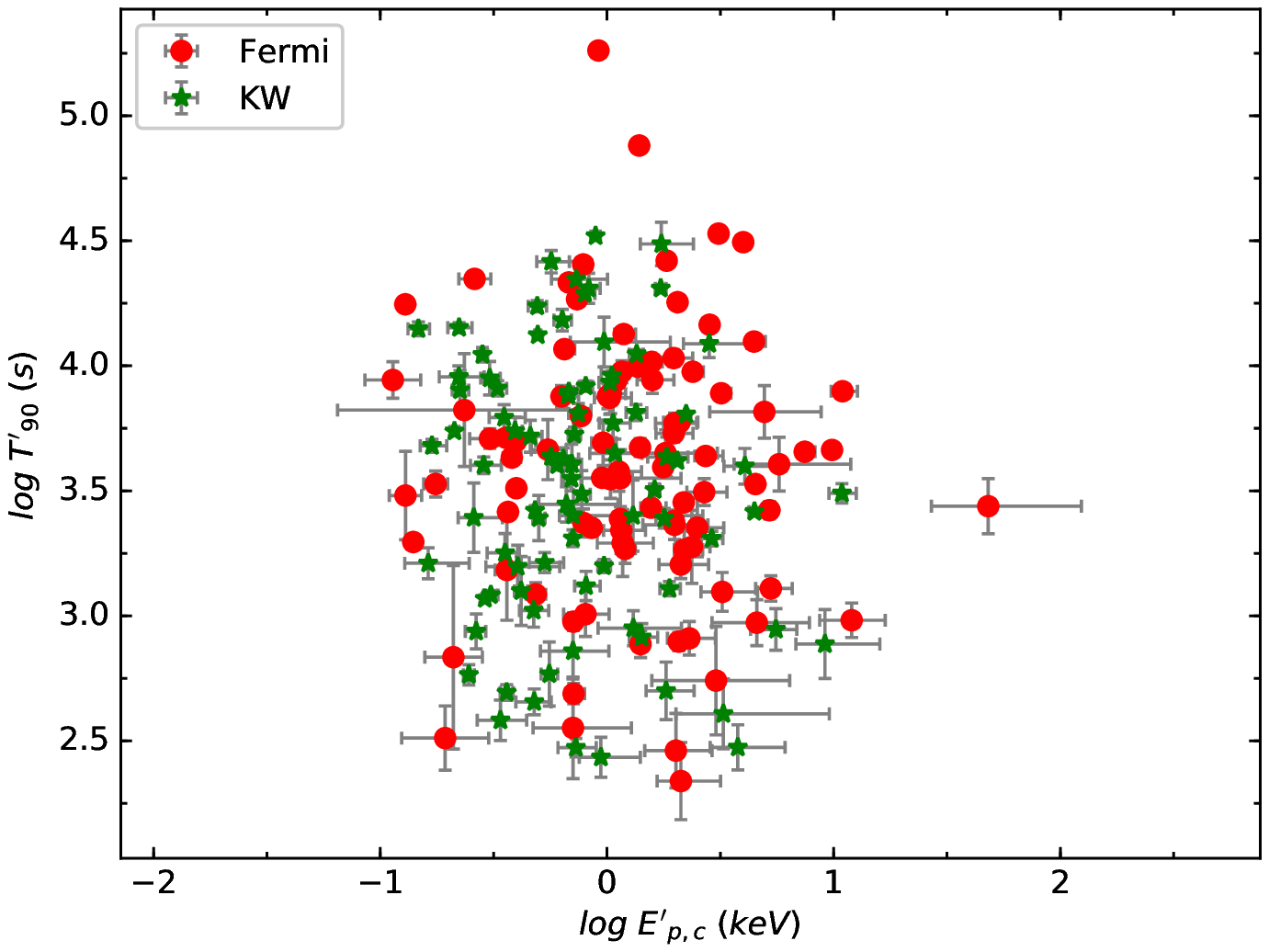}
\includegraphics[angle=0,scale=0.52]{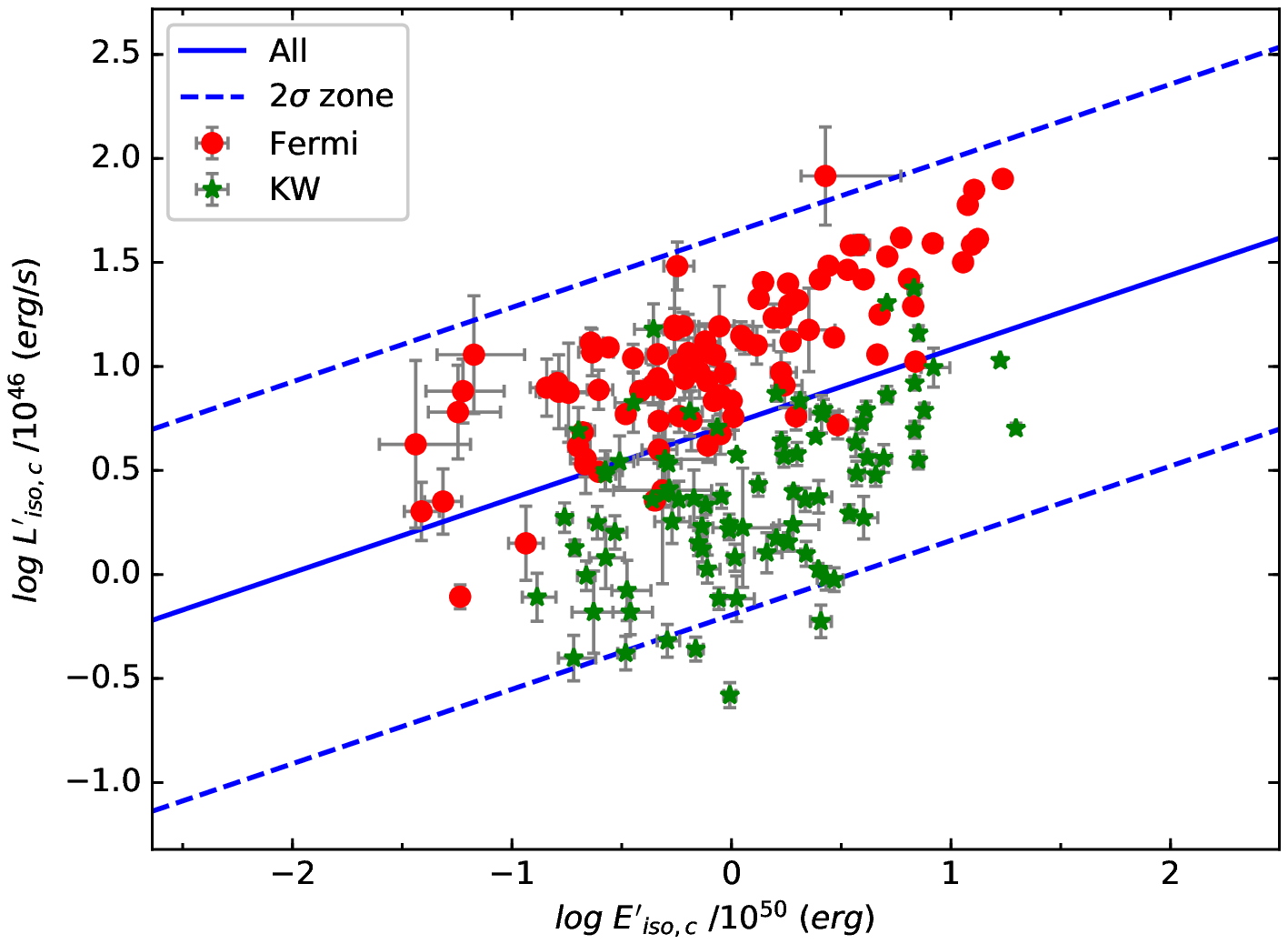}
\caption{Pair correlations among the comoving frame quantities $E'_{p,c}$, $T'_{90}$, $E'_{iso,c}$ and $L'_{iso,c}$, where the redshift evolution are corrected. The other symbols are the same as Figure 4.}
\end{figure*}

\begin{figure*}
\label{figure14}
\includegraphics[angle=0,scale=0.52]{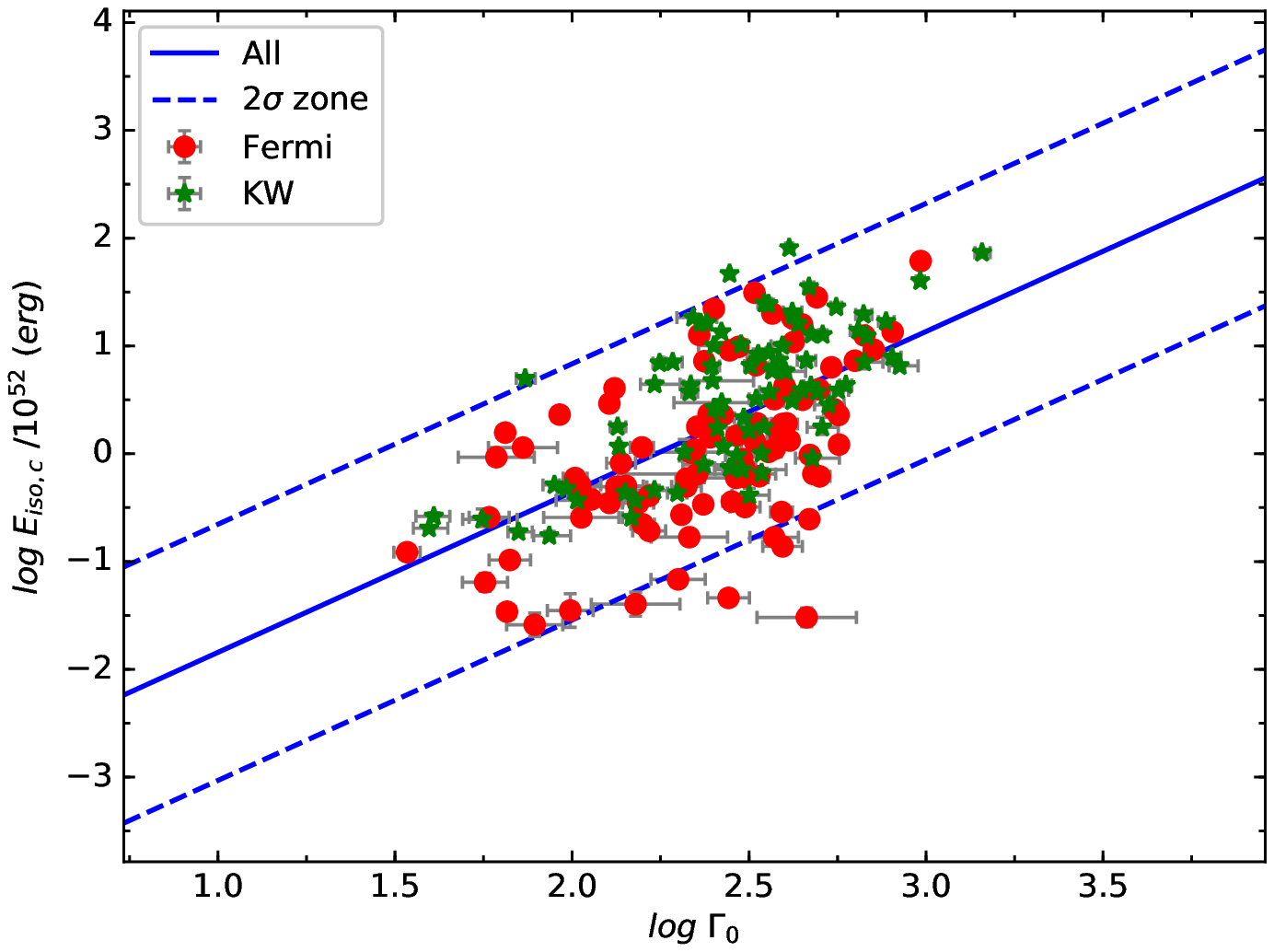}
\includegraphics[angle=0,scale=0.52]{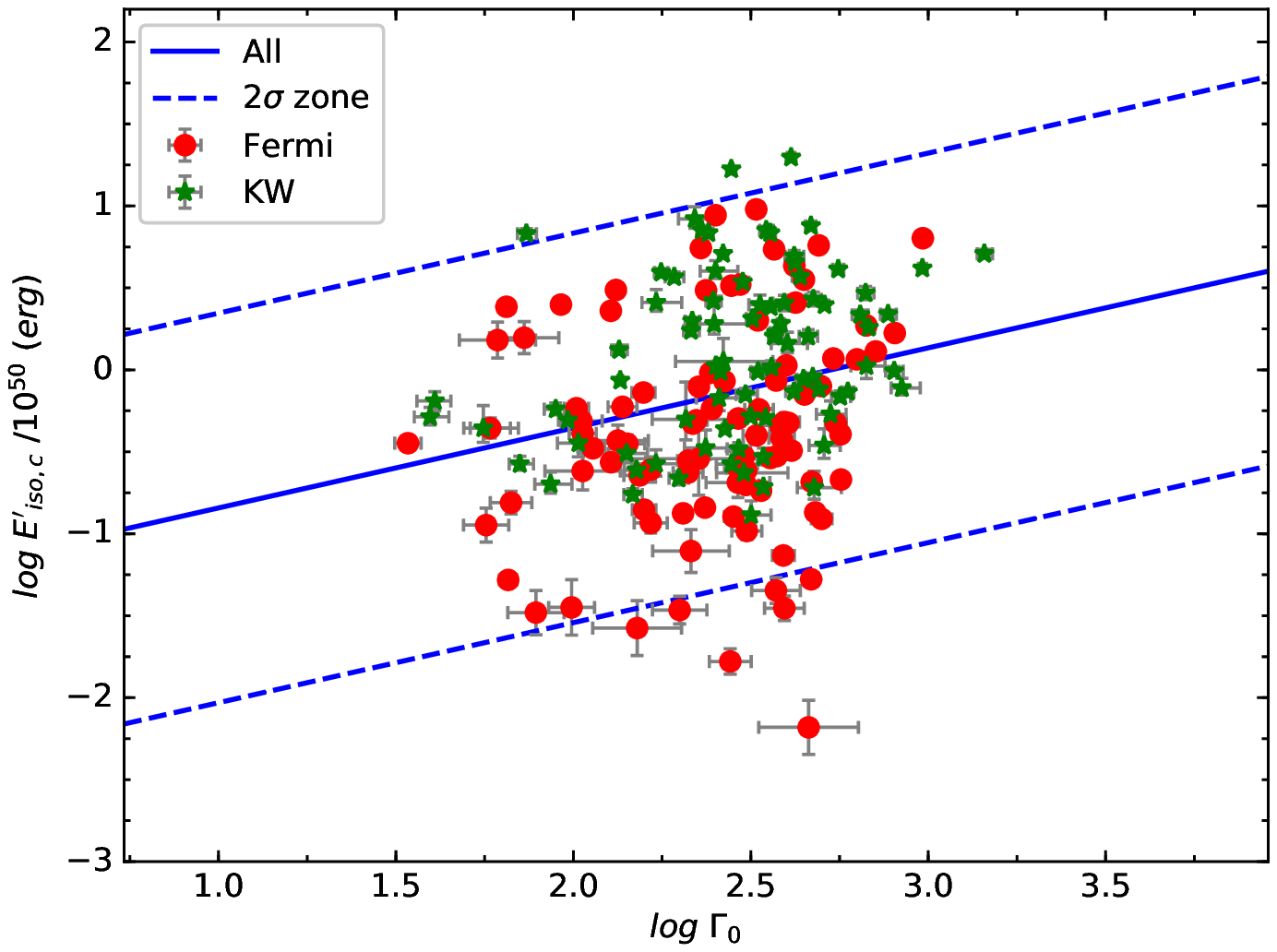}
\includegraphics[angle=0,scale=0.52]{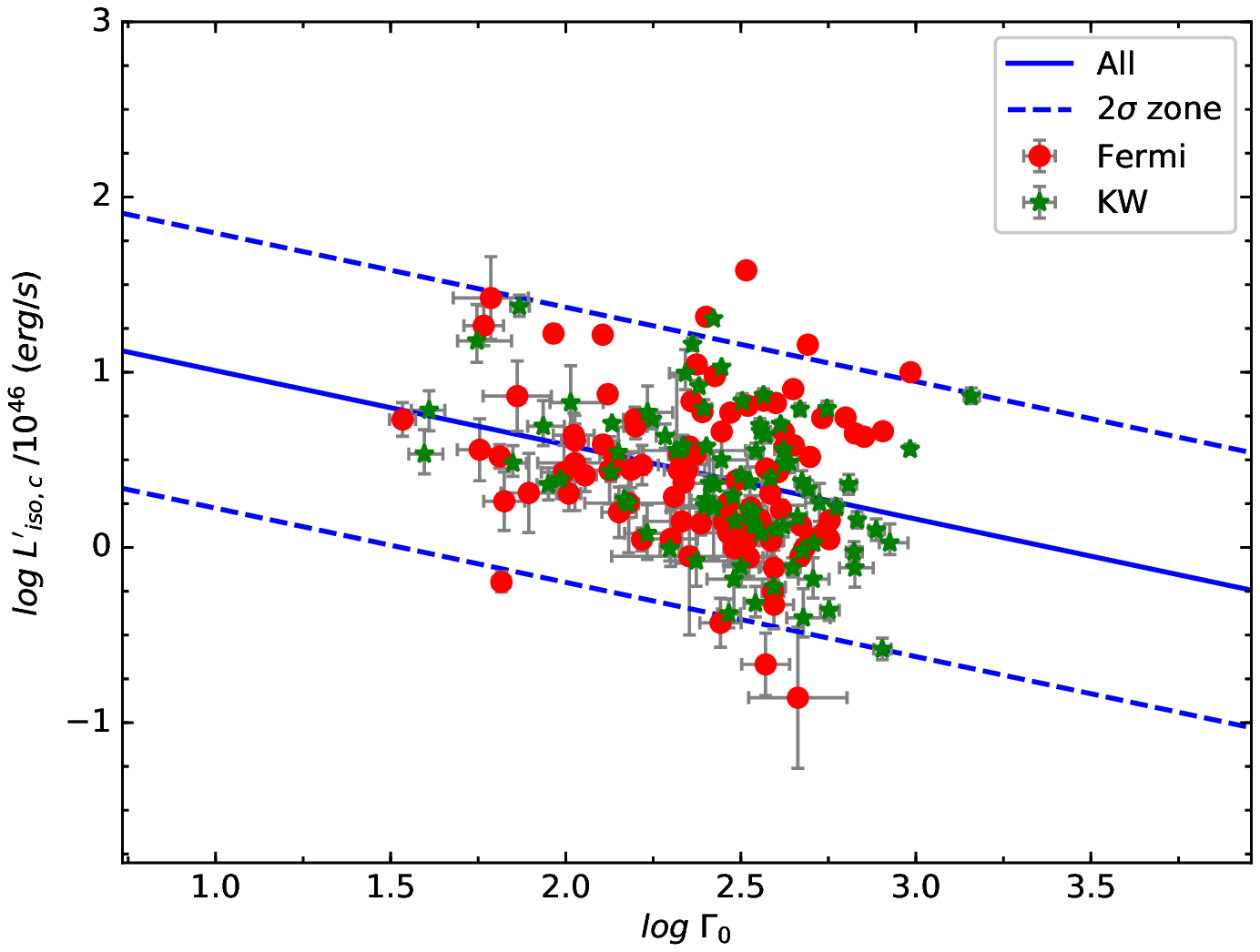}
\includegraphics[angle=0,scale=0.52]{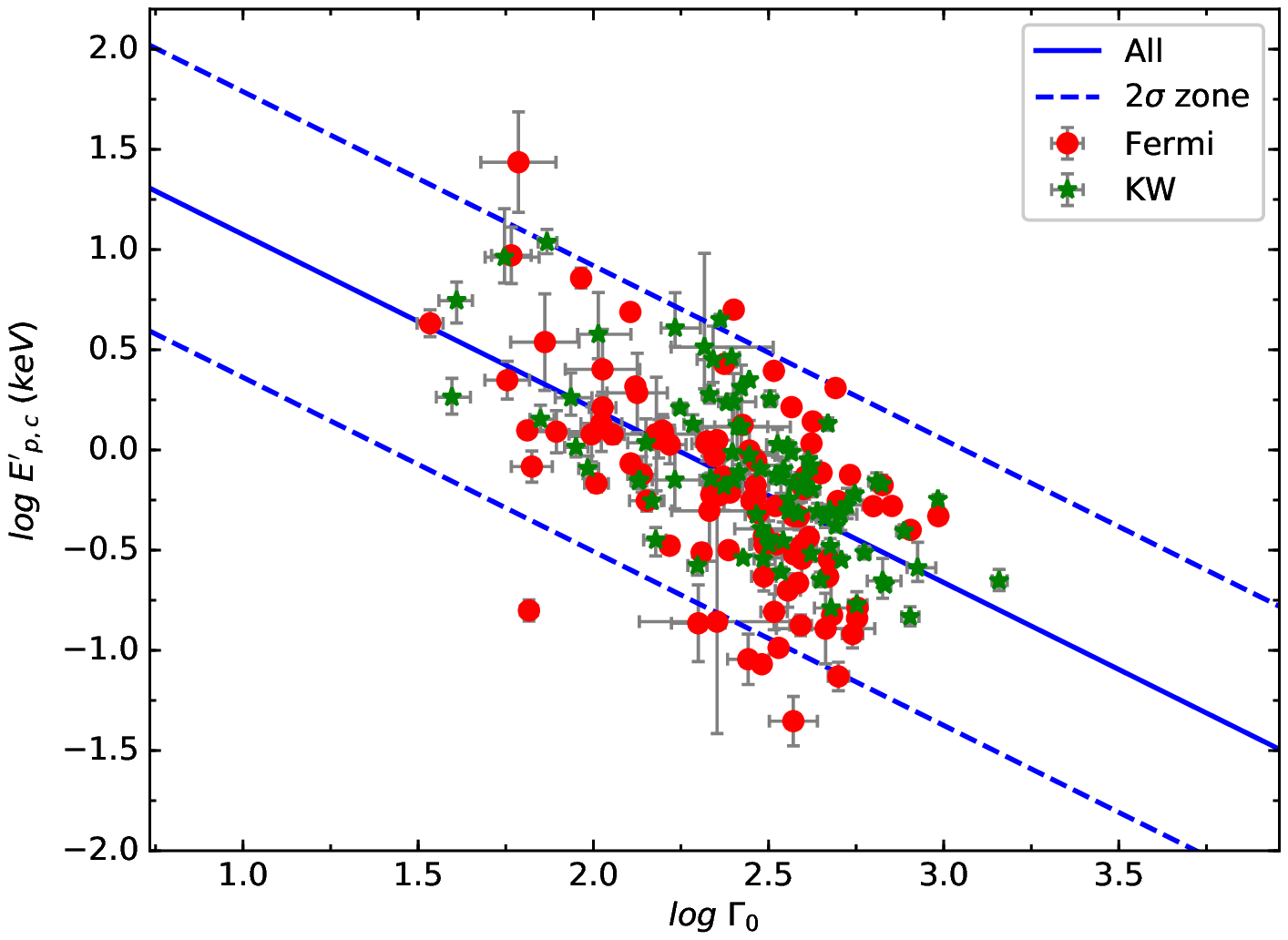}
\caption{Relations between $\Gamma_0$ and $E_{\rm iso,c}$, $E'_{\rm iso,c}$, $L'_{\rm iso,c}$ and $E'_{\rm p,c}$. The other symbols are the same as Figure 4. }
\end{figure*}

\clearpage

\begin{longrotatetable}
\startlongtable
\begin{deluxetable*}{llcccccccr}
\tablecaption{Observed data of Fermi sample\label{tab1:dataFM}}
\tablewidth{700pt}
\tabletypesize{\scriptsize}
\tablehead{
\colhead{GRB} & \colhead{$z$} & \colhead{$T_{\rm 90,obs}$} & \colhead{$E_{\rm p,obs}$} &
\colhead{$\alpha$} & \colhead{$\beta$} & \colhead{$F_{p}$} & \colhead{$S_{\gamma}$} & \colhead{emin} & \colhead{emax}\\
\colhead{} & \colhead{} & \colhead{s} & \colhead{keV} & \colhead{} & \colhead{} &
\colhead{($\times {10^{-7}}$ erg cm$^{-2}$ s$^{-1}$)} & \colhead{($\times {10^{-7}}$ erg cm$^{-2}$)} & \colhead{} & \colhead{}\\
}
\decimalcolnumbers
\startdata
    080916A & 0.689 & 46.34 $\pm$ 7.17 & 106.62 $\pm$ 19.25 & -0.82 $\pm$ 0.13 & -1.78 $\pm$ 0.05 & 8.42 $\pm$ 0.97 & 126.73 $\pm$ 3.95 & 10    & 1000 \\
    081109 & 0.9787 & 58.37 $\pm$ 5.22 & 32.44 $\pm$ 38.3 & -0.67 $\pm$ 1.85 & -1.65 $\pm$ 0.03 & 3.63 $\pm$ 0.5 & 66.75 $\pm$ 2.78 & 10    & 1000 \\
    081121 & 2.512 & 41.99 $\pm$ 8.51 & 160.84 $\pm$ 16.56 & -0.43 $\pm$ 0.13 & -2.09 $\pm$ 0.09 & 15.42 $\pm$ 0.94 & 147.54 $\pm$ 4.2 & 10    & 1000 \\
    081221 & 2.26  & 29.7 $\pm$ 0.41 & 86.94 $\pm$ 1.41 & -0.9 $\pm$ 0.02 & -3.86 $\pm$ 0.48 & 24.69 $\pm$ 0.9 & 301.64 $\pm$ 5.26 & 10    & 1000 \\
    081222 & 2.77  & 18.88 $\pm$ 2.32 & 142.72 $\pm$ 9.58 & -0.86 $\pm$ 0.05 & -2.31 $\pm$ 0.12 & 16.68 $\pm$ 0.88 & 117.92 $\pm$ 3.4 & 10    & 1000 \\
    090323 & 3.57  & 135.17 $\pm$ 1.45 & 632.9 $\pm$ 40.83 & -1.29 $\pm$ 0.01 & -2.44 $\pm$ 0.17 & 32.01 $\pm$ 1.36 & 1277.04 $\pm$ 14.7 & 10    & 1000 \\
    090328 & 0.736 & 61.7 $\pm$ 1.81 & 639.66 $\pm$ 45.71 & -1.09 $\pm$ 0.02 & -2.37 $\pm$ 0.18 & 39.98 $\pm$ 1.26 & 508.9 $\pm$ 8.15 & 10    & 1000 \\
    090423 & 8.26  & 7.17 $\pm$ 2.42 & 66.01 $\pm$ 16.18 & -0.59 $\pm$ 0.5 & -2.67 $\pm$ 0.74 & 1.52 $\pm$ 1.01 & 6.83 $\pm$ 1.36 & 10    & 1000 \\
\enddata
\end{deluxetable*}
\end{longrotatetable}

\clearpage

\begin{longrotatetable}
\startlongtable
\begin{deluxetable*}{llcccccccr}
\tablecaption{Observed data of Konus-Wind sample\label{tab2:dataKW}}
\tablewidth{700pt}
\tabletypesize{\scriptsize}
\tablehead{
\colhead{GRB} & \colhead{$z$} & \colhead{$T_{\rm 90,obs}$} & \colhead{$E_{\rm p,obs}$} &
\colhead{$\alpha$} & \colhead{$\beta$} & \colhead{$F_{p}$} & \colhead{$S_{\gamma}$} & \colhead{emin} & \colhead{emax}\\
\colhead{} & \colhead{} & \colhead{s} & \colhead{keV} & \colhead{} & \colhead{} &
\colhead{($\times {10^{-7}}$ erg cm$^{-2}$ s$^{-1}$)} & \colhead{($\times {10^{-7}}$ erg cm$^{-2}$)} & \colhead{} & \colhead{}\\
}
\decimalcolnumbers
\startdata
    970228 & 0.69  & 53.44 $\pm$ 2.89 & 159$_{ -36}^{ + 38}$ & -1.24$_{ -0.22}^{ + 0.31}$ & -2.82$_{ -7.18}^{ + 0.6}$ & 22.4$_{ -3.4}^{ + 4.3}$ & 80.7$_{ -4.1}^{ + 4.9}$ & 10    & 10000 \\
    970828 & 0.96  & 66.21 $\pm$ 2.78 & 271$_{ -22}^{ + 24}$ & -0.73$_{ -0.06}^{ + 0.06}$ & -2.18$_{ -0.15}^{ + 0.11}$ & 43$_{ -4.4}^{ + 4.6}$ & 1018$_{ -36}^{ + 35}$ & 10    & 10000 \\
    971214 & 3.42  & 15.89 $\pm$ 2.13 & 154$_{ -19}^{ + 20}$ & -0.32$_{ -0.22}^{ + 0.26}$ & -2.39$_{ -0.35}^{ + 0.23}$ & 6.04$_{ -0.68}^{ + 0.71}$ & 57.2$_{ -2.2}^{ + 2.4}$ & 10    & 10000 \\
    990123 & 1.6   & 62.02 $\pm$ 1.18 & 675$_{ -45}^{ + 40}$ & -0.92$_{ -0.02}^{ + 0.02}$ & -2.53$_{ -0.34}^{ + 0.25}$ & 256$_{ -11}^{ + 11}$ & 3106$_{ -78}^{ + 80}$ & 10    & 10000 \\
    990506 & 1.31  & 128.61 $\pm$ 0.65 & 296$_{ -25}^{ + 27}$ & -1.19$_{ -0.04}^{ + 0.05}$ & -2.09$_{ -0.08}^{ + 0.07}$ & 93.2$_{ -4.6}^{ + 4.7}$ & 2610$_{ -88}^{ + 88}$ & 10    & 10000 \\
    990510 & 1.62  & 55.89 $\pm$ 8.11 & 136$_{ -10}^{ + 12}$ & -1.35$_{ -0.08}^{ + 0.08}$ & -2.82$_{ 0}^{ + 0}$ & 35.6$_{ -3.3}^{ + 3.5}$ & 217.2$_{ -8.2}^{ + 8.9}$ & 10    & 10000 \\
    990705 & 0.84  & 33.23 $\pm$ 1.12 & 299$_{ -9}^{ + 9}$ & -0.72$_{ -0.03}^{ + 0.03}$ & -2.68$_{ -0.15}^{ + 0.12}$ & 48$_{ -4.2}^{ + 4.4}$ & 1090$_{ -38}^{ + 38}$ & 10    & 10000 \\
    990712 & 0.43  & 16.63 $\pm$ 1.78 & 103$_{ -11}^{ + 14}$ & -1.44$_{ -0.13}^{ + 0.13}$ & -3.47$_{ -6.53}^{ + 0.81}$ & 8.02$_{ -0.87}^{ + 1.18}$ & 62.5$_{ -3.3}^{ + 3.8}$ & 10    & 10000 \\
\enddata
\tablenotetext{a}{Peak energy flux calculated on the 64-ms time scale}
\end{deluxetable*}
\end{longrotatetable}

\clearpage

\begin{longrotatetable}
\startlongtable
\begin{deluxetable*}{llccccccccr}
\tablecaption{The intrinsic properties of Fermi sample\label{tab3:FM}}
\tablewidth{700pt}
\tabletypesize{\scriptsize}
\tablehead{
\colhead{GRB} & \colhead{$z$} & \colhead{$E_{iso}$} & \colhead{$L_{iso}$} &
\colhead{$E_{p,z}$} & \colhead{$T_{90,z}$} & \colhead{$E'_{iso}$} & $L'_{iso}$  & $E'_{p}$  & $T'_{90}$ & \colhead{$\Gamma_0$}\\
\colhead{} & \colhead{} & \colhead{($\times {10^{52}}$ erg)} & \colhead{($\times {10^{52}}$ erg s$^{-1}$)} &
\colhead{(keV)} & \colhead{(s)} & \colhead{($\times {10^{50}}$ erg)} &
 \colhead{($\times {10^{46}}$ erg s$^{-1}$)} & \colhead{(keV)} & \colhead{(s)} & \colhead{}\\
}
\decimalcolnumbers
\startdata
    171222A & 2.409 & 3.35 $\pm$ 0.3 & 1.02 $\pm$ 0.27 & 59.8 $\pm$ 14.1 & 23.58 $\pm$ 1.35 & 0.9 $\pm$ 0.16 & 5.54 $\pm$ 2.28 & 0.16 $\pm$ 0.05 & 8776.99 $\pm$ 1473.85 & 372.22 $\pm$ 58.74 \\
    171010A & 0.3285 & 25.32 $\pm$ 0.06 & 1.01 $\pm$ 0.01 & 182.88 $\pm$ 1.9 & 80.74 $\pm$ 0.61 & 11.07 $\pm$ 0.08 & 14.48 $\pm$ 0.25 & 0.8 $\pm$ 0.01 & 18465.91 $\pm$ 187.13 & 228.7 $\pm$ 1.55 \\
    170705A & 2.01  & 17.88 $\pm$ 0.54 & 11.38 $\pm$ 0.4 & 294.61 $\pm$ 23.01 & 7.57 $\pm$ 0.46 & 3.15 $\pm$ 0.15 & 26.49 $\pm$ 2.18 & 0.52 $\pm$ 0.04 & 4295.37 $\pm$ 304.94 & 567.54 $\pm$ 21.13 \\
    170607A & 0.557 & 1.03 $\pm$ 0.02 & 0.27 $\pm$ 0.01 & 174.06 $\pm$ 14.06 & 14.51 $\pm$ 1.83 & 0.81 $\pm$ 0.04 & 12.5 $\pm$ 1.26 & 1.36 $\pm$ 0.12 & 1853.23 $\pm$ 246.86 & 127.74 $\pm$ 5.44 \\
    170214A & 2.53  & 392.53 $\pm$ 2.56 & 30.32 $\pm$ 0.85 & 1699.3 $\pm$ 39.63 & 34.81 $\pm$ 0.21 & 93.6 $\pm$ 1.65 & 129.31 $\pm$ 5.59 & 4.05 $\pm$ 0.12 & 14597.94 $\pm$ 254.27 & 419.35 $\pm$ 6.87 \\
    170113A & 1.968 & 2.39 $\pm$ 0.41 & 1.54 $\pm$ 0.36 & 333.92 $\pm$ 174.49 & 16.56 $\pm$ 1.39 & 1.12 $\pm$ 0.34 & 25.14 $\pm$ 13.79 & 1.56 $\pm$ 0.9 & 3550.59 $\pm$ 931.95 & 214.4 $\pm$ 53.3 \\
    161129A & 0.645 & 1.73 $\pm$ 0.07 & 0.24 $\pm$ 0.03 & 240.84 $\pm$ 70.09 & 21.94 $\pm$ 0.44 & 1.64 $\pm$ 0.23 & 16.37 $\pm$ 4.73 & 2.29 $\pm$ 0.73 & 2312.55 $\pm$ 314.13 & 105.39 $\pm$ 14.16 \\
    161117A & 1.549 & 22.88 $\pm$ 0.62 & 1.29 $\pm$ 0.14 & 205.62 $\pm$ 7.76 & 47.93 $\pm$ 0.26 & 9.41 $\pm$ 0.55 & 16.33 $\pm$ 2.45 & 0.85 $\pm$ 0.05 & 11657.44 $\pm$ 611.49 & 243.21 $\pm$ 12.69 \\
\enddata
\end{deluxetable*}
\end{longrotatetable}

\clearpage
\begin{longrotatetable}
\startlongtable
\begin{deluxetable*}{llccccccccr}
\tablecaption{The intrinsic properties of Konus-Wind sample\label{tab4:KW}}
\tablewidth{700pt}
\tabletypesize{\scriptsize}
\tablehead{
\colhead{GRB} & \colhead{$z$} & \colhead{$E_{iso}$} & \colhead{$L_{iso}$} &
\colhead{$E_{p}$} & \colhead{$T_{90}$} & \colhead{$E'_{iso}$} & $L'_{iso}$  & $E'_{p}$  & $T'_{90}$ & \colhead{$\Gamma_0$}\\
\colhead{} & \colhead{} & \colhead{($\times {10^{52}}$ erg)} & \colhead{($\times {10^{52}}$ erg s$^{-1}$)} &
\colhead{(keV)} & \colhead{(s)} & \colhead{($\times {10^{50}}$ erg)} &
 \colhead{($\times {10^{46}}$ erg s$^{-1}$)} & \colhead{(keV)} & \colhead{(s)} & \colhead{}\\
}
\decimalcolnumbers
\startdata
    970228 & 0.69  & 1.08$_{-  0.05}^{+ 0.07}$ & 0.51$_{-  0.08}^{+ 0.1}$ & 268.71$_{-  60.84}^{+ 64.22}$ & 31.62 $\pm$ 1.71 & 0.77$_{-  0.1}^{+ 0.11}$ & 19.09$_{-  5.41}^{+ 6.34}$ & 1.9$_{-  0.49}^{+ 0.52}$ & 4466.88$_{-  587.19}^{+ 651.69}$ & 141.26$_{-  16.92}^{+ 19.14}$ \\
    970828 & 0.96  & 22.71$_{-  0.8}^{+ 0.78}$ & 1.88$_{-  0.19}^{+ 0.2}$ & 531.16$_{-  43.12}^{+ 47.04}$ & 33.78 $\pm$ 1.42 & 11.8$_{-  0.81}^{+ 0.84}$ & 38.08$_{-  5.92}^{+ 6.25}$ & 2.76$_{-  0.28}^{+ 0.3}$ & 6499.83$_{-  468.64}^{+ 487.97}$ & 192.42$_{-  11.28}^{+ 11.97}$ \\
    971214 & 3.42  & 12.59$_{-  0.48}^{+ 0.53}$ & 5.88$_{-  0.66}^{+ 0.69}$ & 680.68$_{-  83.98}^{+ 88.4}$ & 3.6 $\pm$ 0.48 & 4.31$_{-  0.36}^{+ 0.38}$ & 51.63$_{-  9.61}^{+ 10.06}$ & 2.33$_{-  0.34}^{+ 0.35}$ & 1050.39$_{-  161.06}^{+ 162.92}$ & 292.14$_{-  21.66}^{+ 22.71}$ \\
    990123 & 1.6   & 184.25$_{-  4.63}^{+ 4.75}$ & 39.48$_{-  1.7}^{+ 1.7}$ & 1755$_{-  117}^{+ 104}$ & 23.85 $\pm$ 0.45 & 39.45$_{-  1.69}^{+ 1.63}$ & 135.79$_{-  11.11}^{+ 10.52}$ & 3.76$_{-  0.28}^{+ 0.25}$ & 11138.88$_{-  441.9}^{+ 416.99}$ & 466.99$_{-  16.26}^{+ 15.06}$ \\
    990506 & 1.31  & 105.57$_{-  3.56}^{+ 3.56}$ & 8.71$_{-  0.43}^{+ 0.44}$ & 683.76$_{-  57.75}^{+ 62.37}$ & 55.67 $\pm$ 0.28 & 30.21$_{-  1.65}^{+ 1.71}$ & 53.49$_{-  5.29}^{+ 5.57}$ & 1.96$_{-  0.19}^{+ 0.2}$ & 19454.23$_{-  839.08}^{+ 892.02}$ & 349.43$_{-  14.97}^{+ 15.92}$ \\
    990510 & 1.62  & 16.31$_{-  0.62}^{+ 0.67}$ & 7$_{-  0.65}^{+ 0.69}$ & 356.32$_{-  26.2}^{+ 31.44}$ & 21.33 $\pm$ 3.09 & 3.9$_{-  0.25}^{+ 0.28}$ & 30.01$_{-  4.23}^{+ 4.61}$ & 0.85$_{-  0.08}^{+ 0.09}$ & 8924.45$_{-  1378.67}^{+ 1397.8}$ & 418.37$_{-  22.21}^{+ 24.7}$ \\
    990705 & 0.84  & 19.91$_{-  0.69}^{+ 0.69}$ & 1.61$_{-  0.14}^{+ 0.15}$ & 550.16$_{-  16.56}^{+ 16.56}$ & 18.06 $\pm$ 0.61 & 11.27$_{-  0.62}^{+ 0.63}$ & 38.77$_{-  4.72}^{+ 4.93}$ & 3.11$_{-  0.16}^{+ 0.17}$ & 3190.77$_{-  172.52}^{+ 177.12}$ & 176.67$_{-  7.47}^{+ 7.79}$ \\
\enddata
\end{deluxetable*}
\end{longrotatetable}

\clearpage

\begin{table}\scriptsize
\begin{center}
\caption{The properties of distributions}
\tabletypesize{\scriptsize}
\begin{tabular}{llllllllllrr}
\hline
Parameter & \multicolumn3c{Median(log10)} & \multicolumn3c{$\sigma$} & \multicolumn3c{Median(real)} & \multicolumn2c{p\tablenotemark{a}} \\
          & Fermi   & KW     & All      & Fermi   & KW     & All      & Fermi   & KW     & All      & Fermi   & KW     \\
\hline
\bm{$T_{\rm 90,obs}$} (s) & 1.61  & 1.48  & 1.53  & 0.47  & 0.50  & 0.49  & 41.07  & 30.00  & 34.18 & 0.27 & 0.006 \\
\bm{$E_{\rm p,obs}$}(keV) & 2.21  & 2.36  & 2.30  & 0.41  & 0.30  & 0.36  & 161.58  & 230.64  & 198.96 & 0.75 & 0.72 \\
\bm{$S_{\rm \gamma}$} (erg cm$^{-2}$) & -4.79  & -4.38  & -4.55  & 0.71  & 0.64  & 0.70  & 1.62E-05 & 4.20E-05 & 2.82E-05 & 0.11 & 0.10 \\
\bm{$F_{\rm p}$}(erg cm$^{-2}$ s$^{-1}$) & -5.80  & -5.32  & -5.52  & 0.61  & 0.55  & 0.63  & 1.58E-06 & 4.83E-06 & 3.03E-06 & 0.01 & 0.04 \\
\bm{$z$\tablenotemark{b}} & 0.14  & 0.12  & 0.13  & 0.31  & 0.30  & 0.30  & 1.37  & 1.31  & 1.33 &   &     \\
\bm{$z$\tablenotemark{c}} & \multicolumn3c{0.17}  & \multicolumn3c{0.38} & \multicolumn3c{1.48} &  & \\
\hline
\end{tabular}
\end{center}
\tablenotetext{a}{P-Value of normal distribution test.}
\tablenotetext{b}{Redshift of our sample.}
\tablenotetext{c}{Redshift of 412 long GRBs.}
\label{tab5}
\end{table}

\clearpage

\begin{table*}
\begin{center}
\caption{The results of regression analysis for correlations in the rest frame and comoving frame, in which $r$ is the correlation coefficient, $p$ is the chance probability, and $\sigma$ is the dispersion}
\begin{tabular*}{\linewidth}{lp{10cm}p{2.1cm}p{2.1cm}r}
\hline
\hline
Relations & Expressions & $r$ & $p$ & $\sigma$ \\
\hline
$E_{\rm p}(E_{\rm iso})$ & $\log E_{\rm p} = (2.28 \pm 0.04) + (0.36 \pm 0.03) \times \log E_{\rm iso,52}$ & 0.71 & $< 10^{-4}$ & 0.58 \\
$E_{\rm p}(L_{\rm iso})$ & $\log E_{\rm p} = (2.50 \pm 0.03) + (0.34 \pm 0.03) \times \log L_{\rm iso,52}$ & 0.72 & $< 10^{-4}$ & 0.57 \\
$T_{\rm 90}(E_{\rm iso})$ & $\log T_{\rm 90} = (1.03 \pm 0.06) + (0.10 \pm 0.05) \times \log E_{\rm iso,52}$ & 0.17 & 0.023 & 0.99 \\
$T_{\rm 90}(L_{\rm iso})$ & $\log T_{\rm 90} = (1.20 \pm 0.04) + (-0.12 \pm 0.04) \times \log L_{\rm iso,52}$ & -0.20 & 0.008 & 0.98 \\
$L_{\rm iso}(E_{\rm iso})$ & $\log L_{\rm iso,52} = (-0.51 \pm 0.05) + (0.94 \pm 0.03) \times \log E_{\rm iso,52}$ & 0.90 & $< 10^{-4}$ & 0.75 \\
\hline
$E_{\rm iso}(\Gamma_0)$ & $\log E_{\rm iso,52} = (-4.10 \pm 0.33) + (2.15 \pm 0.13) \times \log \Gamma_0$ & 0.77 & $< 10^{-4}$ & 1.04 \\
$T_{\rm 90}(\Gamma_0)$ & $\log T_{\rm 90} = (2.20 \pm 0.30) + (-0.44 \pm 0.12) \times \log \Gamma_0$ & -0.26 & 0.001 & 0.97 \\
\hline
\end{tabular*}
\end{center}
\label{tab6}
\end{table*}

\begin{table*}
\begin{center}
\caption{The results of regression analysis for correlations in the rest frame and comoving frame after eliminating the redshit evolution, in which $r$ is the correlation coefficient, $p$ is the chance probability, and $\sigma$ is the dispersion}
\begin{tabular*}{\linewidth}{lp{9.5cm}p{2.1cm}p{2.1cm}r}
\hline
\hline
Relations & Expressions & $r$ & $p$ & $\sigma$ \\
\hline
$E_{\rm p,c}(E_{\rm iso,c})$ & $\log E_{\rm p,c} = (2.18 \pm 0.02) + (0.30 \pm 0.03) \times \log E_{\rm iso,c,52}$ & 0.62 & $< 10^{-4}$ & 0.56 \\
$E_{\rm p,c}(L_{\rm iso,c})$ & $\log E_{\rm p,c} = (2.48 \pm 0.03) + (0.34 \pm 0.04) \times \log L_{\rm iso,c,52}$ & 0.58 & $< 10^{-4}$ & 0.58 \\
$T_{\rm 90}(E_{\rm iso,c})$ & $\log T_{\rm 90} = (1.08 \pm 0.04) + (0.21 \pm 0.05) \times \log E_{\rm iso,c,52}$ & 0.31 & $< 10^{-4}$ & 0.95 \\
$T_{\rm 90,c}(L_{\rm iso,c})$ & $\log T_{\rm 90,c} = (1.13 \pm 0.05) + (-0.01 \pm 0.06) \times \log L_{\rm iso,c,52}$ & -0.02 & 0.839 & 1.00 \\
$L_{\rm iso,c}(E_{\rm iso,c})$ & $\log L_{\rm iso,c,52} = (-0.81 \pm 0.03) + (0.68 \pm 0.03) \times \log E_{\rm iso,c,52}$ & 0.83 & $< 10^{-4}$ & 0.68 \\
\hline
$E_{\rm iso,c}(\Gamma_0)$ & $\log E_{\rm iso,c,52} = (-3.33 \pm 0.37) + (1.49 \pm 0.15) \times \log \Gamma_0$ & 0.59 & $< 10^{-4}$ & 1.19 \\
$L_{\rm iso,c}(\Gamma_0)$ & $\log L_{\rm iso,c,52} = (-4.44 \pm 0.25) + (1.58 \pm 0.10) \times \log \Gamma_0$ & 0.76 & $< 10^{-4}$ & 0.79 \\
\hline
\end{tabular*}
\end{center}
\label{tab7}
\end{table*}

\end{document}